\documentclass[11pt]{article}

\usepackage[utf8]{inputenc}
\usepackage[T1]{fontenc}
\usepackage[english]{babel}
\usepackage{lmodern}
\usepackage{xcolor}

\usepackage[colorlinks=False]{hyperref}  % https://en.wikibooks.org/wiki/LaTeX/Hyperlinks

\usepackage{comment}

\usepackage{amsmath}  % use [intlimits]?
\usepackage{amsfonts,dsfont}
\usepackage{amssymb}
\usepackage{amsthm}
\usepackage{mathtools}
\usepackage{thmtools}
\usepackage{bm} % bold math symbols, especially greek

\numberwithin{equation}{section} % equation number: (s.n) where s is section number, n is counter

\newlength{\bredde}
\def\slash#1{\settowidth{\bredde}{$#1$}\ifmmode\,\raisebox{.15ex}{/}
\hspace*{-\bredde} #1\else$\,\raisebox{.15ex}{/}\hspace*{-\bredde} #1$\fi}
\renewcommand{\epsilon}{\varepsilon} % replace the ugly epsilon with the nice epsilon
\newcommand{\be}{\begin{equation}}
\newcommand{\ee}{\end{equation}}

\newcommand{\Tr}{{\rm Tr}\,}

\usepackage{extpfeil}
\newextarrow{\xrightrightarrows}{{5}{8}{0}{0}}
{\bigRelbar\bigRelbar{\bigtwoarrowsleft\rightarrow\rightarrow}}

\DeclareMathOperator{\erf}{erf} % error function
\DeclareMathOperator{\erfc}{erfc} % complementary error function
\DeclareMathOperator{\erfi}{erfi} % error function in imaginary direction (erfi(z) := erf(i z) / i)
\DeclareMathOperator{\Pf}{Pf} % pfaffian of a skew-symmetric matrix
\declaretheorem[numberwithin=section]{proposition}
\declaretheorem[numberlike=proposition]{theorem}

\makeatletter
\newcommand*{\doublerightarrow}[2]{\mathrel{
  \settowidth{\@tempdima}{$\scriptstyle#1$}
  \settowidth{\@tempdimb}{$\scriptstyle#2$}
  \ifdim\@tempdimb>\@tempdima \@tempdima=\@tempdimb\fi
  \mathop{\vcenter{
    \offinterlineskip\ialign{\hbox to\dimexpr\@tempdima+1em{##}\cr
    \rightarrowfill\cr\noalign{\kern.5ex}
    \rightarrowfill\cr}}}\limits^{\!#1}_{\!#2}}}
\makeatother

\begin{document}
\title{
Complex symmetric, self-dual, and  Ginibre  random matrices: Analytical results for three classes of bulk and edge statistics 
}
\author{{\sc Gernot Akemann$^{1,2}$\footnote{akemann@physik.uni-bielefeld.de}, Noah Ayg\"un$^{1,3}$\footnote{noah.ayguen@uni-bielefeld.de}, Mario Kieburg$^{1,4}$\footnote{corresponding author: m.kieburg@unimelb.edu.au}, 
Patricia P\"a{\ss}ler$^{1}$\footnote{patricia@physik.uni-bielefeld.de}}\\~\\
$^1$Faculty of Physics, Bielefeld University, PO-Box 100131, D-33501 Bielefeld, Germany\\
$^2$School of Mathematics, University of Bristol, Fry Building, Woodland Road, Bristol,\\ BS8 1UG, United Kingdom\\
$^3$Department of Mathematical Sciences, 
Seoul National University, Seoul 151-747,\\ Republic of Korea\\
$^4$School of Mathematics and Statistics, University of Melbourne, 813 Swanston Street,\\ Parkville, Melbourne VIC 3010, Australia
}
%\keywords{}

\date{}

\maketitle

\begin{abstract}

Recently, a conjecture about the local bulk statistics of complex eigenvalues has been made based on numerics. It claims that there are only three universality classes, which have all been observed in open chaotic quantum systems. Motivated by these new insights, we compute and compare the expectation values of $k$ pairs of complex conjugate characteristic polynomials in three ensembles of Gaussian non-Hermitian random matrices representative for the three classes: the well-known complex Ginibre ensemble, complex symmetric and complex self-dual matrices. In the Cartan classification scheme of non-Hermitian random matrices they are labelled as class A, AI$^\dag$ and AII$^\dag$, respectively. Using the technique of Grassmann variables, we derive explicit expressions for a single pair of expected characteristic polynomials for finite as well as infinite matrix dimension. For the latter we consider the global limit as well as zoom into the edge and the bulk of the spectrum, providing new analytical results for classes AI$^\dag$ and AII$^\dag$. For general $k$, we derive the effective Lagrangians corresponding to the non-linear $\sigma$-models in the respective physical systems. Interestingly, they agree for all three ensembles, while the corresponding Goldstone manifolds, over which one has to perform the remaining integrations, are different and equal the three classical compact groups in the bulk. In particular, our analytical results show that these three ensembles have indeed different local bulk and edge spectral statistics, corroborating the conjecture further.

\end{abstract}

%\tableofcontents
%%%%%%%%%%%%%%%%%%%%%%%%%%%%%%%%%%%%%%%%%%%%
\section{Introduction}\label{intro}

It is one of the goals of statistical mechanics to derive simple, analytical models that describe universal aspects of physical systems in an approximative way. One class of such models is given by ensembles of random matrices. Their application in physics is two-fold. Firstly, they may model physical operators such as the Hamilton and the Dirac operator. Secondly, they typically belong to the wider class of static Coulomb gases with logarithmic interaction. From this point of view and at specific values of the inverse temperature $\beta$, these models can be interpreted as statistical ensembles of eigenvalues of matrices. 
In the present work, we will focus on random matrices with complex eigenvalues, thus relating to a two-dimensional Coulomb gas. A prominent example is the Ginibre ensemble~\cite{Ginibre} with complex matrix elements, corresponding to $\beta=2$.

The Ginibre ensemble, non-Hermitian random matrices and Coulomb gases in general 
have seen a recent revival of interest in physics and mathematics, see recent reviews~\cite{Non-Herm-Phys,PeterSungsoo,Sylvia}. One of the reasons is that they describe open quantum mechanical systems in the regime when they become chaotic. This goes back to the work of Grobe, Haake and Sommers~\cite{GHS}, generalising the quantum chaos conjecture for closed systems relating to Hermitian random matrices in the chaotic regime, and Poisson statistic for generic integrable quantum systems. 
Such and similar applications of non-Hermitian random matrices include the kicked top with damping~\cite{GHS,Pandey}, Hamiltonian systems with imaginary vector potential~\cite{Efetov2}, quantum field theory with quark chemical potential~\cite{Tilo,Misha, James,Tilo2} as well as with lattice corrections~\cite{DSV10,Kie12,KVZ13}, quantum spin chains with dissipative disorder~\cite{Hamazaki,Sa,Rubio,ABC}, quantum maps~\cite{BCSZ09,AKMP,SRCP20} and time-lagged correlation matrices~\cite{TB08,NT17}.

A crucial question to be addressed before applying random matrices is the identification of the corresponding symmetry class of the physical system under consideration. Only then the benchmarks obtained from random matrix theory can be expected to be an approximation. In the Hermitian case with real spectra, this has been facilitated through the symmetry classification by  Altland and Zirnbauer~\cite{AZ} in the so-called ten-fold way, utilising the Cartan classification from semi-simple Lie algebras. As the authors emphasise, this is not a classification of the possible limiting eigenvalue statistics in the limit of large matrix size, but rather the classes of possible $\sigma$-models and their effective Lagrangians. For example, it turns out that for Hermitian random matrices in the bulk of the spectrum generically only three different bulk statistics result, that coincide with Dyson's earlier classification, the three-fold way~\cite{Dyson}.
In contrast, at the hard edge more limiting statistics emerge, for references and an overview over all 10 classes in the context of quantum field theory and their naive lattice discretisations see~\cite{MarioTim}.

What is the situation for non-Hermitian random matrices? Based on a symmetry classification of non-Hermitian random matrices by Bernard and LeClair~\cite{BLC}, a matrix representation 
was proposed by Magnea~\cite{Ulrika}. Recently, the classification~\cite{BLC} was revisited by Kawabata et al.~\cite{Kawabata} through the lens of topological properties, finding a total of 38 symmetry classes of non-Hermitian random matrices\footnote{Cf. the revised version v2 of \cite{BLC} that was submitted after \cite{Kawabata} appeared.}. Much less is known to date about the complex eigenvalue statistics of these 38 classes. Only for 7 classes the joint probability density of the eigenvalues of the corresponding Gaussian ensembles is known for finite matrix size $N$, the three Ginibre ensembles with real, complex and quaternion matrix entries~\cite{Ginibre,LS}, their chiral counterparts~\cite{James,A05,APS} and the pseudo-Hermitian matrix ensemble (also known as the Wilson-Dirac random matrix in QCD)~\cite{KVZ13}. All 7 ensembles have been solved for finite-$N$ and in various large-$N$ limits, as they represent determinantal and Pfaffian point processes, see~\cite{APh,KVZ13} for a list of the corresponding kernels in these 7 classes.

It came as a surprise that only 3 different kinds of local bulk statistics have been identified in~\cite{Hamazaki} amongst 9 of the 38 classes. This analysis was based on numerical computations of the nearest-neighbour spacing distribution of complex eigenvalues in radial distance away from the edge and symmetry axes and points (such as the real axis and the origin) and on $N=2$ computations. The latter are actually not good approximations to the large-$N$ limit~\cite{AMP}\footnote{This is in stark contrast to the Wigner surmise for  
Hermitian ensembles, as pointed out for class A 
already 
in~\cite{GHS}.}. The authors of~\cite{Hamazaki} identify a transposition symmetry to be responsible for the deformation of the bulk statistics in a subset of 9 classes. They also show numerically that these 3 bulk statistics are shared by random matrices with Gaussian and Bernoulli distributed elements and, hence, show universal characteristics.  
The difference between these 3 universality classes has furthermore been observed in numerical computations of their respective complex spacing ratios \cite{Sa,Tilo2} introduced in \cite{Sa}.

Due to the insights of~\cite{Hamazaki,AMP}, there is a conjecture that only three universality classes of the bulk statistics for complex eigenvalues exist.
Partial analytical results exist about an agreement for the limiting eigenvalue statistics within different non-Hermitian ensembles. Tao and Vu~\cite{TaoVu} have proven for a large class of complex Wigner ensembles that the local statistics in the bulk and at the edge agree with the complex Ginibre ensemble. It was also shown for the real and quaternion Ginibre ensemble that away from the real axis their limiting eigenvalue statistics agree with that of the simpler complex Ginibre ensemble in the bulk~\cite{BS,AKMP} and at the edge~\cite{Rider} of the spectrum (while they differ along the real line).
Likewise, the complex Ginibre ensemble and its chiral counterpart where shown to share the same limiting statistics at the edge~\cite{ABender}. Hence, at least 6 of the 38 classes follow the local spectral statistics of the complex Ginibre ensemble in the bulk and at least 4 at the edge of the spectrum.

In the present work, we analytically confirm the conjecture~\cite{Hamazaki,AMP} that two symmetry classes,  
complex symmetric matrices (class AI$^\dag$) and the $2N$-dimensional complex self-dual random matrices (class AII$^\dag$), do {\it not} follow the local statistics of the complex Ginibre ensemble (class A). These ensembles are all Gaussian to keep the analysis simple. 
For the complex Ginibre ensemble the nearest-neighbour spacing distribution in the bulk (as well as all other complex eigenvalue correlation functions) is well known,  
see~\cite{GHS}. However, there are no analytical results known so far for the symmetry classes AI$^\dag$ and AII$^\dag$
in the large-$N$ limit, 
except for the 
spectral density at weak non-Hermiticity for class AI$^\dag$ in~\cite{SFT}. 
It is given in terms of a triple integral.  

It is the goal of this article to start filling this gap, by providing analytical results,  
in particular 
in the global limit, local bulk and local edge scaling limit for classes  
AI$^\dag$ and AII$^\dag$. We will show that in all these limits the difference amongst the 3 classes prevails. Due to the lack of an explicit expression for the joint probability density of the complex eigenvalues for the two classes AI$^\dag$ and AII$^\dag$, we have chosen to study expectation values of products of characteristic polynomials to demonstrate their differences. For those quantities, a direct computation based on their matrix representation is feasible with the help of supersymmetric techniques, see~\cite{Wegner} for an introduction of this method. 
An alternative approach uses diffusion equations \cite{Liu}, which has been applied to the 9 classes considered in \cite{Hamazaki}, including classes AI$^\dag$ and AII$^\dag$. In~\cite{Liu}, a duality was derived expressing the average over $k$ pairs of characteristic polynomials with matrix size $N$ by matrix averages of dimension $k$, without taking the large-$N$ limit though. We will recover these expressions using supersymmetry.

Characteristic polynomials are useful quantities in mathematics and physics, and we refer to~\cite{Brezin-Hikami} for a review when studying Hermitian and unitary ensembles. In non-Hermitian ensembles characteristic polynomials have been employed in the replica method~\cite{NK,Kanzieper,KimJac}, offering access to the effective Lagrangian. For finite matrix size, general expressions exist for products~\cite{AV03} and ratios~\cite{AP,MB2,determinant} of characteristic polynomials in general determinantal ensembles, which have been generalised to Pfaffian ensembles for products in~\cite{ABa,Pfaffian,AKP10,APS,Kie12,Simm2}. 
Such relations are important for algebraic aspects of  the corresponding determinantal and Pfaffian point processes, providing representations for the respective (skew-)orthogonal polynomials and their kernels, see~\cite{Heine,Szego}. 
However,  such direct relations do not (yet) exist for products of characteristic polynomials in classes AI$^\dag$ and AII$^\dag$.
At large-$N$ and away from the real line it was shown for products that the real and symplectic Ginibre ensemble agree~\cite{Simm2} with the much simpler complex Ginibre ensemble~\cite{Simm1}. 
The universality of products of characteristic polynomials has been explored for independent matrix entries using supersymmetric~\cite{FKS} and probabilistic methods~\cite{Afanasiev}.

Our findings presented in the ensuing sections are as follows. In Section~\ref{results-N}, we introduce the ensembles and our notation. The main result in this section is Proposition~\ref{Prop_char_exp} about the analytical expressions for a single pair of characteristic polynomials for finite matrix size $N$ which are proven in Subsection~\ref{results-k1N}. They are given most explicitly 
in terms of sums of truncated exponentials in all three ensembles. Section~\ref{asymptot} is devoted to several large-$N$ limits, also for $k=1$. We begin with the global limit, which is compared to the rescaled quantities from Subsection~\ref{results-k1N}, plotted at various values of $N$. 
To set the stage for the local scaling limits, we briefly recall the limiting results for the complex Ginibre ensemble class A in the bulk and at the edge of the spectrum. Our intention is to highlight the intimate relation between the expected pair of characteristic polynomials and the kernel as well as the spectral density, which allows to make such an identification. 
In Subsection~\ref{edge-lim},  we derive new analytical expressions in class AI$^\dag$ and AII$^\dag$ for the edge of the spectrum which are summarised in Theorem~\ref{Thm_lim_char_exp}. 
The local bulk limit is covered in Proposition~\ref{Prop_lim_char_exp} in Subsection~\ref{bulk-lim}. While both results follow easily from finite matrix dimension $N$ in class A and AI$^\dag$, class AII$^\dag$ is more involved, requiring a detailed analysis. 
A differential equation for its integral representation is presented in Appendix~\ref{App:fNdiff}.  

In Section~\ref{k>1}, we consider expectation values of products of 
$k>1$ pairs of characteristic polynomials.
Both limiting expressions when zooming into in the bulk (Theorem~\ref{thm:bulk}) and at the edge~(Theorem~\ref{thm:edge}) are new and can be found in Subsections~\ref{bulkk} and~\ref{edgek}, respectively. The former is given in terms of  Itzykson--Zuber integrals~\cite{IZ}, which are compact group integrals  over the unitary, unitary-symplectic, and orthogonal group of dimensions $k$, respectively $2k$. At the edge, the resulting group integrals of the same dimensions become non-compact, over complex, quaternion and real valued matrices.  In class A these group integrals can be performed and a comparison based on previous results on orthogonal polynomials is made in Appendix~\ref{App:Ak>1}. 
Our conclusions and open questions are presented in Section~\ref{conclusio}.

%%%%%%%%%%%%%%%%%%%%%%%%%%%%%%%%%%%%%%%%%%%%
\section{Finite-$N$ results for a single pair of characteristic polynomials}\label{results-N}

\subsection{Definition of the ensembles of random matrices of class A, AI$^\dag$ and AII$^\dag$}\label{def-As}

To get an idea of our techniques and results as well as have very explicitly formulas, we consider the expectation value of a product of two characteristic polynomials. One characteristic polynomial depends on the non-Hermitian random matrix $J$ and the other one on its complex conjugate $J^*$, i.e., the main object will be
\begin{equation}
D_N(z,w^*):=\langle \det[z\,\mathbf{1}_{d_N}-J]\det[w^*\,\mathbf{1}_{d_N}-J^*]\rangle , 
\label{def-char-pol}
\end{equation}
with $z,w\in\mathbb{C}$ fixed, $\mathbf{1}_{d_N}$ is the identity matrix of matrix dimension $d_N$, and $J$ is drawn from a Gaussian probability density
\begin{equation}
P(J)=\frac{1}{Z_N}\exp[-\Tr (JJ^\dag)],
\end{equation}
where the partition function $Z_N$ (normalisation constant) is specified below. The latter certainly strongly depends on from which non-Hermitian matrix space $J$ is taken. The average over an observable $\mathcal{O}(J)$ is denoted for all ensembles by 
\begin{equation}
\langle \mathcal{O}(J)\rangle
:= \frac{1}{Z_N} 
\int[dJ]\,\mathcal{O}(J)
\exp[-\Tr (JJ^\dag)] ,
\label{vevdef}
\end{equation}
where $[dJ]$ denotes the flat Lebesgue measure 
over all independent complex matrix elements. We underline that the measure $[dJ]$ looks different for the three ensembles when writing it out explicitly, as the possible symmetries relate the matrix entries. This leads to $N^2$ complex independent elements for class $A$, $N(N+1)/2$ for class AI$^\dag$  and $2N^2-N$ for class AII$^\dag$. The measure of a complex number $z = x + iy$ is chosen as $d^2z = dxdy$.  The Hermitian conjugate of $J$ is denoted by $J^\dagger$.

 The first and  simplest ensemble, we study, has been introduced by Ginibre~\cite{Ginibre}. It is given by $N\times N$ random matrices with independent identically normal distributed complex  matrix elements, $J_{ij}\in\mathcal{N}_{\mathbb{C}}(0,1)$ for $i,j=1,\ldots,N$. We adapt the notation of~\cite{AZ,Kawabata,Non-Herm-Phys} and label it with the Cartan class notation A. Its partition function reads 
\begin{equation}
Z_N^{\rm A} :=\int[dJ]\exp[-\Tr (JJ^\dag)] =\prod_{i,j=1}^N\int d^2J_{ij}\exp[-|J_{ij}|^2] = \pi^{N^2}.
\label{ZA}
\end{equation} 

The second ensemble we consider consists of complex symmetric $N\times N$ matrices $J=J^T$. As in the complex Ginibre ensemble, the requirement of  invariance under the group action $J\to UJU^T$ for any unitary matrix $U\in{\rm U}(N)$ fixes the weight up to a scaling, when assuming that all independent matrix elements are Gaussian. This however also implies that the diagonal and off-diagonal matrix elements have a different variance. We label this ensemble with AI$^\dag$, see~\cite{AZ,Kawabata,Non-Herm-Phys}, and its partition function is  
\begin{equation}
Z_N^{\rm AI^\dag} :=\!\int[dJ]\exp[-\Tr (JJ^\dag)] =
\prod_{l=1}^N\int d^2J_{ll}\exp[-|J_{ll}|^2] 
\prod_{1\leq i<j\leq N}\int d^2J_{ij}\exp[-2|J_{ij}|^2] = \pi^{N}\!\left( \frac{\pi}{2}\right)^{\frac{N(N-1)}{2}}\!\!\!\!. 
\label{ZCS}
\end{equation} 

The third ensemble we consider is that of a complex self-dual $2N\times 2N$ random matrix $J$. In a $2N$-dimensional complex representation it consists of 3 complex non-Hermitian matrices $A,B,C$ of size $N\times N$, where the latter two are anti-symmetric, $B^T=-B$, $C^T=-C$:
\begin{equation}
J=\left(
\begin{array}{cc}
A & B\\
C& A^T\\
\end{array}
\right),\quad \mbox{with}\ 
J^\dag=\left(
\begin{array}{cc}
A^\dag & -C^*\\
-B^*& A^*\\
\end{array}
\right).
\label{Jqs-def}
\end{equation}
 Choosing the following embedding of the second Pauli matrix 
\begin{equation}
\Sigma_y:=\left(
\begin{array}{cc}
0 & -i \mathbf{1}_N\\
i\mathbf{1}_N& 0\\
\end{array}
\right),\quad \Sigma_y^{-1}=\Sigma_y\ ,
\label{Sigmay}
\end{equation}
we have the following symmetry on $J$ 
\begin{equation}
\Sigma_y J^T\Sigma_y=J.
\label{QS}
\end{equation}
This symmetry is called self-duality, and we follow~\cite{AZ,Kawabata,Non-Herm-Phys} by labelling the ensemble with AII$^\dag$. The corresponding partition function, thus, reads 
\begin{eqnarray}
Z_N^{\rm AII^\dag} &:=&\int[dJ]\exp[-\Tr (JJ^\dag)] 
\label{ZQS}\\
&=&\prod_{i,j=1}^N\int d^2A_{ij}\exp[-2|A_{ij}|^2] 
\prod_{1\leq n<l\leq N}\int d^2B_{nl}d^2C_{nl}\exp[-2|B_{nl}|^2-2|C_{nl}|^2] 
= \left( \frac{\pi}{2}\right)^{2N^2-N}\nonumber
\end{eqnarray}

It is not difficult to see from the definition of the expectation value \eqref{vevdef}, that for $zw^*\to\infty$ the right hand side of 
\eqref{def-char-pol} asymptotically becomes $\sim(zw^*)^N$ for the Ginibre and AI$^\dag$ ensembles (matrix dimension is $d_N=N$) while it will be $\sim(zw^*)^{2N}$ for the AII$^\dag$ ensemble because the matrices have dimension $d_N=2N$. 

In order to take the large-$N$ limit later, we will adjust the normalisation, where we divide~\eqref{def-char-pol} by its value at vanishing arguments,
\begin{equation}
D_N(0,0)=\langle \det[J]\det[J^*] \rangle .
\label{norm-char-pol}
\end{equation}
It is some natural choice of a reference point, despite that it will not cancel all $N$-dependent terms.

\subsection{Grassmann variables and special functions}\label{preliminaries}

Below we will employ anticommuting complex  Grassmann variables $\xi,\psi$, i.e., $\{\xi,\psi\}=\xi\psi+\psi\xi=0$, in order to write determinants as integrals.
We follow the conventions of Haake~\cite{Haake}, and we refer to~\cite{Wegner} for an introduction into superalgebra and superanalysis.
Complex conjugation  is chosen to satisfy the following relations
$(\xi^*)^*=-\xi$ for single and $(\xi\psi)^*=\xi^*\psi^*$ for a product of Grassmann variables. Complex conjugated variables anticommute equally,   $\{\xi,\xi^*\}= \{\xi,^*\psi\}=0$. The integration convention of each Grassmann variable is chosen to be
\begin{equation}
\int d\xi =0\ , \quad \int d\xi\ \xi=1.
\label{Grassdef}
\end{equation}
For a complex Grassmann vector of size $N$, $\xi^T=(\xi_1,\ldots,\xi_N)$ we, then, obtain for the determinant of an arbitrary $N\times N$ matrix $C$
\begin{equation}
\det[C ]=
\int [d\xi]\exp[\xi^\dag C\xi] .
\label{Grassdet}
\end{equation}
The integration measure is defined as $[d\xi]=\prod_{i=1}^N d\xi_id\xi_i^*$. 
A second identity is needed for the Pfaffian determinant of an arbitrary even dimensional antisymmetric matrix $D$ of size $2N\times 2N$, 
\begin{equation}
\Pf[D]=(-1)^{N(N+1)/2}
\int[d\xi]\exp\left[\frac12(\xi^T,\xi^{*\, T}) D
\left(\!
\begin{array}{c}
\xi \\
\xi^* \\
\end{array}
\!\right)
\right].
\label{GrassPf}
\end{equation}

Finally, we would like to introduce some special functions that we will encounter repeatedly.
It is well-known~\cite{Ginibre} that the truncated exponential
\begin{equation}
E_N(x):= \sum_{j=0}^N\frac{x^j}{j!}=e^x \frac{\Gamma(N+1,x)}{\Gamma(N+1)}= e^x Q(N+1,x)\ ,
\label{def-eN}
\end{equation}
 plays an important role in the complex Ginibre ensemble (class A). It actually has its appearance for the other two ensembles studied by us.
The truncated exponential can be rewritten in terms of the upper incomplete Gamma-function~\cite[8.4.10]{NIST}
\begin{equation}
\Gamma(a,x):= \int_x^\infty dt t^{a-1}e^{-t}\ ,
\label{def-Gax}
\end{equation}
as well as its normalised version
\begin{equation}
Q(a,x):= \frac{\Gamma(a,x)}{\Gamma(a)}\ .
\label{def-Q}
\end{equation}

At the edge of the complex Ginibre ensemble, 
one encounters the complementary error function
\begin{equation}
\erfc(x)= \frac{2}{\sqrt{\pi}}\int_x^\infty dt e^{-t^2} ,
\label{erfc-def}
\end{equation}
whose relation to  the error function $\erf$ is 
\begin{equation}
\erfc(y)=1-\erf(y)=1+\erf(-y)=2-\erfc(-y) ,
\label{erfc-rel}
\end{equation}
and to the imaginary error function
\begin{equation}
\label{erfi}
\erfi(z)=-i\erf(iz)=\frac{2}{\sqrt{\pi}}\int_0^z ds\,e^{s^2}.
\end{equation}
Those functions can be also found in the case of complex symmetric matrices (AI$^\dag$) while for the case of complex self-dual matrices (AII$^\dag$) we 
encounter Owen's $T$-function \cite{OwenT} defined as follows,
\begin{equation}
T(x,a):=\frac{1}{2\pi}\int_0^a dt \frac{e^{-\frac12 x^2(1+t^2)}}{1+t^2}, \quad x,a\in\mathbb{R}.
\label{OwenT-def}
\end{equation}
We actually need the function analytically continued  to the imaginary axis in its second argument,
\begin{equation}
\label{OTi}
T\left(x,ib\right)=\frac{i}{2\pi}\int_0^b du\frac{e^{-\frac12 x^2(1-u^2)}}{1-u^2} , \quad x\in\mathbb{R}\quad{\rm and}\quad b\in(-1,1).
\end{equation}
When going beyond the specified integral one needs to be aware of the emerging branch cuts on $b\in\mathbb{R}\setminus(-1,1)$. We will not run into this problem, though, as we avoid those cuts.

We will mention some properties of all these special functions where necessary in the computations.

%%%%%%%%%%%%%%%%%%%%%%%%%%%%%%%%%%%%%%%
\subsection{Statement and derivation for a pair of characteristic polynomials}\label{results-k1N}

We are now ready to state our first results for the expectation value of one pair of characteristic polynomials at finite matrix dimension $N$ for all three ensembles.

\begin{proposition}
[One pair of expected characteristic polynomials for finite-$N$]
\label{Prop_char_exp}\

Employing 
definitions~\eqref{def-char-pol} and~\eqref{norm-char-pol} as well as~\eqref{def-eN} and~\eqref{def-Q}, it holds for any finite $N\geq 1$
\begin{eqnarray}
\frac{D_N^{\rm A}(z,w^*)}{D_N^{\rm A}(0,0)}
&=&  E_N(zw^*)=e^{zw^*} Q(N+1,zw^*),
\label{DN-A}\\
\frac{D_N^{\rm AI^\dag}(z,w^*)}{D_N^{\rm AI^\dag}(0,0)}&=& E_N(2zw^*)
-\frac{2zw^*}{N+1}E_{N-1}(2zw^*)\nonumber\\
&=& e^{2zw^*}\left(Q(N+1,2zw^*)-\frac{2zw^*}{N+1}Q(N,2zw^*)\right),
\label{DN-CS}\\
\frac{D_N^{\rm AII^\dag}(z,w^*)}{D_N^{\rm AII^\dag}(0,0)}&=& 
\frac{N!}{(2N)!}\sum_{j=0}^N\frac{(2j)!}{j!}(4zw^*)^{N-j}E_{2j}(2zw^*)
\nonumber\\
&=&e^{2zw^*} \frac{N!}{(2N)!}\sum_{j=0}^N\frac{(2j)!}{j!}(4zw^*)^{N-j}Q(2j+1,2zw^*).
\label{DN-QS}
\end{eqnarray}
\end{proposition}

The combination of truncated exponentials in~\eqref{DN-CS} for AI$^\dagger$ is precisely the one appearing in the computation of the eigenvector statistics for class A, see~\cite[Eq.~(2.16)]{ATTZ}. This may be a total coincidence, but it is worth mentioning.

The expectation \eqref{DN-A} is well-known and only a special case of a more general theorem in \cite{AV03}. 
It was derived using the fact that the complex Ginibre ensemble (class A) is a determinantal point process, and planar orthogonal polynomials can be utilised, cf., eq.~\eqref{K-DN-A} on the relation to the kernel.
For pedagogical reasons we will also derive~\eqref{DN-A} in class A using the supersymmetry method that generalises to the two other ensembles naturally. 

We also mention in eqs.~\eqref{DN-A} and~\eqref{DN-CS} the expression in terms of truncated exponentials~\eqref{def-eN} which become useful when taking the large-$N$ limit in the bulk of the spectrum, see Subsection~\ref{bulk-lim}. The second form of these expectation values in terms of the normalised incomplete Gamma-function~\eqref{def-Q} will prove useful in the edge scaling limit, shown in Subsection~\ref{edge-lim}. Last but not least there exists a double integral representation~\eqref{fNx-int2} together with~\eqref{QS-int} for the result~\eqref{DN-QS} of class AII$^\dag$, which is needed to derive the corresponding large-$N$ limits at the edge and in the bulk.

{\bf Class A:} We begin with deriving~\eqref{DN-A} for the complex Ginibre ensemble. Using~\eqref{Grassdet}, the expectation value~\eqref{def-char-pol} can be written as 
\begin{eqnarray}
D_N^{\rm A}(z,w^*)=\int\frac{[dJ]}{\pi^{N^2}}\int[d\xi][d\psi]
\exp\left[ \sum_{j,l=1}^N\left(-|J_{jl}|^2+\xi_j^*(z\delta_{jl}-J_{jl})\xi_l+\psi_j^*(w^*\delta_{jl}-J_{jl}^*)\psi_l\right)\right].
\label{Gin-G1}
\end{eqnarray}
Splitting up $J_{jl}=J^1_{jl}+iJ^2_{jl}$ into its real and imaginary part, the $J$-dependent part of the exponent can be written as 
\begin{eqnarray}
&&-\sum_{j,l=1}^N\left((J^1_{jl})^2+ J^1_{jl}(\xi^*_j\xi_l+\psi_j^*\psi_l)+ 
(J^2_{jl})^2+ iJ^2_{jl}(\xi^*_j\xi_l-\psi_j^*\psi_l)\right)
\nonumber\\
=&&-\sum_{j,l=1}^N\left(\left(J^1_{jl}+\frac12 (\xi^*_j\xi_l+\psi_j^*\psi_l)\right)^2
-\frac14 (\xi^*_j\xi_l+\psi_j^*\psi_l)^2\right)\nonumber\\
&&-\sum_{j,l=1}^N\left(
\left(J^2_{jl}+ \frac{i}{2}(\xi^*_j\xi_l-\psi_j^*\psi_l)\right)^2
+\frac14(\xi^*_j\xi_l-\psi_j^*\psi_l)^2\right).
\label{Gin-G2}
\end{eqnarray}
As the product of two Grassmann variables is again a commuting but nilpotent variable, we may shift the real and imaginary parts of the matrix elements $J_{jl}$ individually and integrate them out, to obtain 
\begin{eqnarray}
D_N^{\rm A}(z,w^*)&=&\int[d\xi][d\psi]
\exp\left[ \sum_{j,l=1}^N(\xi_j^*z\delta_{jl}\xi_l+\psi_j^*w^*\delta_{jl}\psi_l
+\xi^*_j\xi_l\psi_j^*\psi_l)
\right]
\nonumber\\
&=&\int[d\xi][d\psi]\frac{1}{\pi}\int d^2a
\exp\left[ -|a|^2+\sum_{j=1}^N(\xi_j^*z\xi_j+\psi^*_jw^*\psi_j+ia\xi_j^*\psi_j^*+ia^*\xi_j\psi_j)
\right].\nonumber\\
\label{Gin-G3}
\end{eqnarray}
In the first step, we multiplied out the squares of Grassmann variables. In order to do the integrals, we  linearised the quartic term in the Grassmann variables  with an auxiliary variable $a\in\mathbb{C}$  integrated over the complex  plane, i.e.,
\begin{equation}
1=\frac{1}{\pi}\int d^2a\exp\left[-\left(a-i\sum_{l=1}^N\xi_l\psi_l\right)\left(a^*-i\sum_{j=1}^N\xi^*_j\psi^*_j\right)\right].
\label{aux-int}
\end{equation}
The single sum in the exponent in~\eqref{Gin-G3} is the scalar product
\begin{equation}
\label{Gin-matrix}
\sum_{j=1}^N(\xi_j^*z\xi_j+\psi^*_jw^*\psi_j+ia\xi_j^*\psi_j^*+ia^*\xi_j\psi_j)=(\xi^\dag,\psi^T)
\left(
\begin{array}{cc}
z\mathbf{1}_N & ia \mathbf{1}_N\\
-ia^*\mathbf{1}_N& -w^*\mathbf{1}_N\\
\end{array}
\right)
\left(\!
\begin{array}{c}
\xi \\
\psi^* \\
\end{array}
\!\right).
\end{equation}
It is not difficult to see that in the integral \eqref{Grassdet} an exchange of variables $\xi\leftrightarrow\xi^*$ leads to a Jacobian of $(-1)^N$. This change of variables will put the scalar product in \eqref{Gin-matrix} into the form that \eqref{Grassdet} can be applied, in this case for a $2N\times 2N$ matrix. This leads to the result\footnote{Because the integrals over Grassmann variables are in the end always integrals over polynomials in these variables, they commute with the Gaussian integral(s).}
\begin{eqnarray}
D_N^{\rm A}(z,w^*)&=&\frac{(-1)^N}{\pi}\int d^2a\ e^{-|a|^2}
\det\left[
\begin{array}{cc}
z\mathbf{1}_N & ia \mathbf{1}_N\\
-ia^*\mathbf{1}_N& -w^*\mathbf{1}_N\\
\end{array}
\right] = \frac{1}{\pi}\int d^2a \exp[-|a|^2]\left[zw^*+ |a|^2 \right]^N
\nonumber\\
&=&\exp[zw^*]\int_{zw^*}^\infty dt\exp[-t]t^N = e^{zw^*} \Gamma(N+1,zw^*) = N!\ E_N(zw^*)\ .
\end{eqnarray}
Here, we changed to polar coordinates $a=\sqrt{t} e^{i\phi}$ with $t$ the squared radius of $a$ and exploited~\eqref{def-eN} and~\eqref{def-Gax}. 
Because the upper incomplete Gamma-function~\eqref{def-Gax}  is analytic in its second argument we are allowed to employ it for complex $zw^*$, as well.

The normalisation can be readily obtained from this result, 
\begin{equation}
\label{DNA00}
D_N^{\rm A}(0,0)=N!
\end{equation}
finishing the proof of~\eqref{DN-A} for the complex Ginibre ensemble (class A).

%%%%%%%%%%%%%%%%%%%%%%%%%%%%%%%
%%%%AI^+

{\bf Class AI$^\dag$:} Next, we show~\eqref{DN-CS} for the  ensemble of complex symmetric matrices. The average of the determinants can be expressed as in \eqref{Gin-G1}, however, we now have to take into account the symmetry $J_{jl}=J_{lj}$,
\begin{eqnarray}
D_N^{\rm AI^\dag}(z,w^*)&=&\int\frac{[dJ]}{Z_N^{\rm AI^\dag}}\int[d\xi][d\psi]
\exp\left[ \sum_{j=1}^N(-|J_{jj}|^2+\xi_j^*(z-J_{jj})\xi_j+\psi_j^*(w^*-J_{jj}^*)\psi_j)\right]
\nonumber\\
&&\quad\quad\times\exp\left[ -\sum_{1\leq j<l\leq N}2(|J_{jl}|^2+J_{jl}(\xi_j^*\xi_l+\xi_l^*\xi_j) +J^*_{jl}(\psi_j^*\psi_l+\psi_l^*\psi_j))\right].
\label{CS-G1}
\end{eqnarray}
Splitting again the matrix entries into real and imaginary part, $J_{jl}=J^1_{jl}+iJ^2_{jl}$, the diagonal matrix elements in the exponent become,
\begin{equation}
\sum_{j=1}^N\left(-\left(J^1_{jj}+\frac12(\xi_j^*\xi_j+\psi^*_j\psi_j)\right)^2+\frac14 (\xi_j^*\xi_j+\psi^*_j\psi_j)^2
-\left(J^2_{jj}+\frac{i}{2}(\xi_j^*\xi_j-\psi^*_j\psi_j)\right)^2-\frac14 (\xi_j^*\xi_j-\psi^*_j\psi_j)^2\right),
\label{CS-G2}
\end{equation}
while the upper triangular matrix elements take the form
\begin{eqnarray}
\sum_{1\leq n<l\leq N}\left(-2\left(J^1_{nl}+\frac14(\xi_n^*\xi_l+\xi_l^*\xi_n+\psi_n^*\psi_l+\psi_l^*\psi_n)\right)^2+\frac18 (\xi_n^*\xi_l+\xi_l^*\xi_n+\psi_n^*\psi_l+\psi_l^*\psi_n)^2\right.
\nonumber\\
\quad\quad\   \left.-2\left(J^2_{nl}+\frac{i}{4}(\xi_n^*\xi_l+\xi_l^*\xi_n-\psi_n^*\psi_l-\psi_l^*\psi_n)\right)^2-\frac18 (\xi_n^*\xi_l+\xi_l^*\xi_n-\psi_n^*\psi_l-\psi_l^*\psi_n)^2\right).
\label{CS-G3}
\end{eqnarray}
Integrating out the shifted Gaussian matrix elements and multiplying out the quartic Grassmann terms, we obtain for~\eqref{CS-G1} after some simplifications
\begin{eqnarray}
D_N^{\rm AI^\dag}(z,w^*)&=&\int[d\xi][d\psi]\ 
e^{\sum_{j=1}^N( \xi_j^*z\xi_j+\psi_j^*w^*\psi_j+\xi^*_j\xi_j\psi^*_j\psi_j)
+\sum_{1\leq n<l\leq N}\frac12 (\xi_n^*\xi_l+\xi_l^*\xi_n)(\psi_n^*\psi_l+\psi_l^*\psi_n)
}
\nonumber\\
&=&\int[d\xi][d\psi]
\exp\!\left[ \sum_{j=1}^N (\xi_j^*z\xi_j+\psi_j^*w^*\psi_j) 
+\frac12 \sum_{n,l=1}^N(\xi^*_n\psi_n \xi_l\psi^*_l-\xi^*_n\psi^*_n\xi_l\psi_l)
\right].
\label{CS-G4}
\end{eqnarray}
In the second step, we have combined diagonal and upper triangular sums to obtain two fully factorised sums over all Grasmann variables, using their antisymmetry. 

In the next step, we apply~\eqref{aux-int} twice, though with different choices of factors, and obtain
\begin{eqnarray}
D_N^{\rm AI^\dag}(z,w^*)&=&\int[d\xi][d\psi]
\int \frac{d^2a\, d^2b}{\pi^2}\exp\left[-|a|^2-|b|^2+\sum_{j=1}^N
(\xi_j^*z\xi_j+\psi_j^*w^*\psi_j)\right]
\nonumber\\
&&\quad\quad\quad\quad\times 
\exp\left[\frac{1}{\sqrt{2}}\sum_{j=1}^N\left(i(a\xi_j\psi_j+a^*\xi_j^*\psi^*_j)+b\xi_j\psi^*_j+b^*\xi_j^*\psi_j\right)
\right].
\label{CD-G5}
\end{eqnarray}
In this form we can identify the scalar product
\begin{equation}
\label{CS-matrix}
\begin{split}
&\sum_{j=1}^N
(\xi_j^*z\xi_j+\psi_j^*w^*\psi_j)+\frac{1}{\sqrt{2}}\sum_{j=1}^N\left(i(a\xi_j\psi_j+a^*\xi_j^*\psi^*_j)+b\xi_j\psi^*_j+b^*\xi_j^*\psi_j\right)\\
=&
(\xi^\dag,\xi^T,\psi^\dag,\psi^T)
\left(
\begin{array}{cccc}
0&\frac{1}{2}z\mathbf{1}_N&  \frac{i}{2\sqrt{2}}a^* \mathbf{1}_N&  \frac{1}{2\sqrt{2}}b^* \mathbf{1}_N\\
-\frac{1}{2}z\mathbf{1}_N& 0& \frac{1}{2\sqrt{2}}b \mathbf{1}_N&  \frac{i}{2\sqrt{2}}a \mathbf{1}_N\\
\frac{-i}{2\sqrt{2}}a^* \mathbf{1}_N&  \frac{-1}{2\sqrt{2}}b \mathbf{1}_N&0&\frac{1}{2}w^*\mathbf{1}_N  \\
\frac{-1}{2\sqrt{2}}b^* \mathbf{1}_N&  \frac{-i}{2\sqrt{2}}a \mathbf{1}_N&-\frac{1}{2}w^*\mathbf{1}_N& 0 \\
\end{array}
\right)
\left(\!
\begin{array}{c}
\xi^* \\
\xi\\
\psi^* \\
\psi\\
\end{array}
\!\right)
\end{split}
\end{equation}
This allows us to recognise and exploit the identity~\eqref{GrassPf} once we have changed $\xi\leftrightarrow
 \xi^*$ and $\psi\leftrightarrow\psi^*$ so that
 we arrive at
\begin{equation}
D_N^{\rm AI^\dag}(z,w^*)=
\int \frac{d^2a\, d^2b}{\pi^2}\exp\left[-|a|^2-|b|^2\right]\Pf\left[
\begin{array}{cccc}
0&{z}\mathbf{1}_N&  \frac{i}{\sqrt{2}}a^* \mathbf{1}_N&  \frac{1}{\sqrt{2}}b^* \mathbf{1}_N\\
-{z}\mathbf{1}_N& 0& \frac{1}{\sqrt{2}}b \mathbf{1}_N&  \frac{i}{\sqrt{2}}a \mathbf{1}_N\\
\frac{-i}{\sqrt{2}}a^* \mathbf{1}_N&  \frac{-1}{\sqrt{2}}b \mathbf{1}_N&0&{w^*}\mathbf{1}_N  \\
\frac{-1}{\sqrt{2}}b^* \mathbf{1}_N&  \frac{-i}{\sqrt{2}} a\mathbf{1}_N&-{w^*}\mathbf{1}_N& 0 \\
\end{array}
\right].
\label{CS-G6}
\end{equation}
This duality formula with $k=1$ pair of integration variables over a Pfaffian determinant of size $4N$ was independently derived in \cite{Liu} using a diffusion equation rather than supersymmetry. Here, we further simplify the duality formula \eqref{CS-G6}. 
Using the fact that the Pfaffian determinant of an antisymmetric matrix $G$ relates to the determinant as $\Pf[G]=\sqrt{\det[G]}$, as well as the evaluation of determinants with block matrices $A,B,C,D$ ($A$ must be invertible) via Schur complement,
\begin{equation}
\det\left[
\begin{array}{cc}
A& B\\
C& D\\
\end{array}
\right]=\det\left[AD-ACA^{-1}B\right], 
\label{det-Id}
\end{equation}
we can compute the Pfaffian in~\eqref{CS-G6} which is
\begin{equation}
\det\left[
\begin{array}{cc}
-\frac12(2zw^*+|a|^2+|b|^2)\mathbf{1}_N& 0\\
0& -\frac12(2zw^*+|a|^2+|b|^2)\mathbf{1}_N\\
\end{array}
\right]^{\frac12}=\frac{1}{2^N}(2zw^*+|a|^2+|b|^2)^N.
\end{equation}
This leads to the following double integral expression for the expected characteristic polynomials
\begin{equation}
D_N^{\rm AI^\dag}(z,w^*)=
\int \frac{d^2a\, d^2b}{\pi^2}\exp\left[-|a|^2-|b|^2\right]\frac{1}{2^N}(2zw^*+|a|^2+|b|^2)^N
=\frac{1}{2^N}\int_0^\infty dt\ e^{-t}t(2zw^*+t)^N.
\end{equation}
In the second equality we have employed spherical coordinates with $t=|a|^2+|b|^2$ the squared radius. Once we shift $t\to t-2zw^*$ we can anew identify the truncated exponential~\eqref{def-eN} and \eqref{def-Gax} so that
\begin{eqnarray}
D_N^{\rm AI^\dag}(z,w^*)=\frac{N!}{2^N}\left( 
(N+1)E_N(2zw^*)-2zw^*E_{N-1}(2zw^*)
\right).
\end{eqnarray}
This result also implies the normalisation
\begin{equation}
\label{DNAI00}
D_N^{\rm AI^\dag}(0,0)= \frac{N!}{2^N}(N+1)\ ,
\end{equation}
finishing the derivation of~\eqref{DN-CS} for complex symmetric matrices class AI$^\dag$, 
after inserting~\eqref{def-eN}. 

%%%%%%%%%%%%%%%%%%%%%%%%%%%%%%%%%%%%%

{\bf Class AII$^\dag$:} Finally, we show~\eqref{DN-QS} for complex self-dual  matrices. Because of the doubling of the dimension, we introduce two sets of $N$-dimensional Grassmann vectors $\xi_1^T=(\xi_{11},\ldots,\xi_{1N})$ and $\xi_2^T=(\xi_{21},\ldots,\xi_{2N})$ and likewise the  vectors $\psi_1^T$ and $\psi_2^T$. Applying the relation \eqref{Grassdet} and observing the block-structure of the complex self-dual matrix $J$ in \eqref{Jqs-def} we thus obtain
\begin{eqnarray}
D_N^{\rm AII^\dag}(z,w^*)&=&\int\!\frac{[dA][dB][dC]}{Z_N^{\rm AII^\dag}}\!\int[d\xi_1][d\xi_2][d\psi_1][d\psi_2]
\nonumber\\
&&\times e^{\sum_{i,j=1}^N[-2|A_{ij}|^2+\xi_{1i}^*(z\delta_{ij}-A_{ij})\xi_{1j}+\xi_{2i}^*(z\delta_{ij}-A_{ji})\xi_{2j}+\psi_{1i}^*(w^*\delta_{ij}-A_{ij}^*)\psi_{1j}+\psi_{2i}^*(w^*\delta_{ij}-A_{ji}^*)\psi_{2j}]}\nonumber\\
&&\times e^{-\sum_{1\leq l<n\leq N}[2|B_{nl}|^2+(\xi_{1n}^*\xi_{2l}-\xi_{1l}^*\xi_{2n})B_{nl}+(\psi_{1n}^*\psi_{2l}-\psi_{1l}^*\psi_{2n})B_{nl}^*]}
\nonumber\\
&&\times e^{-\sum_{1\leq l<n\leq N}[2|C_{nl}|^2+(\xi_{2n}^*\xi_{1l}-\xi_{2l}^*\xi_{1n})C_{nl}+(\psi_{2n}^*\psi_{1l}-\psi_{2l}^*\psi_{1n})C_{nl}^*]}.
\label{QS-G1}
\end{eqnarray}
Splitting again the matrix elements $A_{nl}=A^1_{nl}+iA^2_{nl}$ into real and imaginary part, we can write the exponent of all $A$-dependent parts as follows,
\begin{eqnarray}
&&-2\sum_{n,l=1}^N\left((A^1_{nl})^2+(A^2_{nl})^2 +\frac12 \Xi_{nl}^+A^1_{nl}+i\frac12 \Xi_{nl}^-A^2_{nl}\right)
\nonumber\\
&=& -2\sum_{n,l=1}^N\left(\left( A^1_{nl}+\frac14 \Xi_{nl}^+\right)^2- \frac{1}{16}(\Xi_{nl}^+)^2
+\left( A^2_{nl}+i\frac14 \Xi_{nl}^-\right)^2+ \frac{1}{16}(\Xi_{nl}^-)^2
\right),
\label{QS-A}
\end{eqnarray}
where we have defined two bilinears of Grassmann variables
\begin{eqnarray}
\Xi_{nl}^+:= \xi_{1n}^*\xi_{1l}+\xi_{2l}^*\xi_{2n}+\psi_{1n}^*\psi_{1l}+\psi_{2l}^*\psi_{2n}\ ,
\quad\Xi_{nl}^-:=\xi_{1n}^*\xi_{1l}+\xi_{2l}^*\xi_{2n}-(\psi_{1n}^*\psi_{1l}+\psi_{2l}^*\psi_{2n}).\quad
\label{Xi-def}
\end{eqnarray} 
After integrating out all matrix elements of $A$, we find anew quartic terms in the Grassmann variables which are
\begin{eqnarray}
&&\frac18\sum_{n,l=1}^N\left( (\Xi_{nl}^+)^2-(\Xi_{nl}^-)^2\right) 
=\frac12\sum_{n,l=1}^N(\xi_{1n}^*\xi_{1l}+\xi_{2l}^*\xi_{2n})(\psi_{1n}^*\psi_{1l}+\psi_{2l}^*\psi_{2n})\nonumber\\
&=&\frac12\sum_{n,l=1}^N(-\xi_{1n}^*\psi_{1n}^*\xi_{1l}\psi_{1l}+\xi_{1n}^*\psi_{2n}\xi_{1l}\psi_{2l}^*+\xi_{2n}\psi_{1n}^*\xi_{2l}^*\psi_{1l}-\xi_{2n}\psi_{2n}\xi_{2l}^*\psi_{2l}^*).
\label{QS-A4}
\end{eqnarray}
Likewise, all terms depending on the matrix elements of $B_{nl}=B^1_{nl}+iB^2_{nl}$ in \eqref{QS-G1} read
\begin{equation}
-2\sum_{1\leq l<n\leq N}\left(\left( B^1_{nl}+\frac14 \Psi_{nl}^+\right)^2- \frac{1}{16}(\Psi_{nl}^+)^2
+\left( B^2_{nl}+i\frac14 \Psi_{nl}^-\right)^2+ \frac{1}{16}(\Psi_{nl}^-)^2
\right),
\label{QS-B}
\end{equation}
with 
\begin{eqnarray}
\Psi_{nl}^+:=\xi_{1n}^*\xi_{2l}-\xi_{1l}^*\xi_{2n}+\psi_{1n}^*\psi_{2l}-\psi_{1l}^*\psi_{2n},\quad
\Psi_{nl}^-:=\xi_{1n}^*\xi_{2l}-\xi_{1l}^*\xi_{2n}-(\psi_{1n}^*\psi_{2l}-\psi_{1l}^*\psi_{2n}) .\quad
\label{Psi-def}
\end{eqnarray} 
The integrals over all matrix elements of $B$, thus, yield 
\begin{equation}
\frac18\sum_{1\leq l<n\leq N}\left( (\Psi_{nl}^+)^2-(\Psi_{nl}^-)^2\right) =
\frac12\sum_{1\leq l<n\leq N}(\xi_{1n}^*\xi_{2l}-\xi_{1l}^*\xi_{2n})(\psi_{1n}^*\psi_{2l}-\psi_{1l}^*\psi_{2n}).
\label{QS-B4}
\end{equation}
Similarly, the terms depending on the matrix elements of 
$C_{nl}=C^1_{nl}+iC^2_{nl}$ in~\eqref{QS-G1} give
\begin{equation}
-2\sum_{1\leq l<n\leq N}\left(\left( C^1_{nl}+\frac14 \Phi_{nl}^+\right)^2- \frac{1}{16}(\Phi_{nl}^+)^2
+\left( C^2_{nl}+i\frac14 \Phi_{nl}^-\right)^2+ \frac{1}{16}(\Phi_{nl}^-)^2
\right),
\label{QS-C}
\end{equation}
with 
\begin{eqnarray}
\Phi_{nl}^+:= \xi_{2n}^*\xi_{1l}-\xi_{2l}^*\xi_{1n}+\psi_{2n}^*\psi_{1l}-\psi_{2l}^*\psi_{1n} ,
\quad\Phi_{nl}^-:=\xi_{2n}^*\xi_{1l}-\xi_{2l}^*\xi_{1n}-(\psi_{2n}^*\psi_{1l}-\psi_{2l}^*\psi_{1n}) ,\quad
\label{Phi-def}
\end{eqnarray} 
which leads to
\begin{equation}
\frac18\sum_{1\leq l<n\leq N}\left( (\Phi_{nl}^+)^2-(\Phi_{nl}^-)^2\right) =
\frac12\sum_{1\leq l<n\leq N}(\xi_{2n}^*\xi_{1l}-\xi_{2l}^*\xi_{1n})(\psi_{2n}^*\psi_{1l}-\psi_{2l}^*\psi_{1n}) ,
\label{QS-C4}
\end{equation}
after the integration over all matrix elements of $C$.

The sum of eqs.~\eqref{QS-B4} and~\eqref{QS-C4} can be combined into
\begin{equation}
\frac12\sum_{\substack{1\leq n, l\leq N\\ n\neq l}}(-\xi_{1n}^*\psi_{1n}^*\xi_{2l}\psi_{2l}-\xi_{1n}^*\psi_{2n}\xi_{2l}\psi_{1l}^*- \xi_{2n}^*\psi_{2n}^*\xi_{1l}\psi_{1l}
-\xi_{2n}^*\psi_{1n}\xi_{1l}\psi_{2l}^*
) ,
\label{QS-BC4}
\end{equation}
 once we relabel indices.
Because at $n=l$ all terms cancel as the Grassmann variables anti-commute, we can remove the condition $n\neq l$ in the sum. 

Adding all quartic terms of Grassmann variables together,~\eqref{QS-A4} and~\eqref{QS-BC4}, we can write the exponential of the double sum in the following factorised form
\begin{eqnarray}
&&\!\!\!\exp\left[-\frac12\sum_{n,l=1}^N\left((\xi_{1n}^*\psi_{1n}^*+\xi_{2n}^*\psi_{2n}^*)(\xi_{1l}\psi_{1l}+\xi_{2l}\psi_{2l})
+(\xi_{1n}^*\psi_{2n}-\xi_{2n}^*\psi_{1n})(\xi_{2l}\psi_{1l}^*-\xi_{1l}\psi_{2l}^*)\right)\right]
\nonumber\\
&=&\int\frac{d^2p}{\pi}\exp\left[-|p|^2+\frac{i}{\sqrt{2}}p^*\sum_{n=1}^N(\xi_{1n}^*\psi_{1n}^*+\xi_{2n}^*\psi_{2n}^*)
+\frac{i}{\sqrt{2}}p\sum_{l=1}^N(\xi_{1l}\psi_{1l}+\xi_{2l}\psi_{2l})
\right]
\nonumber\\
&&\times \int\frac{d^2q}{\pi}\exp\left[-|q|^2+\frac{i}{\sqrt{2}}q^*\sum_{n=1}^N
(\xi_{1n}^*\psi_{2n}-\xi_{2n}^*\psi_{1n})
+\frac{i}{\sqrt{2}}q\sum_{l=1}^N(\xi_{2l}\psi_{1l}^*-\xi_{1l}\psi_{2l}^*)
\right].
\label{pq-int}
\end{eqnarray}
In the second step, we have again exploited the integral~\eqref{aux-int} to obtain bilinear Grassmann terms. These bilinear terms can be combined with the $z$- and $w^*$-dependent  terms in~\eqref{QS-G1} to identify the scalar product
\begin{eqnarray}
\label{QS-matrix}
\left(\!
\begin{array}{c}
\xi_1^* \\
\xi_1\\
\xi_2^* \\
\xi_2\\
\psi_1^* \\
\psi_1\\
\psi_2^* \\
\psi_2\\
\end{array}
\!\right)^T
\left(
\begin{array}{cccccccc}
0&\frac{z}{2}\mathbf{1}_N&0&0&  \frac{ip^*}{2\sqrt{2}} \mathbf{1}_N& 0&0& \frac{iq^*}{2\sqrt{2}} \mathbf{1}_N\\
-\frac{z}{2}\mathbf{1}_N&0&0&0& 0& \frac{ip}{2\sqrt{2}}p \mathbf{1}_N&\frac{-iq}{2\sqrt{2}} \mathbf{1}_N&0\\
0&0&0&\frac{z}{2}\mathbf{1}_N&0&\frac{-iq^*}{2\sqrt{2}}\mathbf{1}_N&\frac{ip^*}{2\sqrt{2}} \mathbf{1}_N&0\\
0&0&-\frac{z}{2}\mathbf{1}_N&0&  \frac{iq}{2\sqrt{2}} \mathbf{1}_N& 0&0& \frac{ip}{2\sqrt{2}} \mathbf{1}_N\\
\frac{-ip^*}{2\sqrt{2}} \mathbf{1}_N&0&0&\frac{-iq}{2\sqrt{2}} \mathbf{1}_N&0&\frac{w^*}{2}\mathbf{1}_N&0&0\\
0&\frac{-ip}{2\sqrt{2}} \mathbf{1}_N&\frac{iq^*}{2\sqrt{2}} \mathbf{1}_N&0&-\frac{w^*}{2}\mathbf{1}_N&0&0&0\\
0&\frac{iq}{2\sqrt{2}} \mathbf{1}_N&\frac{-ip^*}{2\sqrt{2}} \mathbf{1}_N&0&0&0&0&\frac{w^*}{2}\mathbf{1}_N\\
\frac{-iq^*}{2\sqrt{2}}\mathbf{1}_N&0&0&\frac{-ip}{2\sqrt{2}} \mathbf{1}_N&0&0&-\frac{w^*}{2}\mathbf{1}_N&0\\
\end{array}
\right)
\left(\!
\begin{array}{c}
\xi_1^* \\
\xi_1\\
\xi_2^* \\
\xi_2\\
\psi_1^* \\
\psi_1\\
\psi_2^* \\
\psi_2\\
\end{array}
\!\right).\nonumber\\
\end{eqnarray}
The Grassmann integration can thus be performed using~\eqref{GrassPf} to arrive at
\begin{eqnarray}
&&D_N^{\rm AII^\dag}(z,w^*)=
\int \frac{d^2p\, d^2q}{\pi^2}\exp\left[-|p|^2-|q|^2\right]
\nonumber\\
&&\times \Pf
\left[
\begin{array}{cccccccc}
0&z\mathbf{1}_N&0&0&  \frac{i}{\sqrt{2}}p^* \mathbf{1}_N& 0&0& \frac{i}{\sqrt{2}}q^* \mathbf{1}_N\\
-z\mathbf{1}_N&0&0&0& 0& \frac{i}{\sqrt{2}}p \mathbf{1}_N&\frac{-i}{\sqrt{2}}q \mathbf{1}_N&0\\
0&0&0&z\mathbf{1}_N&0&\frac{-i}{\sqrt{2}}q^* \mathbf{1}_N&\frac{i}{\sqrt{2}}p^* \mathbf{1}_N&0\\
0&0&-z\mathbf{1}_N&0&  \frac{i}{\sqrt{2}}q \mathbf{1}_N& 0&0& \frac{i}{\sqrt{2}}p \mathbf{1}_N\\
\frac{-i}{\sqrt{2}}p^* \mathbf{1}_N&0&0&\frac{-i}{\sqrt{2}}q \mathbf{1}_N&0&w^*\mathbf{1}_N&0&0\\
0&\frac{-i}{\sqrt{2}}p \mathbf{1}_N&\frac{i}{\sqrt{2}}q^* \mathbf{1}_N&0&-w^*\mathbf{1}_N&0&0&0\\
0&\frac{i}{\sqrt{2}}q \mathbf{1}_N&\frac{-i}{\sqrt{2}}p^* \mathbf{1}_N&0&0&0&0&w^*\mathbf{1}_N\\
\frac{-i}{\sqrt{2}}q^* \mathbf{1}_N&0&0&\frac{-i}{\sqrt{2}}p \mathbf{1}_N&0&0&-w^*\mathbf{1}_N&0\\
\end{array}
\right].\qquad
\end{eqnarray}
This duality formula for class AII$^\dag$ was also derived in \cite{Liu}.
Once again we can exploit the block structure of the matrix and apply \eqref{det-Id}, to obtain for the Pfaffian determinant
\begin{equation}
\begin{split}
&\det\!\left[\!
\begin{array}{cccc}
-\left[zw^*+\frac{|p|^2+|q|^2}{2}\right]\mathbf{1}_N &0&0& -pq^*\mathbf{1}_N\\
0& -\left[zw^*+\frac{|p|^2+|q|^2}{2}\right]\mathbf{1}_N&p^*q\mathbf{1}_N&0\\
0&pq^*\mathbf{1}_N&-\left[zw^*+\frac{|p|^2+|q|^2}{2}\right]\mathbf{1}_N&0\\
-p^*q\mathbf{1}_N&0&0&-\left[zw^*+\frac{|p|^2+|q|^2}{2}\right]\mathbf{1}_N\\
\end{array}\!\!
\right]^{\frac12}\\
&=\left(\left(zw^*+\frac{|p|^2+|q|^2}{2}\right)^2-|p|^2|q|^2
\right)^N\\
&=\frac{1}{4^N}\left((2zw^*)^2+4zw^*(|p|^2+|q|^2)+(|p|^2-|q|^2)^2
\right)^N.
\end{split}
\label{QS-Pfresult}
\end{equation}
Thence, we are left with computing the following integral
\begin{eqnarray}
D_N^{\rm AII^\dag}(z,w^*)&=&
\int \frac{d^2p\, d^2q}{\pi^2}\exp\left[-|p|^2-|q|^2\right]\frac{1}{4^N}\left((2zw^*)^2+4zw^*(|p|^2+|q|^2)+(|p|^2-|q|^2)^2\right)^N
\nonumber\\
&=&\frac{1}{4^N}f_N(x)\ , \quad \mbox{with} \quad x=zw^*\quad \mbox{and}
\label{QS-int}\\
f_N(x)&:=&
\int_0^\infty dr\int_0^\infty ds\exp\left[-r-s\right]\left(4x^2+4x(r+s)+(r-s)^2\right)^N,
\label{fNx-def}
\end{eqnarray}
after changing to polar coordinates for $q$ and $p$.
 
 It is useful to introduce so-called light cone variables at this stage, the sum and difference of the integration variables,
\begin{eqnarray}
v:=r+s,\quad u:=r-s,\quad \mbox{with}\ v\geq0,\ u\in[-v,v].
\label{lightcone}
\end{eqnarray}
The Jacobian is $dr\,ds=\frac12 dv\,du$ so that we  get, due to Fubini's theorem,
\begin{eqnarray}
f_N(x)
&=& \frac12 \int_0^\infty dv\ e^{-v}\int_{-v}^vdu
\left(4x^2+4xv+u^2\right)^N.
\label{fNx-int1}
\end{eqnarray}
We note in passing that this integral satisfies 
an inhomogenous first order differential equation as is explained in Appendix~\ref{App:fNdiff}.

The integral can be further simplified as follows. Using the evenness of the integrand with respect to variable $u\to-u$,
\begin{eqnarray}
f_N(x)
&=& \int_0^\infty dv\int_0^vdu\ e^{-v}
\left(4x^2+4xv+u^2\right)^N\nonumber\\
&=& \int_0^\infty du\int_u^\infty dv \ e^{-v}
\left(4x^2+4xv+u^2\right)^N
\nonumber\\
&=& \int_0^\infty du\int_0^\infty dv \ e^{-v-u}
\left(4xv+(2x+u)^2\right)^N.
\label{fNx-int2}
\end{eqnarray}
In the second equality we reordered the integrals and in the third one we shifted $v\to v'=v-u$.

This expression~\eqref{fNx-int2} leads to a simple summation formula at finite-$N$, as well as to the large-$N$ asymptotic limits performed in the next section. From a simple expansion of the binomial we obtain 
\begin{eqnarray}
f_N(x)
&=& \int_0^\infty du\int_0^\infty dv \ e^{-v-u}\sum_{j=0}^N\frac{N!}{j!(N-j)!}(4xv)^{N-j}(2x+u)^{2j}
\nonumber\\
&=& \int_{2x}^\infty du' \ e^{-u'+2x} \sum_{j=0}^N\frac{N!}{j!}(4x)^{N-j}(u')^{2j}
\nonumber\\
 &=&e^{2x} \sum_{j=0}^N\frac{N!}{j!}(4x)^{N-j}\,\Gamma(2j+1,2x),
\label{fNx-sum}
\end{eqnarray}
after shifting again $u'=2x+u$ and using the definition of the incomplete Gamma-function~\eqref{def-Gax}. 

At $x=zw^*=0$ we only have a contribution from $j=N$, and hence with $\Gamma(2j+1,0)=(2j)!$ we obtain
\begin{equation}
f_N(0)=(2N)!
\end{equation}
and, thus, together with~\eqref{QS-int}
\begin{equation}
\label{DNAII00}
D_N^{\rm AII^\dag}(0,0)=\frac{(2N)!}{4^N}.
\end{equation}
Inserting this into the definition on the left hand side of~\eqref{DN-QS} together with~\eqref{def-Q} and~\eqref{def-eN}, this proves both right hand sides of~\eqref{DN-QS}.

%%%%%%%%%%%%%%%%%%%%%%%%%%%%%%%%%%%%%%%%%%%%
%%%%%%%%%%%%%%%%%%%%%%%%%%%%%%%%%%%%%%%%%%%%
\section{Asymptotic global and local large-$N$ limits for a single pair}\label{asymptot}

%%%%%%%%%%%%%%%%%%%%%%%%%%%%%%%%%%
\subsection{Scaling with $N$ and global limit for a single pair}\label{plot}

In this subsection, we start exploring the possible large-$N$ limits.
First, we present plots for the rescaled characteristic polynomials  at different values of $N$ over the entire range of the spectrum, helping us in identifying different scaling regimes. In particular, we present the global limit for all three classes, before zooming into the local edge and bulk regime in the subsequent subsections.  To deal with those in a unifying way, it is useful to introduce the dimension
\begin{equation}\label{def.dim}
d_N=N\quad{\rm for}\quad {\rm A}\ {\rm and}\ {\rm AI}^\dag\qquad{\rm and}\qquad d_N=2N\quad{\rm for}\quad {\rm AII}^\dag\ ,
\end{equation}
and the scaling
\begin{equation}\label{def.scale}
s=1\quad{\rm for}\quad {\rm A} \qquad{\rm and}\qquad s=\frac{1}{\sqrt{2}}\quad{\rm for}\quad {\rm AI}^\dag\ {\rm and}\  {\rm AII}^\dag.
\end{equation}
The scaling $s$ properly normalises the limiting support of the expectation value~\eqref{def-char-pol} as well as the eigenvalues to the unit disc for all three ensembles. The latter we have checked numerically  which also implies that the local mean level spacing of the eigenvalues is normalised to the same distance so that the local statistics become comparable.

Our results for the global limit can be summarised as follows.

\begin{proposition}
[Global limit for a pair of characteristic polynomials]\label{Prop_global}\

Let $r\geq0$. For the rescaled characteristic polynomials 
\begin{equation}
\label{F-def}
F_N(r):=\left.\exp[-d_N zw^*] \frac{D_N\left(\sqrt{d_N}sz,\sqrt{d_N}s w^*\right)}{D_N(0,0)}\right|_{zw^*=r^2},
\end{equation}
cf., Proposition~\ref{Prop_char_exp}, we obtain the following global limits 
\begin{eqnarray}
D_{\rm global}^{\rm A}(r)&:=&\lim_{N\to\infty}F_N^{\rm A}(r)\ \ \ =\Theta(1-r)\ ,\nonumber\\
D_{\rm global}^{\rm AI^\dag}(r)&:=&\lim_{N\to\infty}F_N^{\rm AI^\dag}(r)\ =(1-r^2)\Theta(1-r)\ ,
\nonumber\\
D_{\rm global}^{\rm AII^\dag}(r)&:=&\lim_{N\to\infty}F_N^{\rm AII^\dag}(r)
=\frac{1}{1-r^2}\Theta(1-r)\ ,
\label{Dgobal-all}
\end{eqnarray}
where $\Theta(1-r)$ is the Heaviside step-function restricting the support to the interval $[0,1]$.
\end{proposition}

\begin{figure}[t!]
	\centering
	\includegraphics[width=\textwidth]{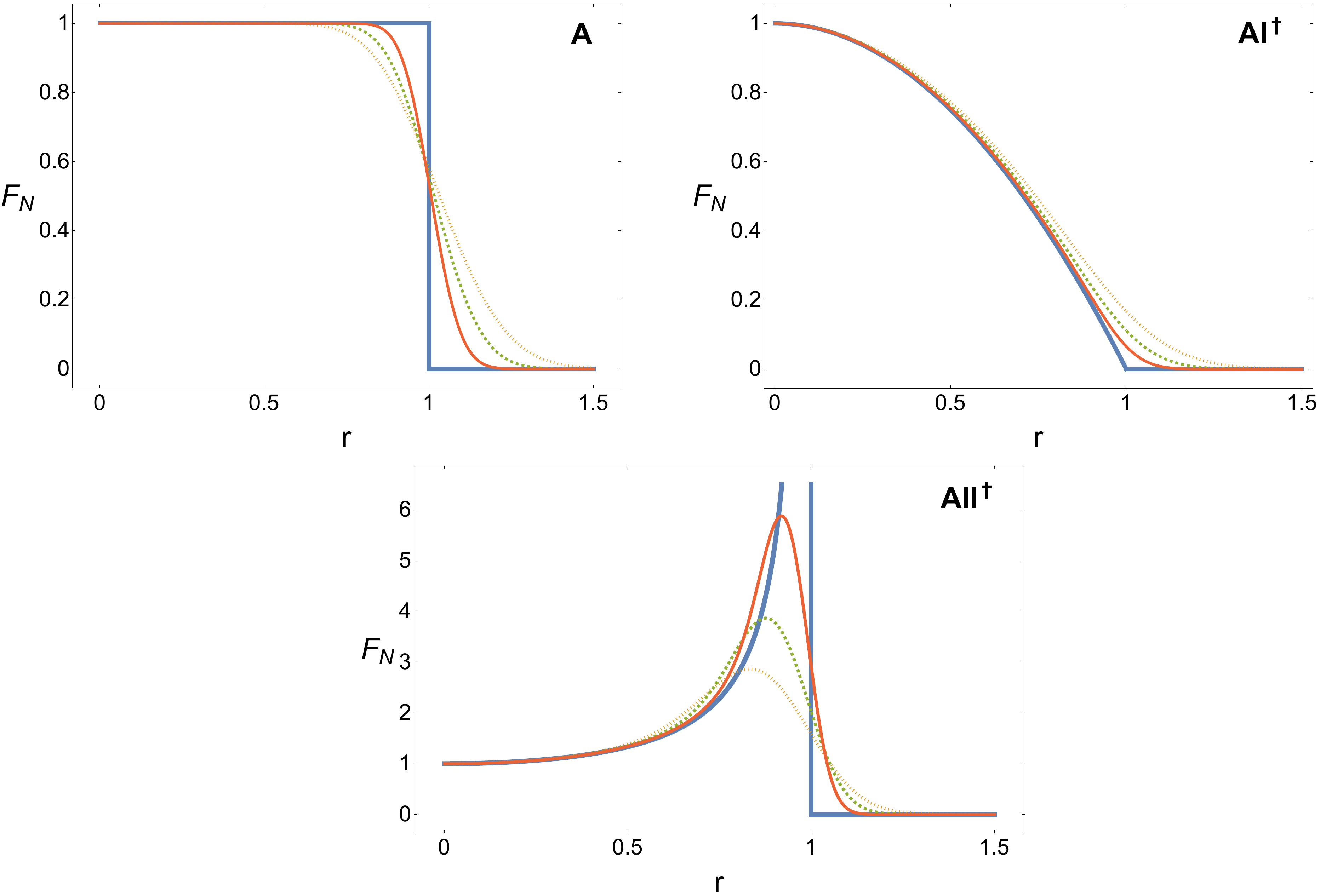}
	\caption{Plots of the rescaled characteristic polynomials for finite-$N$ as a function of $r=zw^*\geq0$ for $N=10,20,50$ (dotted orange, dashed green, solid red, respectively) and the global limit $N\to\infty$ (thick blue). 
	{\bf Top left:} rescaled density of the complex Ginibre ensemble~\eqref{Gin-NF} (class A);
	{\bf Top right:} rescaled characteristic polynomials for the complex symmetric ensemble~\eqref{CS-NF} (class AI$^\dagger$);
	 {\bf Bottom:} rescaled characteristic polynomials for the complex self-dual ensemble~\eqref{QS-NF}  (class AII$^\dagger$).
	 The position of the edge of the global limit is rescaled to unity.}
	\label{Fig-rescaled}
\end{figure}

{\bf Class A:} Anew, we begin with the derivation for the complex Ginibre ensemble, where it is well known that it converges to the circular law, since $F_N^{\rm A}(r)$ is up to a normalisation factor the global spectral density of the eigenvalues in the complex plane when $z=w$, see~\cite{Ginibre}. The rescaled quantity~\eqref{F-def} can be rewritten in terms of the ratio~\eqref{def-Q} due to~\eqref{DN-A}, i.e.,
\begin{equation}
F_N^{\rm A}(r)=Q(N+1,Nr^2).
\label{Gin-NF}
\end{equation}
The exponential factor in \eqref{DN-A} is removed, as it is well known that the remaining normalised incomplete Gamma-function has a finite limit. 
The global limit of this function
is obtained from the convergence of the normalised lower incomplete Gamma-function~\cite[8.11.6 \& 8.11.7]{NIST}
\begin{equation}
D_{\rm global}^{\rm A}(r)=\Theta(1-r).
\label{Dglobal-Gin}
\end{equation}
The corrections are exponentially suppressed except when $r\approx 1$ where it becomes algebraic due to~\cite[8.11.12]{NIST}.
Later, in the local edge scaling limit the step of the function $F_N^{\rm A}(r)$ at $r=1$ will be studied at a finer resolution, cf., eqs.~\eqref{EN-P-relation} and~\eqref{P-lim}. 
The results~\eqref{Gin-NF} and~\eqref{Dglobal-Gin} are plotted in Fig.~\ref{Fig-rescaled} (top left plot) for $N=10,20,50$ and in the limit $N\to\infty$.

{\bf Class AI$^\dag$:}  We turn to the derivation for complex symmetric matrices. Due to the very similar form, cf., eqs.~\eqref{DN-A} and~\eqref{DN-CS}, the computation goes along the same lines. We would like to emphasise that the joint probability density of the eigenvalues for this ensemble is unknown, and no relation between the expected characteristic polynomials and the eigenvalue density can be expected a priori. The global spectral density is expected to satisfy the circular law with radius $\sqrt{2}$ without introducing the scaling $s$. 

To have a similar situation as in the complex Ginibre ensemble, we also remove the exponential prefactor in~\eqref{DN-CS}. Once we use the result~\eqref{DN-CS} in~\eqref{F-def}, we consider
\begin{eqnarray}
F_N^{\rm AI^\dag}(r)= Q(N+1,Nr^2)-\frac{Nr^2}{N+1}Q(N,Nr^2),
\label{CS-NF}
\end{eqnarray}
 As for class $A$, it is clear from the asymptotic of
 the incomplete Gamma-function~\cite[8.11.6 \& 8.11.7]{NIST} that we obtain for the global limit 
\begin{equation}
D_{\rm global}^{\rm AI^\dag}(r)
=(1-r^2)\Theta(1-r) .
\label{Dglobal-CS}
\end{equation}
As in class A the correction terms are exponentially suppressed when staying away from the vicinity about $r=1$. 
 Plots of~\eqref{CS-NF} for $N=10,20,50$ and~\eqref{Dglobal-CS} for $N\to\infty$ can be found in Fig.~\ref{Fig-rescaled} (top right plot).

{\bf Class AII$^\dag$:}  The case of the complex self-dual ensemble is more involved.  
We start like before with properly scaling the spectral variables $z$ and $w^*$ and then removing the exponential prefactor in~\eqref{DN-QS},  
\begin{eqnarray}
F_N^{\rm AII^\dag}(r)=\sum_{j=0}^N\frac{N!(2N-2j)!}{(2N)!(N-j)!}(4Nr^2)^{j}Q(2N-2j+1,2Nr^2).
\label{QS-NF}
\end{eqnarray}
We have renamed the summation index $j\to N-j$ as the main contribution for $N\to\infty$ comes from a neighbourhood where $j$ is of order one. The summands have indeed a point wise limit
\begin{equation}
\lim_{N\to\infty}\frac{N!(2N-2j)!}{(2N)!(N-j)!}(4Nr^2)^{j}Q(2N-2j+1,2Nr^2)=r^{2j}\Theta(1-r)
\end{equation}
for any fixed $j$, due Stirling's approximation for the factorials and the asymptotic of the incomplete Gamma-function~\cite[8.11.6 \& 8.11.7]{NIST}. One can even relax the fixed condition of $j$ and allow any $j=o(\sqrt{N})$ when considering the Taylor expansion of the summands. We omit here to show that there is actually a sequence of $N$-independent summands whose series converges. This would be needed to be allowed to apply Lebesgue's dominated convergence theorem.

Summarising, we obtain a geometric series
\begin{eqnarray}
D_{\rm global}^{\rm AII^\dag}(r)
&\approx& \sum^{\infty}_{j=0}r^{2j} \Theta(1-r)=\frac{1}{1-r^2}\Theta(1-r),
\label{Dglobal-QS}
\end{eqnarray}
which is drawn in Fig.~\ref{Fig-rescaled} (bottom plot) as well as for the values $N=10,20,50$ of the finite-$N$ expression~\eqref{QS-NF}.

Let us remark again that in both classes AI$^\dag$ and AII$^\dag$ the rescaled characteristic polynomials do not directly relate to the eigenvalue density as it is the case in class A, though some kind of relation will be still present. Therefore, the quantity $D_{\rm global}(r)$ is not normalised  nor it is guaranteed to be normalisable as we can see in the class AII$^\dag$ for which~\eqref{Dglobal-QS} is  not integrable over its support. Indeed, the integrals over their finite-$N$ expressions depend on the matrix dimension.

The question remains how we can interpret the different shapes of the curves in Fig.~\ref{Fig-rescaled}. 
In~\cite{AMP}, the local  nearest-neighbour spacing distribution in radial distance in the bulk of the spectrum was compared for complex symmetric and complex self-dual matrices to the one of the two-dimensional Coulomb gas as a function of inverse temperature $\beta$ (the Dyson index). Fitting the value of $\beta$ accordingly, the following values of $\beta\approx 1.4$ (AI$^\dag$) and $\beta\approx 2.6$ (AII$^\dag$) were found in~\cite{AMP}. This approximate local  description by a two-dimensional Coulomb gas was further corroborated by fitting also the next-to-nearest-neighbour spacing distributions, that confirm these approximate values of $\beta$, see~\cite{ABC}.
Furthermore, perturbation theoretic methods~\cite{CFTW14} as well as recent numerical simulations~\cite{CSA} show that for $\beta > 2$ the spectral density of the two-dimensional Coulomb gas overshoots the $\beta=2$ result in the neighbourhood of the edge and eventually starts to oscillate when increasing $\beta$ even further, cf.,~\cite[Figure~1]{CFTW14} and~\cite[Figure~4]{CSA}. The overshooting is also visible for class AII$^\dag$ for the average of a pair of characteristic polynomials, see bottom plot of Fig.~\ref{Fig-rescaled}. In contrast, for $\beta<2$ the authors of~\cite{CSA} found a damping of the spectral density at the edge compared to $\beta=2$, see Figure~2 therein, which agrees with what we found for the pair of characteristic polynomials in the top right plot of Fig.~\ref{Fig-rescaled} for class AI$^\dag$. Thus, our analytical findings in the present work are consistent with the (approximate) values of $\beta\approx 1.4$ and $\beta\approx2.6$ for the ensembles 
AI$^\dag$ and AII$^\dag$ for their respective local bulk statistics. 

We conclude that although we have to emphasise that the expectation value of two characteristic polynomials does {\it not} have any known relation to the spectral density in the AI$^\dag$ or AII$^\dag$ ensemble, the behaviour just described is consistent with the heuristically determined $\beta$-values in~\cite{AMP,ABC}, and the properties of a two-dimensional Coulomb gas in~\cite{CFTW14,CSA}. For this reason we expect that the expectation value of two characteristic polynomials contains important information about the spectrum, too, at least at the edge of the spectrum. 

%%%%%%%%%%%%%%%%%%%%%%%%%%%%%%%%%%%%%%%%%%%%
%%%%%%%%%%%%%%%%%%%%%%%%%%%%%%%%%%%%%%%%%%%%

\subsection{Complex Ginibre ensemble: Edge and bulk scaling of the density revisited}\label{Ginrev}

In this subsection, we briefly recall the edge and bulk scaling limit of the complex Ginibre ensemble, which is  a determinantal point process with kernel $K_N(z,w^*)$. We choose the convention that $K_N(z,w^*)$ denotes the polynomial part of the kernel, without the corresponding weight functions, which is also know as pre-kernel. In this form it equals the normalised expectation value of two conjugate characteristic polynomials.

 The following  relation is known between this kernel at finite-$N$ and the characteristic polynomials, e.g., see~\cite{Ginibre,AV03},
\begin{equation}
K_N(z,w)=
\frac{D_N^{\rm A}(z,w^*)}{D_N^{\rm A}(0,0)}
=
E_N(zw^*)= e^{
zw^*}Q(N+1,zw^*).
\label{K-DN-A}
\end{equation}
The eigenvalue density is given by the kernel at equal arguments, times the weight function of the Ginibre ensemble, $w(z)=\exp[-|z|^2]$,
\begin{equation}
R_{1,N}(z)=e^{-|z|^2}K_N(z,z^*)=Q(N+1,|z|^2).
\label{Gin-R1}
\end{equation}
We would like to point out that this density is normalised to $N$. 

In the global large-$N$ limit it converges to the circular law on a disc of radius $\sqrt{N}$ in this normalisation, 
\begin{equation}
R_{1,N}(z)\sim\Theta(\sqrt{N}-|z|).
\label{Gin-density}
\end{equation}
A further rescaling $z\to\sqrt{N} z$ maps the support to the unit disc.

The local bulk scaling limit is defined as follows. We pick a point $\sqrt{N}z_0$ in the bulk of the spectrum, that is with $|z_0|<1$. The vicinity of this bulk point is then parametrised as follows
\begin{eqnarray}
z&=&\sqrt{N}z_0+\chi, \quad w^*=\sqrt{N}z_0^*+\eta^*,
\nonumber\\
x&:=&zw^*=N|z_0|^2+\sqrt{2N}\,y+\chi\eta^*,\quad y:=(\chi z_0^*+\eta^*z_0)/\sqrt{2}\ .
\label{zoom}
\end{eqnarray}
The limiting kernel in the bulk is, then, obtained by replacing the truncated sum $E_N(zw^*)$ in \eqref{K-DN-A} by an exponential, leading to
\begin{eqnarray}
K_{\rm bulk}^{\rm A}(\chi,\eta^*)&=&\lim_{N\to\infty}K_N(\sqrt{N}z_0+\chi,\sqrt{N}z_0^*+\eta^*)\Big|_{|z_0|<1}\sim \exp\left[ N|z_0|^2+\sqrt{2N}\,y+\chi\eta^*\right].\qquad
\label{Gin-bulk-kernel}
\end{eqnarray}
Strictly speaking, the pre-kernel does not have a limit. 
Only after multiplying with the weight function, $e^{-\frac12(|z|^2+|w|^2)}K_N(z,w^*)$ and after cancelling cocycles that drop out in the determinant of the $k$-point correlation function, it does lead to the limiting Ginibre kernel. 
This is why below we will remove the exponential factor in~\eqref{K-DN-A} when taking limits. Nevertheless, the bulk asymptotic~\eqref{Gin-bulk-kernel} and edge asymptotic below will be very useful when comparing results for averages over more than one pair of characteristic polynomials, cf., Appendix~\ref{App:Ak>1}.

Turning to the local edge scaling limit, we will choose the same rescaling as in \eqref{zoom}, but require now that the point we zoom in is located on the unit circle, that is $|z_0|=1$.
For the edge scaling we rewrite the truncated exponential function in the following way,
\begin{equation}
E_N(x)=e^xQ(N+1,x)
=e^x(1-P(N+1,x)),
\label{EN-P-relation}
\end{equation}
where 
\begin{equation}
P(N+1,x)=\frac{\gamma(N+1,x)}{\Gamma(N+1)}
\label{Pdef}
\end{equation}
is the regularised lower incomplete Gamma-function
$\gamma(a,x)=\Gamma(a)-\Gamma(a,x)$. It enjoys the following asymptotic \cite[8.11.10]{NIST} for large-$N$, with the edge rescaling~\eqref{zoom} at $|z_0|=1$,
\begin{equation}
P(N+1,x)=\frac12\erfc(-y)+\mathcal{O}(N^{-\frac12}), 
\label{P-lim}
\end{equation}
with the complementary error function~\eqref{erfc-def}. Once, we employ the identity~\eqref{erfc-rel},  we find
\begin{equation}
Q(N+1,x)=\frac12\erfc(y)+\mathcal{O}(N^{-\frac12}), 
\label{Q-lim}
\end{equation}
and hence the edge scaling asymptotic becomes
\begin{eqnarray}
K_{\rm edge}^{\rm A}(\chi,\eta^*)=\left.\lim_{N\to\infty}K_N(\sqrt{N}z_0+\chi,\sqrt{N}z_0^*+\eta^*)\right|_{|z_0|=1}\sim\frac12\exp\left[ N+\sqrt{2N}\,y+\chi\eta^*\right]
\erfc(y).\quad
\label{Gin-edge-kernel}
\end{eqnarray}
After multiplying it with the weight function and removing cocycles, it leads to the universal complementary error function kernel (sometimes called Faddeyeva plasma-kernel~\cite{FT61}).

%%%%%%%%%%%%%%%%%%%%%%%%%%%%%%%%%%%%%%%%%%%%%
\subsection{Local edge scaling limit for one pair of characteristic polynomials}\label{edge-lim}

Because only one of the three ensembles we consider is a determinantal point process, the complex Ginibre ensemble, we do not have the freedom to remove $N$-dependent cocycles from the limiting characteristic polynomials which are proportional to the kernel, nor can we argue why to multiply with a weight function as in the Ginibre class A, which removes the leading contribution $\exp[N|z_0|^2]$ from the expected characteristic polynomials in that case. 

Therefore, we choose a different route in defining the limiting characteristic polynomials, by removing the exponential prefactor $\exp[zw^*]$ that appears in all three ensembles, see eqs.~\eqref{DN-A},~\eqref{DN-CS} and~\eqref{DN-QS}, respectively. Furthermore, as in the expressions for finite matrix dimension $N$ given in Proposition~\ref{Prop_char_exp}, we divide by the value of the respective characteristic polynomials at the origin, $D_N(0,0)$. It turns out that for classes AI$^\dag$ and AII$^\dag$ an additional rescaling in powers of $N^{\,\pm1/2}$ 
is necessary. 

\begin{theorem}[Local limits of a pair of characteristic polynomials at the spectral edge]\label{Thm_lim_char_exp}\

We recall the definitions~\eqref{def.dim},~\eqref{def.scale} as well as Owen's $T$-function~\eqref{OTi} and the results of Proposition~\ref{Prop_char_exp}. The vicinity of edge points is defined as 
\begin{equation}
z=\sqrt{d_N}sz_0+s\chi\ , \quad
w^*=\sqrt{d_N}sz_0^*+s\eta^*, \quad zw^*=s^2(d_N+\sqrt{2d_N}\,y
+\chi\eta^*),
\label{zoom1}
\end{equation}
with $|z_0|=1$ and $y=(z_0^*\chi+z_0\eta^*)/\sqrt{2}$. With this rescaling we have 
\begin{eqnarray}
D_{\rm edge}^{\rm A}(\chi,\eta^*)&:=&\lim_{N\to\infty}
e^{-zw^*}\frac{D_N^{\rm A}(z,w^*)}{D_N^{\rm A}(0,0)}=\frac12\erfc(y),
\label{Dedge-A}\\
D_{\rm edge}^{\rm AI^\dag}(\chi,\eta^*)&:=&\lim_{N\to\infty}
\sqrt{N}\,e^{-2zw^*}
\frac{D_N^{\rm AI^\dag}\left(z,w^*\right)}{D_N^{\rm AI^\dag}(0,0)}
=\frac{1}{\sqrt{2\pi}}e^{-y^2}-\frac{y}{\sqrt{2}}
\erfc(y),
\label{Dedge-CS}
\\
D_{\rm edge}^{\rm AII^\dag}(\chi,\eta^*)&:=&\lim_{N\to\infty}
\frac{1}{\sqrt{2N}}\,e^{-2zw^*}\frac{D_N^{\rm AII^\dag}(z,w^*)}{D_N^{\rm AII^\dag}(0,0)}\nonumber\\
&=&
\left(-\frac{\sqrt{\pi}}{4}\erfc\left(y\right)\erfi(y/\sqrt{2})-i\sqrt{\pi}\,T\left(\sqrt{2}y,i/\sqrt{2}\right)\right) e^{-y^2/2}.
\label{Dedge-QS}
\end{eqnarray}
The corresponding left and right tail asymptotic read (for real $y$)
\begin{eqnarray}
\label{Dtails-A}
D_{\rm edge}^{\rm A}(\chi,\eta^*)&\sim&
\left\{
\begin{array}{cl}
\displaystyle\frac{1}{2\sqrt{\pi}y}\, e^{-y^2}, & {\rm for}\ y\to+\infty,\\
\displaystyle1+\frac{1}{2\sqrt{\pi}y}\,e^{-y^2}, & {\rm for}\ y\to-\infty,
\\
\end{array}
\right.
\\
\label{Dtails-CS}
D_{\rm edge}^{\rm AI^\dag}(\chi,\eta^*)&\sim&
\left\{
\begin{array}{cl}
\displaystyle\frac{1}{\sqrt{8\pi}\,y^2}\, e^{-y^2}, & {\rm for}\ y\to+\infty,\\
\displaystyle-\sqrt{2}\,y+\frac{1}{\sqrt{8\pi}\,y^2}\, e^{-y^2}, & {\rm for}\ y\to-\infty,
\\
\end{array}
\right.
\\
\label{Dtails-QS}
D_{\rm edge}^{\rm AII^\dag}(\chi,\eta^*)&\sim&
\left\{
\begin{array}{cl}
\displaystyle\frac{1}{\sqrt{8\pi}\, y^2}e^{-y^2}, & {\rm for}\ y\to+\infty,\\
\displaystyle-\frac{1}{\sqrt{2}\,y}-\frac{1}{\sqrt{2}\,y^3}, & {\rm for}\ y\to-\infty.\\
\end{array}
\right.
\end{eqnarray}

\end{theorem}

The asymptotic~\eqref{Dedge-CS} also follows from~\cite[Corollary 2]{ATTZ} for the eigenvector statistics as the initial expression~\eqref{DN-CS} has been the same.

From the results~\eqref{Dedge-A}-\eqref{Dedge-QS} we can readily  read off the value of these normalised expectation values at $\chi=\eta^*=0$ which is
\begin{equation}\label{edge.origin}
D_{\rm edge}^{\rm A}(0,0)=\frac{1}{2},\qquad D_{\rm edge}^{\rm AI^\dag}(0,0)=\frac{1}{\sqrt{2\pi}}\qquad{\rm and}\qquad D_{\rm edge}^{\rm AII}(0,0)=\frac{{\rm arctanh}(\frac{1}{\sqrt{2}})}{\sqrt{4\pi}}.
\end{equation}
For AII$^\dag$, we need Owen's $T$-function at $x=0$ in its first argument, see~\eqref{OTi}, which is an elementary integral.
We need the values~\eqref{edge.origin} when going over to an average over $k\geq1$ pairs of characteristic polynomials, see Theorem~\ref{thm:edge}.

We would like to highlight that the edge limits of all three ensembles are translation invariant along the boundary since they only depend on $y$ and not on $\chi$ and $\eta^*$ independently. Indeed, when shifting $(\chi,\eta^*)\to(\chi,\eta^*)+t(z_0,-z_0^*)$ the variable $y$ remains unchanged, see~\eqref{zoom}. This could be expected for a smooth boundary of the global support because in the local  limit the boundary  will be well-approximated by its tangent at this point so that the edge will follow this straight line.

\begin{figure}[t!]
	\centering
	\includegraphics[width=\textwidth]{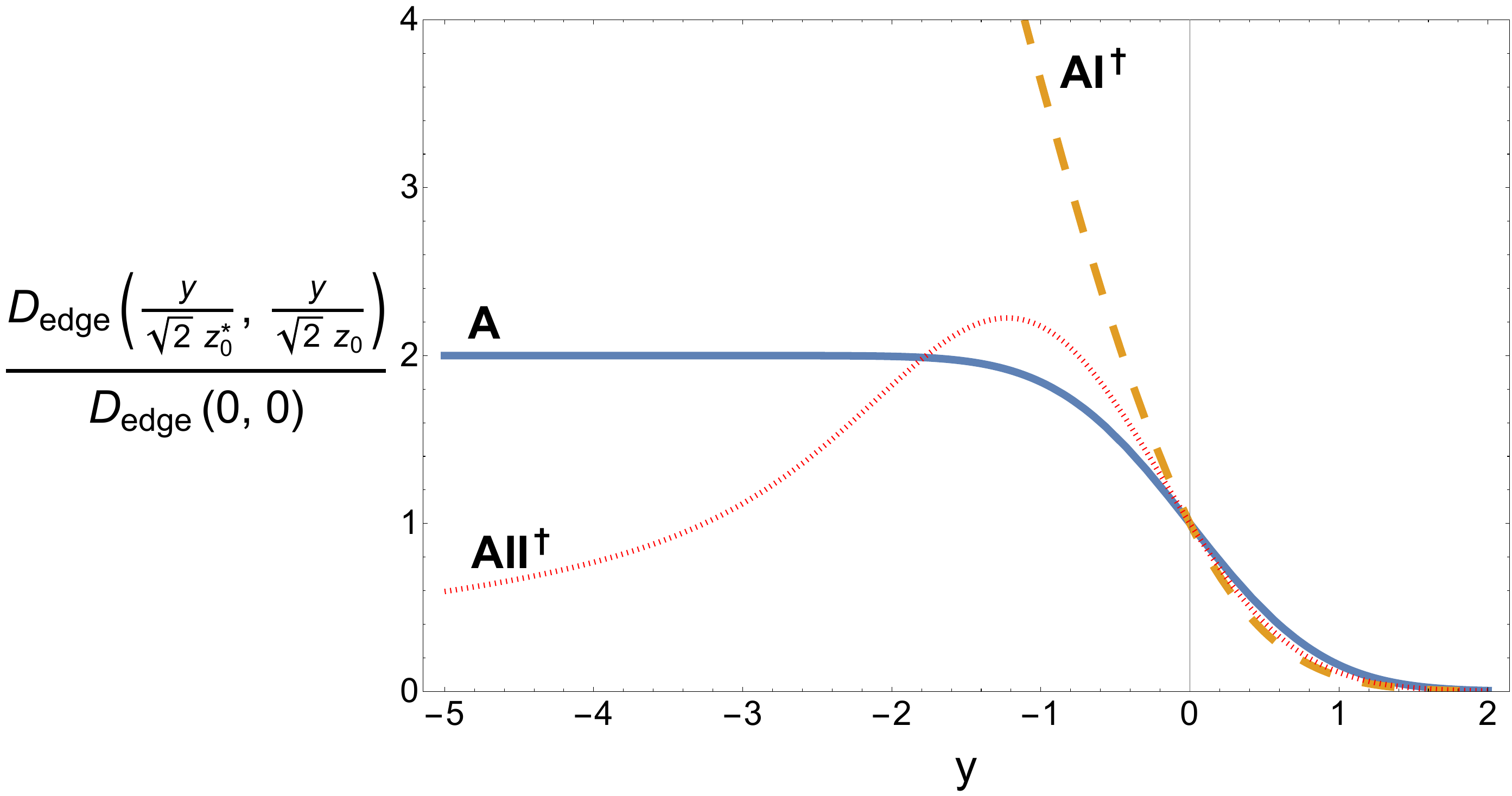}
	\caption{Plots of the expectation value of a pair of characteristic polynomials in the edge scaling limit as a function of a real $y=(z_0^*\chi+z_0\eta^*)/\sqrt{2}$, which can be achieved by setting $\chi=y/\sqrt{2}z_0^*$ and $\eta^*=y/\sqrt{2}z_0$. We normalised the expectation value by its value at $\chi=\eta^*=0$ for comparability which is given by~\eqref{edge.origin}.
	The curve for class A (solid blue) with $D_{\rm edge}^{\rm A}(\chi,\eta^*)$ from~\eqref{Dedge-A} normalised by $D_{\rm edge}^{\rm A}(0,0)$ saturates at $2$. Class AI$^\dag$ (dashed orange) given by~\eqref{Dedge-CS} has a clearly visible linear rise, and  class AII$^\dag$ (dotted red)  following~\eqref{Dedge-QS} 
	shows the decay $\sim-1/\sqrt{2}y$ which starts to be visible for large negative values of $y$.}
	\label{Fig-edge}
\end{figure}

Theorem~\ref{Thm_lim_char_exp} implies that at the edge of the spectrum we indeed found three different classes of limiting characteristic polynomials. They are conjectured to be universal based on numerics and symmetry arguments~\cite{Kawabata}. They are plotted in Fig.~\ref{Fig-edge} for illustration and comparison.

The difference between the three ensembles is obvious. It is perhaps simplest to see in the left tail asymptotic, $y\to-\infty$ for which we have given not only the leading but also the next to leading order. For the complex Ginibre ensemble the limiting characteristic polynomial saturates to unity, the mean value of the spectral density, whereas for the classes AI$^\dag$ and AII$^\dag$ a linear rise and an algebraic decay dominates, respectively, cf., Fig.~\ref{Fig-rescaled}. 
The right tail in the complex Ginibre ensemble was noted first in~\cite{Kanzieper} for the decay of the density at the edge. 

We can only speculate why a different scaling is needed in all three ensembles in the edge scaling limit. For comparison, in the real and quaternion Ginibre ensembles the corresponding skew-symmetric kernel, that relates to the spectral density, is obtained from the expectation value of two characteristic polynomials after multiplying, respectively dividing by the difference of the arguments. When rescaling them with $\sqrt{N}$ as it is done here, this will also give rise to extra powers in $\sqrt{N}$.  Our choice of the numerical part of the scaling is natural as we multiply with $\sqrt{d_N}=\sqrt{N}$ for AI$^\dagger$ and divide by $1/\sqrt{d_N}=1/\sqrt{2N}$.

{\bf Class A:}  The edge limit of the complex Ginibre ensemble is well known and has been summarised in the previous subsection, especially one needs to insert   the expansion~\eqref{P-lim} together with~\eqref{EN-P-relation} into the definition~\eqref{Dedge-A} and~\eqref{DN-A} to arrive at the claim. The tail asymptotic~\eqref{Dtails-A} follows from the expansion of the  complementary error function~\cite[7.12.1]{NIST}
\begin{eqnarray}
\label{erfc-asympt}
\erfc(z)&\sim& 
\left\{
\begin{array}{r}
\displaystyle \frac{1}{\sqrt{\pi}z}e^{-z^2}\left( 1-\frac{1}{2z^2}+\mathcal{O}(z^{-4})\right), \quad \mbox{for}\ z\to+\infty,\\
\displaystyle 2+\frac{1}{\sqrt{\pi}z}e^{-z^2}\left( 1-\frac{1}{2z^2}+\mathcal{O}(z^{-4})\right), \quad \mbox{for}\ z\to-\infty,\\
\end{array}
\right.
\end{eqnarray}
where we give the next-to-leading order, as well.
%%%%%%%%%%%%%%%%%%%%%%%%%%%%%%%%%%%%%%%%%%%%%%%%5

{\bf Class AI$^\dag$:}  For the complex symmetric ensemble, it is preferable to write $E_{N-1}(x)=\partial_xE_N(x)= 
\partial_x(e^xQ(N+1,x))$ in~\eqref{DN-CS}. Let us mention again that the scaling with $s=1/\sqrt{2}$ guarantees the unit disc as the limiting support. 
Proceeding as for the Ginibre ensemble, we multiply~\eqref{DN-CS} with $e^{-2zw^*}$ to cancel the exponential and set $x=2zw^*=N+\sqrt{2N}y+\chi\eta^*$,
\begin{eqnarray}
e^{-2zw^*}
\frac{D_N^{\rm AI^\dag}\left(z,w^*\right)}{D_N^{\rm AI^\dag}(0,0)}=Q(N+1,x)-\frac{x e^{-x}}{N+1}\partial_x\left(e^xQ(N+1,x) \right).
\end{eqnarray}
With help of the edge rescaling \eqref{zoom1} at $|z_0|=1$ and the asymptotic~\eqref{Q-lim} with $y=(z_0^*\chi+z_0\eta^*)/\sqrt{2}$, we arrive at
\begin{eqnarray}
e^{-2zw^*}
\frac{D_N^{\rm AI^\dag}\left(z,w^*\right)}{D_N^{\rm AI^\dag}(0,0)}&=& \frac12\erfc(y)+\mathcal{O}(N^{-\frac12})\nonumber\\
&&-\frac{N+\sqrt{2N}y+\chi\eta^* }{N+1}\left(1+\frac{1}{\sqrt{2N}}\partial_y\right)\left(\frac12\erfc(y)+\mathcal{O}(N^{-\frac12}) \right)
\nonumber\\
&=&-\frac{y}{\sqrt{2N}}\erfc(y)+\frac{1}{\sqrt{2\pi N}}e^{-y^2}+\mathcal{O}\left(\frac{1}{N}\right).
\end{eqnarray}
The two order one terms cancel after expanding the ratio in $1/N$. Eventually, we multiply with $\sqrt{N}$ to find the result~\eqref{Dedge-CS}. 

For the tail asymptotic we insert~\eqref{erfc-asympt} into~\eqref{Dedge-CS} and see that the leading order Gaussian decay cancels for the right tail, $y\to+\infty$ 
\begin{equation}
D_{\rm edge}^{\rm AI^\dag}(\chi,\eta^*)\sim -y\frac{1}{\sqrt{2\pi}y}e^{-y^2}\left(1-\frac{1}{2y^2}+\mathcal{O}(y^{-4})\right)+\frac{1}{\sqrt{2\pi}}e^{-y^2} = \frac{1}{\sqrt{8\pi}\,y^2}e^{-y^2}\left(1+\mathcal{O}(y^{-2})\right).
\end{equation}
The errors are uniform for any compact set where $y$ can be drawn from so that the derivative can indeed be interchanged with the limit $N\to\infty$.

For the left tail with $y\to-\infty$, we obtain from the corresponding asymptotic~\eqref{erfc-asympt} of the complementary error function, 
\begin{eqnarray}
D_{\rm edge}^{\rm AI^\dag}(\chi,\eta^*)&\sim& 
\frac{1}{\sqrt{2\pi}}e^{-y^2}-\frac{y}{\sqrt{2}}\left(2+\frac{1}{\sqrt{\pi}y}e^{-y^2}
\left(1-\frac{1}{2y^2}+\mathcal{O}(y^{-4})\right)
\right)
\nonumber\\
&=&-\sqrt{2}y+\frac{1}{\sqrt{8\pi}\,y^2}e^{-y^2}\left(1+\mathcal{O}(y^{-2})\right),
\end{eqnarray}
a linear rise in $y$ (as $y<0$) as the leading term.
%%%%%%%%%%%%%%%%%%%%%%%%%%%%%%%%%%%%%%%%%%%%%%%%%%%%%%%%%%%%%%%%

{\bf Class AII$^\dag$:}  For the remaining ensemble of complex self-dual matrices, we cannot use standard asymptotic results. Our starting point is the integral representation~\eqref{fNx-int2} of~\eqref{DN-QS} in the derivation for finite $N$, that we repeat for the readers' convenience:
\begin{eqnarray}
\frac{D_N^{\rm AII^\dag}(z,w^*)}{D_N^{\rm AII^\dag}(0,0)}\biggl |_{x=zw^*}&=&\frac{1}{(2N)!}f_N(x),\quad 
f_N(x)= \int_0^\infty dv\int_0^\infty du \ e^{S(u,v)},\quad
\nonumber\\
S(u,v)&:=&-u-v+N\log\left(4xv+(2x+u)^2\right).
\label{DNAII+fNrel}
\end{eqnarray}
This calls for a Laplace approximation at large $N$. Any critical point $(u_*,v_*)$ satisfies
\begin{eqnarray}
0=\partial_v S(u_*,v_*)= -1+\frac{4Nx}{4xv_*+(2x+u_*)^2}\quad{\rm and}\quad 0=\partial_u S(u_*,v_*)= -1+\frac{2N(2x+u_*)}{4xv_*+(2x+u_*)^2}.\quad
\label{SPeqs}
\end{eqnarray}
Inserting the first into the second equation yields $u_*=0$, and reinserted into the first equation gives $v_*=N-x$. 

To figure out whether the point $(u_*,v_*)$ is the global maximum of $S(u,v)$ we need to compare it with the extrema along the two boundaries. For the boundary $u=0$ and $v\geq0$, 
\begin{equation}
0=\partial_v S(0,v)= -1+\frac{N}{v+x},
\end{equation}
we find the same point $(u_*,v_*)$, whereas along the boundary $u\geq0$ and $v=0$ 
\begin{equation}
0=\partial_uS(u_0,0)= -1+\frac{2N}{2x+u_0},
\end{equation}
we find $(u_0,v_0)=(2(N-x),0)$. Taking the second derivative confirms that both extrema  are indeed local maxima. 
However, when inserting the scaling limit~\eqref{zoom1} with $x=N+\sqrt{N}y+\chi\eta^*/2$, since $d_N=2N$ and $s=1/\sqrt{2}$, we see that both maxima $(v_*,u_*)$ and $(u_0,v_0)$ are only within the integration domain $v,u\geq0$ for negative values $y<0$. This can be expected as this is the direction into the bulk of the spectrum which becomes immediate when setting $z=w$.

What is more important is that the leading order of  both extrema is $\sqrt{N}$ which is the same scale as the standard deviations of the Gaussians once we would expand about these points. This means that even an expansion about the origin would cover all relevant contributions from this integral. To pursue this direction we  rescale $v=\sqrt{N}v'$ and $u=2\sqrt{N}u'$,
\begin{equation}
f_N(x)=(2N)^{2N+1}\int_0^\infty dv'\int_0^\infty du'
\exp\left[-\sqrt{N}v'-2\sqrt{N}u'+N\log\left(\frac{xv'}{N^{3/2}}+ \left(\frac{x}{N}+\frac{u'}{\sqrt{N}}\right)^2\right)
\right].
\label{fNexp}
\end{equation}
The logarithm can be expanded as follows,
\begin{equation}
\begin{split}
&\log\left[\frac{xv'}{N^{3/2}}+ \left(\frac{x}{N}+\frac{u'}{\sqrt{N}}\right)^2\right]\\
=&\log\left[1+\frac{v'+2u'+2y}{\sqrt{N}}+\frac{1}{N}\left(yv'+(y+u')^2+\chi\eta^*\right)+\mathcal{O}\left(\frac{v'+u'}{N^{3/2}}\right)\right]\\
=&1+\frac{1}{\sqrt{N}}(v'+2u'+2y)-\frac{1}{N}\left(\frac{(v')^2}{2}+(u')^2+y^2+2v'u'+v'y+2u'y-\chi\eta^*\right)+\mathcal{O}\left(\frac{(v')^3+(u')^3}{N^{3/2}}\right).
\end{split}
\end{equation}
This expansion can be stopped at the second order as long as $(v')^3+(u')^3\ll \sqrt{N}$ because the logarithm is multiplied with $N$. This restriction is, nonetheless, not a problem since already for $v'+u'\geq N^{\epsilon}$ for any fixed $\epsilon>0$ the integrand in~\eqref{fNexp} times $e^{-2\sqrt{N}y}$ is exponentially small which we omit to show here. It is essentially due to the fact that the maximum of the integrand in the region $v'+u'\geq N^{\epsilon}$ is attained on the boundary $v'+u'= N^{\epsilon}$, because the directional derivative of $S(u,v)$ is always negative when the direction vector is non-zero and has non-negative entries, cf. the partial derivatives~\eqref{SPeqs} which are negative in this region.

 Summarising, it is asymptotically
\begin{equation}
f_N(x)\sim(2N)^{2N+1}e^{2\sqrt{N}y-y^2+\chi\eta^*}\int_0^\infty dv\int_0^\infty du\,
e^{-u^2-\frac12v^2-y(v+2u)-2vu},
\end{equation}
when $N\to\infty$.
Consequently, the full expression together with the scaling~\eqref{zoom1} reads to leading order
\begin{equation}
\frac{e^{-2zw^*}}{\sqrt{2N}}\frac{D_N^{\rm AII^\dag}(z,w^*)}{D_N^{\rm AII^\dag}(0,0)}
\sim\frac{(2N)^{2N+1}}{\sqrt{2N}(2N)!}
e^{-2N-y^2}\int_0^\infty dv\int_0^\infty du\,
e^{-u^2-\frac12v^2-y(v+2u)-2vu}.
\end{equation}
After we exploit the Stirling formula $(2N)!\approx \sqrt{2\pi}(2N)^{2N+1/2}e^{-2N}$, we are allowed to take the large-$N$ limit leading to
\begin{eqnarray}
D_{\rm edge}^{\rm AII^\dag}(\chi,\eta^*)&=& \frac{1}{\sqrt{2\pi}}e^{-y^2}\int_0^\infty dv\int_0^\infty du\,
e^{-u^2-\frac12v^2-y(v+2u)-2vu}.
\end{eqnarray}

We can simplify this expression further when substituting $v'=(v+y)/\sqrt{2}$ and $u'=u+v+y$
\begin{eqnarray}
D_{\rm edge}^{\rm AII^\dag}(\chi,\eta^*)=\frac{1}{\sqrt{\pi}}e^{-y^2/2}\int_{y/\sqrt{2}}^\infty dv'\,e^{+\frac12v'^2}\int_{\sqrt{2}v'}^\infty du'\,e^{-u'^2}=\frac{1}{2}e^{-y^2/2}\int_{y/\sqrt{2}}^\infty dv\,e^{v^2}\erfc(\sqrt{2}v),
\label{DN-AII-int1}
\end{eqnarray}
where we employed the definition~\eqref{erfc-def} of the complementary error function. To get a closed expression as stated in~\eqref{Dedge-QS}, we make the ansatz  
\begin{equation}
\label{gdiff}
g'(v)=-\frac12\erfc(\sqrt{2}v)\, e^{v^2},
\end{equation}
cf., Appendix~\ref{App:fNdiff}, which would simplify the integral~\eqref{DN-AII-int1} as follows
\begin{equation}
D_{\rm edge}^{\rm AII^\dag}(\chi,\eta^*)=e^{-y^2/2}(g(y/\sqrt{2})-g(+\infty)).
\label{DAII-g}
\end{equation}
This approach is motivated by the fact that the original integral~\eqref{fNx-int1} satisfies an inhomogenous first order differential equation, as derived in the Appendix~\ref{App:fNdiff}. Here, we directly solve the differential equation~\eqref{gdiff} for $g(v)$. 

For this purpose, we make use of Owen's $T$-function with an imaginary number in its second entry~\eqref{OTi}. Its derivative in its first entry is explicitly given by
\begin{equation}
\partial_x T\left(x,ib\right)=-\frac{ix}{2\pi}\int_0^b du\ e^{-\frac12 x^2(1-u^2)}=-\frac{i}{\sqrt{8\pi}}e^{-x^2/2}{\rm erfi}\left[\frac{bx}{\sqrt{2}}\right].
\end{equation}
Thence, when choosing $x=2v$ and $b=1/\sqrt{2}$, the derivative of
\begin{equation}
g(v)=-\frac{\sqrt{\pi}}{4}\erfc\left(\sqrt{2}\, v\right)\erfi(v)-i\sqrt{\pi}T\left(2v,\frac{i}{\sqrt{2}}\right)
\label{g-sol}
\end{equation}
solves the differential equation~\eqref{gdiff}.

We still have to determine $g(+\infty)$.
While the asymptotic of the complementary error function is given in~\eqref{erfc-asympt}, for the error function at imaginary argument we have~\cite{Dingle}
\begin{equation}
\erfi(v)=\frac{1}{\sqrt{\pi}\,v}e^{+v^2}\left(1+\frac{1}{2v^2}+\mathcal{O}(1/v^4)\right),
\end{equation}
which is even valid for both $y\to\pm\infty$.
Therefore, for the first term in~\eqref{g-sol} we obtain
\begin{equation}
-\frac{\sqrt{\pi}}{4}\erfc\left(\sqrt{2}\, v\right)\erfi(v)
\sim
\left\{
\begin{array}{r}
\displaystyle-\frac{1}{4\sqrt{2\pi}v^2}e^{-v^2}\left( 1+\frac{1}{4v^2}+\mathcal{O}(v^{-4})\right), \quad \mbox{for}\ v\to+\infty,\\
\displaystyle-\frac{1}{2v}e^{+v^2}\left( 1+\frac{1}{2v^2}+\mathcal{O}(v^{-4})\right), \quad \mbox{for}\ v\to-\infty,\\
\end{array}
\right.
\end{equation}
implying that it vanishes for $v\to+\infty$.
For the tail asymptotic, we also give the expansion for $v\to-\infty$.

We did not find the asymptotic of Owen's $T$-function in the literature. Hence, we present a short derivation.  Starting from \eqref{OTi}, the substitution $z=(1-u^2)v^2$ leads to 
\begin{eqnarray}
-i\sqrt{\pi}T\left(2v,i/\sqrt{2}\right)&=&\frac{1}{\sqrt{8\pi}\,v^2}e^{-v^2}\int_0^{v^2}dz\,{e^{-z}}\left(1+\frac{z}{v^2}\right)^{-1}\left(1-\frac{z}{v^2}\right)^{-1/2}
\nonumber\\
&=&\frac{1}{\sqrt{8\pi}\,v^2}e^{-v^2}\int_0^{v^2}dz\,{e^{-z}}\left[1 -\frac12 \frac{z}{v^2}+\mathcal{O}\left(\frac{z^2}{v^4}\right)\right]
\nonumber\\
&=&\frac{1}{\sqrt{8\pi}\,v^2}e^{-v^2}\left[1 -\frac{1}{2v^2}+\mathcal{O}\left(\frac{1}{v^4}\right)\right],
\end{eqnarray}
where in the last step we used elementary integrals. The expansion of the integrand is only uniform when cutting off at $z\leq v^{\epsilon}\ll v^2$ for any constant $\epsilon\in(0,2)$. Like before this is not a problem as the integrand is uniformly bounded from above by $e^{-v^\epsilon}/\sqrt{1-z/v^2}$ when $z\in( v^\epsilon,v^2)$, whose integral thus vanishes exponentially for $v\to\infty$.

Let us remark that the expansion in $1/v$ holds for both $y\to\pm\infty$; in particular, it vanishes in both limits.
 As a result, the asymptotic  yields $g(+\infty)=0$ so that the full solution for the edge scaling limit from~\eqref{DAII-g} is given by~\eqref{Dedge-QS}.

After having evaluated the asymptotic of $g(v)$ already in great detail, we can immediately give the tail estimates. For $y\to+\infty$, it is 
\begin{equation}
D_{\rm edge}^{\rm AII^\dag}(\chi,\eta^*)\sim e^{-y^2}\frac{1}{\sqrt{8\pi}\,y^2}\left( 1-\frac{5}{2y^2}+\mathcal{O}(y^{-4})\right),
\end{equation}
where both terms in \eqref{Dedge-QS} contribute to the same order. For $y\to-\infty$, the asymptotic of the complementary error function \eqref{erfc-asympt} gives the leading contribution while Owen's $T$-function can be neglected. Thence, it is
\begin{equation}
D_{\rm edge}^{\rm AII^\dag}(\chi,\eta^*)\sim -\frac{1}{\sqrt{2}y}\left( 1+\frac{1}{y^2}+\mathcal{O}(y^{-4})\right)
+ e^{-y^2}\frac{1}{\sqrt{2\pi}\,y^2}\left( 1+\mathcal{O}(y^{-2})\right),
\end{equation}
which completes the derivation of~\eqref{Dtails-QS}. We note that for negative $y$ this is a positive function.

All three limiting characteristic polynomials at the edge are plotted in Fig.~\ref{Fig-edge}. The difference is most pronounced at large negative $y$. While class A leads to a constant (density), 
in class AI$^\dag$ there is a linear rise $\sim y$ of the rescaled characteristic polynomial, whereas in class AII$^\dag$ it decays proportional to $\sim1/y$. This is in line with the plots at finite matrix dimension $N$ in Fig.~\ref{Fig-rescaled}, where here we zoom into the edge behaviour.

%%%%%%%%%%%%%%%%%%%%%%%%%%%%%%%%%%%%%%%%%%%%%
\subsection{Bulk scaling limit  for  one pair of characteristic polynomials}\label{bulk-lim}

The limiting expressions for the rescaled characteristic polynomials in the bulk can be defined in a similar way as at the edge.

\begin{proposition}
[Local limits of a pair of characteristic polynomials in the spectral bulk]
\label{Prop_lim_char_exp}\

Recalling Proposition~\ref{Prop_char_exp}, the local scaling in the bulk, defined as 
\begin{equation}
z=\sqrt{d_N}sz_0+s\chi\ , \quad
w^*=\sqrt{d_N}sz_0^*+s\eta^*, \quad zw^*=s^2(d_N|z_0|^2+\sqrt{2d_N}\,y
+\chi\eta^*)
\label{zoom2}
\end{equation}
with $|z_0|<1$ and $y=({z_0^*\chi+z_0\eta^*})/{\sqrt{2}}$ as well as the definitions~\eqref{def.dim} and~\eqref{def.scale}, leads to
\begin{eqnarray}
D_{\rm bulk}^{\rm A}(\chi,\eta^*)&:=&\lim_{N\to\infty}
e^{-zw^*}\frac{D_N^{\rm A}(z,w^*)}{D_N^{\rm A}(0,0)}=1 ,
\label{Dbulk-A}\\
D_{\rm bulk}^{\rm AI^\dag}(\chi,\eta^*)&:=&\lim_{N\to\infty}
e^{-2zw^*}\frac{D_N^{\rm AI^\dag}\left(z,w^*\right)}{D_N^{\rm AI^\dag}(0,0)} =1-|z_0|^2  ,
\label{Dbulk-CS}
\\
D_{\rm bulk}^{\rm AII^\dag}(\chi,\eta^*)&:=& 
\lim_{N\to\infty}e^{-2zw^*}\frac{D_N^{\rm AII^\dag}(z,w^*)}{D_N^{\rm AII^\dag}(0,0)}=\frac{1}{1-|z_0|^2} .
\label{Dbulk-QS}
\end{eqnarray}
\end{proposition}

All three results are consistent with the respective global limit eqs.~\eqref{Dglobal-Gin},~\eqref{Dglobal-CS} and~\eqref{Dglobal-QS} with $r=|z_0|$. In particular, they give unity at the origin, by setting $|z_0|=0$ on the right hand sides, which is consistent with their definition. This agreement and its independence of $\chi$ and $\eta^*$ reflect that all three ensembles satisfy  locally translation invariance and isotropy as it is well-known for the three Ginibre ensembles.

{\bf Class A:} Looking at \eqref{DN-A}, the result for the Ginibre ensemble is trivial, as the corrections to $e^{-x}E_N(x)\to 1$ are exponentially suppressed in this kind of large-$N$ limit.

{\bf Class AI$^\dag$:} Likewise, the result~\eqref{DN-CS} for complex symmetric matrices yields for the limit \eqref{Dbulk-CS}
\begin{eqnarray}
\lim_{N\to\infty} e^{-2zw^*}\left(E_N(2zw^*)-\frac{2zw^*}{N+1}E_{N-1}(2zw^*)\right)
&=&1-\frac{N|z_0|^2+\sqrt{2N}\,y+\chi\eta^*}{N+1}
\nonumber\\
&=& 1-|z_0|^2-\sqrt{\frac{2}{N}}\,y + \mathcal{O}(1/N),
\end{eqnarray}
as both truncated exponentials give unity.
Evidently, the subleading term contains non-trivial information about the local spectral fluctuation $y$, that deviates from a constant behaviour.

{\bf Class AII$^\dag$:} Let us turn to the ensemble of complex self-dual matrices, where we return to~\eqref{SPeqs} determining the critical points. We note that this time the bulk scaling~\eqref{zoom2} is equal to $x=zw^*=N|z_0|^2+\sqrt{N}y+\chi\eta^*/2$, with $0<|z_0|<1$. Because of that, the maximum at $(u_*,v_*)=(0,N-x)$ (and maximum at the boundary $(0,v>0)$) as well the maximum $(u_0,v_0)=(2(N-x),0)$ at the boundary $(u>0,0)$ are both inside the integration domain $u,v\geq0$.

We need to compare the integrand in $f_N(x)$, see~\eqref{fNx-int2}, at these two points to decide which of those is the global maximum. For this purpose, we ask for the leading asymptotic of the integrand $\exp[S(u,v)]$, with
\begin{equation}
S(u,v)=-u-v+N\log[4xv+(2x+u)^2],
\end{equation}
 at these two points which is 
\begin{eqnarray}
 e^{S(u_*,v_*)}\sim (2N|z_0|)^{2N}\exp[-N(1-|z_0|^2)]
\quad{\rm and}\quad e^{S(u_0,v_0)}\sim(2N)^{2N}\exp[-2N(1-|z_0|^2)].
\end{eqnarray}
Thus, we obtain for the difference
\begin{equation}
S(u_*,v_*)-S(u_0,v_0)\sim N(1-|z_0|^2+2\log|z_0|)
\end{equation}
The function $h(a)=1-a+\log a$ is strictly 
increasing 
for $a=|z_0|^2\in(0,1)$ and takes its maximum at $a=1$ which is $h(1)=0$. Therefore, it is $S(u_*,v_*)<S(u_0,v_0)$ for suitable large $N$ and $(u_0,v_0)=(2(N-x),0)$ is the position of the global maximum.

Consequently, we substitute $u=2N(1-|z_0|^2)+2\sqrt{N}u'$ while the variable $v$ remains unchanged. Plugging this into the integral~\eqref{DNAII+fNrel} we arrive at
\begin{eqnarray}
f_N(x)&=&\! 2\sqrt{N}\!\int_0^\infty\!\! dv\!\int_{-\sqrt{N}(1-|z_0|^2)}^\infty\!\!\! du'
e^{-2N(1-|z_0|^2)-2\sqrt{N}u'-v} 
\!\left[ 4xv+ (2x+ 2N(1-|z_0|^2)+2\sqrt{N}u')^2\right]^N 
\nonumber\\
&=&\! \frac{(2N)^{2N+1}}{\sqrt{N}}e^{-2N(1-|z_0|^2)}\int_0^\infty dv\int_{-\sqrt{N}(1-|z_0|^2)}^\infty du'e^{-2\sqrt{N}u'-v} 
\left[1+\frac{2}{\sqrt{N}}(u'+y)
\right.
\nonumber\\
&&\left.
+\frac{1}{N}\left(v|z_0|^2+\chi\eta^*+(u'+y)^2\right)
 +\frac{1}{N^{\frac32}}\left(vy+\chi\eta^*(u'+y)\right)
+\mathcal{O}\left(\frac{v+(u')^2}{N^{2}}\right)\right]^N.
\end{eqnarray}
Here, we have inserted the bulk scaling \eqref{zoom2} and expanded far enough to keep the first subleading order in the final result. 
In the next step, we have to exponentiate the bracket to the $N$th power, and then expand the logarithm up to third order, $\log[1+z]=z-\frac{z^2}{2}+\frac{z^3}{3}+\mathcal{O}(z^4)$. In addition, from the definition~\eqref{Dbulk-QS} we have to appropriately normalise according to~\eqref{DNAII+fNrel}. Inserting all together a lot of terms cancel, and we arrive at 
\begin{eqnarray}
e^{-2x}\frac{f_N(x)}{(2N)!}&=& 
\frac{(2N)^{2N+1}e^{-2N}}{\sqrt{N}(2N)!}\int_0^\infty dv\int_{-\sqrt{N}(1-|z_0|^2)}^\infty du'\exp\left[-v(1-|z_0|^2)-(u'+y)^2\right]
\nonumber\\
&&\times\left[1+\frac{1}{\sqrt{N}}\left( vy-2(u'+y)\left(vy+\frac12\chi\eta^*-\frac13(u'+y)^2\right)\right)+\mathcal{O}\left(\frac{v^2+(u')^4}{N}\right)\right]. \qquad
\end{eqnarray}
Applying the Stirling formula simplifies the prefactor to $1/\sqrt{\pi}$ when $N\to\infty$. The error in the integral is uniform when we restrict the integration to $v^2+(u')^4\ll N$. As before, one can show that this does not cause any problems since the integrand is exponentially small compared to the main contribution at the origin for any choice $\epsilon>0$ of $v+|u'|\geq N^\epsilon$, since the maximum of $S(u,v)$ in this region lies on the boundary $v+|u'|= N^\epsilon$. Therefore, Lebesgue's dominated convergence theorem is applicable.

A similar statement would also allow us to extend the integral to $-\infty$ for which we skip the rigorous mathematical details. Once done, we can exploit that all the odd moments in the Gaussian variable $u'+y$ vanish, i.e., 
\begin{eqnarray}
e^{-2x}\frac{f_N(x)}{(2N)!}&=& 
\int_0^\infty dv\exp\left[-v(1-|z_0|^2)\right]\left[1+\frac{vy}{\sqrt{N}}+\mathcal{O}\left(\frac{v^2}{N}\right)\right].
\end{eqnarray}
Here, we carried out the integral over $u'$ yielding $\sqrt{\pi}$ up to order $1/N$. We are left with the elementary integral over the variable $v$ which is
\begin{eqnarray}
e^{-2x}\frac{f_N(x)}{(2N)!}= \frac{1}{1-|z_0|^2} +\frac{1}{\sqrt{N}}\frac{y}{(1-|z_0|^2)^2}+\mathcal{O}(N^{-1}). 
\end{eqnarray}
This ends the derivation of the bulk limit for class AII$^\dag$ in \eqref{Dbulk-QS}, including the first $y$-dependent correction term.

%%%%%%%%%%%%%%%%%%%%%%%%%%%%%%%%%%%%%%%%%%%%%%%%%%%%%%%%%%%%%
%%%%%%%%%%%%%%%%%%%%%%%%%%%%%%%%%%%%%%%%%%%%
\section{Expectation values for $k$ pairs of characteristic polynomials}\label{k>1}

As the classes AI$^\dag$ and AII$^\dag$ lack the insight of an explicit joint probability density of the complex eigenvalues we need to circumvent this obstacle. The aim is to derive an effective Lagrangian representation, similar to those for real eigenvalue statistics of Hermitian random matrix ensembles. Those were very helpful in explaining the applicability of random matrix theory to quantum field theories~\cite{chiral.Lag} and condensed matter theory~\cite{Efetovbook}.

In these applications, the main object is usually an average over products or ratios of characteristic polynomials. For instance, we consider  the quantity
\begin{eqnarray}\label{D.N.k}
D_{N,k}(Z,W^*)&:=&\frac{\left\langle\prod_{j=1}^k\det[J - z_j\mathbf{1}_{d_N}]\det[J^\dagger - w^*_j\mathbf{1}_{d_N}]\right\rangle}{\left\langle\left|\det[J - \sqrt{d_N}s z_0\mathbf{1}_{d_N}]\right|^{2k}\right\rangle}
\\
&=&\frac{\left\langle\det[J\otimes\mathbf{1}_k-\mathbf{1}_{d_N}\otimes Z]\det[J^\dagger\otimes\mathbf{1}_k-\mathbf{1}_{d_N}\otimes W^*]\right\rangle}{\left\langle\left|\det[J\otimes\mathbf{1}_k - \sqrt{d_N}s z_0\mathbf{1}_{d_N}\otimes\mathbf{1}_k]\right|^2\right\rangle},
\nonumber
\end{eqnarray}
with $Z={\rm diag}(z_1,\ldots,z_k)$ and $W^*={\rm diag}(w^*_1,\ldots,w^*_k)$.
We recall the definitions~\eqref{def.dim} and~\eqref{def.scale}, in particular that the dimension $d_N$ and the scaling $s $ depend on the ensemble. We employed tensor notation to write everything in a compact form, and we normalised the expectation value  to simplify the calculations below so that we can easily keep track of the normalisation constants; especially it holds $D_{N,k}(\sqrt{d_N}s z_0\mathbf{1}_{d_N},\sqrt{d_N}s z_0\mathbf{1}_{d_N})=1$.

In the physics literature, e.g., see~\cite{chiral.Lag,Efetovbook,Penco}, expectation values as those in the numerator of~\eqref{D.N.k} can be usually cast for large-$N$ into the form
\begin{equation}
\left\langle\det[J\otimes\mathbf{1}_k-\mathbf{1}_{d_N}\otimes Z]\det[J^\dagger\otimes\mathbf{1}_k-\mathbf{1}_{d_N}\otimes W^*]\right\rangle\sim C_N\int_{\mathcal{S}} d\mu(U) \exp[-\mathcal{L}_{\rm eff}(U,Z,W^*)],
\end{equation}
where $C_N(Z,W^*)$ is some $N$-dependent constant, $\mathcal{S}$ is a saddle point manifold (also known as Goldstone manifold) whose size depends on $k$ but is independent of $N$, $d\mu(U)$ is the uniform measure  on $\mathcal{S}$, and $\mathcal{L}_{\rm eff}(U,Z,W^*)$ is the effective Lagrangian depending on the source variables $Z$ and $W^*$ as well as on $U$. The spurion analysis~\cite{Penco} gives some low order expansion for $\mathcal{L}_{\rm eff}(U,Z,W^*)$ in terms of its arguments. We note that the set $\mathcal{S}$ as well as the Lagrangian usually depend on the points where one zooms into. In our case, this zooming is given  by 
\begin{eqnarray}
z_j=\sqrt{d_N}s z_0+ s\chi_j \qquad {\rm and}\qquad w^*_j&=&\sqrt{d_N}s z_0+s \eta^*_j\ ,
\label{zoomk}
\end{eqnarray}
with $\chi=\mbox{diag}(\chi_1,\ldots,\chi_k)$ and $\eta=\mbox{diag}(\eta_1,\ldots,\eta_k)$. We will indeed see that both, the saddle point manifold and the Lagrangian, change depending on whether $z_0$ lies in the bulk or at the spectral edge. Here,  $z_0$ (fixed in $N$) is chosen to be either at the edge of the spectrum when $|z_0|=1$, or in the bulk of the spectrum
$|z_0|<1$. 
The parameters $\chi_j$ and $\eta^*_j$ are also fixed and probe the spectral fluctuations on the scale of the local mean level spacing. 
We recall that we average over Gaussian random matrices of the form $\exp[-\Tr JJ^\dagger]$ for any  symmetry class. Hence, we need to adjust $d_N$ and $s$ for any of those.

This procedure works rather well in quantum field theory and condensed matter theory. It is also very effective for a zero-dimensional field theories which are essentially random matrix theories. We will see in the derivations of Theorems~\ref{thm:bulk} and~\ref{thm:edge} that all three ensembles we consider,  with the  normalisation~\eqref{D.N.k}, exhibit the very same effective Lagrangian
 in the bulk ($|z_0|<1$)
\begin{equation}\label{Lag.bulk}
\mathcal{L}_{\rm eff}^{\rm bulk}(U,Z,W^*)=\sqrt{d_N}\Tr(z_0^*\chi+z_0\eta^*)+\frac{1}{2}\Tr({\rm diag}(\chi,\chi) U^\dagger{\rm diag}(\eta^*,\eta^*) U)
\end{equation}
as well as at the edge ($|z_0|=1$)
\begin{equation}\label{Lag.edge}
\begin{split}
\mathcal{L}_{\rm eff}^{\rm edge}(A,Z,W^*)=&\sqrt{d_N}\Tr(z_0^*\chi+z_0\eta^*)-\frac{1}{2}\Tr[(z_0^*\chi)^2+(z_0\eta^*)^2]\\
&-\frac{z_0^*}{\sqrt{2}}\Tr A^\dagger A\,{\rm diag}(\chi,\chi)-\frac{z_0}{\sqrt{2}}\Tr AA^\dagger\, {\rm diag}(\eta^*,\eta^*)-\frac{1}{2}\Tr( A^\dagger A)^2.
\end{split}
\end{equation}
Only the saddle point manifolds are varying from ensemble to ensemble which are the most important indicator that the spectral statistics are different for the three ensembles.
At the edge the set $\mathcal{S}$ is actually a vector space so that the uniform measure is the Lebesgue measure on this vector space, while in the bulk we will see that the saddle point manifolds are classical Lie groups.

We would like to emphasise that we have checked and compared with existing literature~\cite{BS,AKMP,Rider} that the real and quaternion Ginibre ensemble yield the very same Lagrangians and saddle point manifolds as the complex one (class A) when staying away from the real axis. Only when zooming in on the real axis $\mathcal{S}$ will be different from class $A$ though the Lagrangian still looks the same. We will, however, not show this here and postpone it to future work where we want to discuss all symmetry classes of non-Hermitian random matrices in this way.

Coming back to the three ensembles we discuss here, we would like to remark that the normalisation with respect to the denominator of~\eqref{D.N.k} helps to keep track of the normalisation though it is not enough to have a well-defined limit $N\to\infty$. This can be seen in the resulting Lagrangians~\eqref{Lag.bulk} and~\eqref{Lag.edge}, due to the $d_N$-dependent term. To take care of that we normalise the ratio as follows
\begin{equation}
\widehat{D}_{N,k}(Z,W^*):=\frac{D_{N,k}(Z,W^*)}{\prod_{j=1}^kD_{N,1}(z_j,w_j^*)}.
\label{hatDNk}
\end{equation}
The pairings into arguments $(z_j,w_j^*)$ in the product of expectation values in the denominator is introduced {\it ad hoc} but becomes natural when choosing $w_j^*=z_j^*$. The latter choice gives the source variables an interpretation of some complex conjugate pairs of  ``test-charges'' that explore the spectrum of the random matrix $J$.

%%%%%%%%%%%%%%%%%%%%%%%%%%%%%%%%%%%%%%%%%%%%%%5
\subsection{Bulk limit for $k>1$ pairs of characteristic polynomials}\label{bulkk}
 
\begin{theorem}
[Bulk limit for $k>1$]
\label{thm:bulk}\

We recall the definitions~\eqref{D.N.k} and~\eqref{hatDNk}. In the bulk limit~\eqref{zoomk} with $|z_0|<1$
we obtain the following  
expressions for
\begin{equation}\label{lim.bulk}
\begin{split}
D_{{\rm bulk}, k}^{\rm R}(\chi,\eta^*):=&\lim_{N\to\infty}\widehat{D}_{N,k}^{\rm R}(Z,W^*)\\
=&\exp\left[-\Tr\chi\eta^*\right]\int_{\mathcal{G}}d\mu(U)\exp\left[\frac{1}{2}\Tr({\rm diag}(\chi,\chi) U^\dagger{\rm diag}(\eta^*,\eta^*) U)\right].
\end{split}
\end{equation}
In all three cases, the saddle point manifolds are Lie groups given by $\mathcal{G}={\rm U}(k)$ for class R=A (complex Ginibre), $\mathcal{G}={\rm USp}(2k)$ for class R= AI$^\dag$ (complex symmetric), and $\mathcal{G}={\rm O}(2k)$ for class R=AII$^\dag$ (complex self-dual), where $d\mu(U)$ is their normalised Haar measure. For class A, we embed ${\rm U}(k)$ into the $2k\times2k$ matrices as follows, $U={\rm diag}(\widetilde{U},\widetilde{U})$ with $\widetilde{U}\in{\rm U}(k)$, to write the result for all three classes in a unifying way.

\end{theorem}

The expression~\eqref{lim.bulk} is translation and rotation invariant, particularly it is
\begin{equation}
D_{{\rm bulk}, k}^{\rm R}(\chi,\eta^*)=D_{{\rm bulk}, k}^{\rm R}(e^{i\vartheta}\chi+\lambda\mathbf{1}_k,e^{-i\vartheta}\eta^*+\lambda^*\mathbf{1}_k)
\end{equation}
 for any fixed $\vartheta\in\mathbb{R}$ and  $\lambda\in\mathbb{C}$. As aforementioned, this is one of the main features of a bulk 
point. Actually the symmetry group is even larger in this particular case as the invariance even holds for $(\chi,\eta^*)\to(e^{i\vartheta}\chi+\lambda\mathbf{1}_k,e^{-i\vartheta}\eta^*+\mu^*\mathbf{1}_k)$ for any $\vartheta,\lambda,\mu\in\mathbb{C}$. This is due to the analyticity in both arguments, $\chi$ and $\eta^*$, individually.

The result~\eqref{lim.bulk} for class A was previously derived in~\cite[Theorem~1.15]{Simm1}. It was shown in \cite{Simm2} that away from the real axis the same result prevails for the real and quaternion Ginibre ensemble.  We have checked this statement independently using the same technique as below. The group integral is actually the Harish-Chandra--Itzykson--Zuber integral~\cite{HC,IZ} which is explicitly known
\begin{equation}\label{lim-bulk-Gin-det}
\begin{split}
D_{{\rm bulk}, k}^{\rm A}(\chi,\eta^*)=&\exp\left[-\Tr\chi\eta^*\right]\int_{{\rm U}(k)}d\mu(U)\exp\left[\Tr(\chi \widetilde U^\dagger\eta^* \widetilde U)\right]\\
=&\left(\prod_{j=0}^{k-1}j!\right)\frac{\exp\left[-\Tr\chi\eta^*\right]}{\Delta_k(\chi)\Delta_k(\eta^*)}\det[\exp(\chi_a\eta_b^*)]_{a,b=1,\ldots,k},
\end{split}
\end{equation}
where
\begin{equation}\label{Vand}
\Delta_k(x)=\prod_{1\leq a<b\leq k}(x_b-x_a)
\end{equation}
is the Vandermonde determinant. This agrees with a derivation based on planar orthogonal polynomials presented in Appendix~\ref{App:Ak>1}.

The group integrals in~\eqref{lim.bulk} for AI$^\dag$ and AII$^\dag$ are the other two Itzykson-Zuber integrals~\cite{IZ} which are not known explicitly, and only recurrence relations exist~\cite{HeinerThomas}. In particular, the authors  observed that the recurrence relations are incompatible with determinantal or Pfaffian structures.

{\bf Class A:} We start with the derivation for the complex Ginibre ensemble, in order to be self-contained. Let us recall that it is $d_N=N$ and $s=1$ in this case. 

To deal with the characteristic polynomials in~\eqref{D.N.k} we introduce two $N\times k$ dimensional complex Grassmann matrices
\begin{equation}
\xi=\left[\begin{array}{ccc} \xi_{11} & \cdots & \xi_{1k} \\ \vdots & & \vdots \\ \xi_{N1} & \cdots & \xi_{Nk} \end{array}\right]\qquad{\rm and}\qquad \psi=\left[\begin{array}{ccc} \psi_{11} & \cdots & \psi_{1k} \\ \vdots & & \vdots \\ \psi_{N1} & \cdots & \psi_{Nk} \end{array}\right]
\end{equation}
via the standard identity
\begin{equation}\label{Gauss.id}
\begin{split}
&\det[J\otimes\mathbf{1}_k-\mathbf{1}_{N}\otimes Z]\det[J^\dagger\otimes\mathbf{1}_k-\mathbf{1}_{N}\otimes W^*]\\
=&\int [d\xi] [d\psi]\exp\left[\Tr J\xi\xi^\dagger+\Tr J^\dagger\psi\psi^\dagger +\Tr Z \xi^\dagger\xi+\Tr W^*\psi^\dagger \psi\right]
\end{split}
\end{equation}
with $[d\xi] [d\psi]=\prod_{l=1}^{N}\prod_{j=1}^{k}d\xi_{lj}d\xi^*_{lj}d\psi_{lj}d\psi^*_{lj}$ and $\psi^\dagger$ being the transpose and complex conjugate of $\psi_l$ which is the generalisation of~\eqref{Grassdet}. We would like to underline that we have exploited the anti-commutation relations of Grassmann variables when interchanging the matrices $\psi$ and $\psi^\dagger$. 
For the determinant in the denominator in \eqref{D.N.k} we only need to set $\chi_j=\eta_j=0$ in the above expression.

For the complex Ginibre ensemble the matrix $J$ does not satisfy any  symmetries so that we can take the expectation of it without further ado, 
\begin{equation}\label{D.N.k.Gin.a}
\begin{split}
&D_{N,k}^{\rm A}(Z,W^*)=\\
=&\frac{\int [d\xi] [d\psi]\exp\left[\Tr \xi\xi^\dagger\psi\psi^\dagger +\Tr Z \xi^\dagger\xi+\Tr W^*\psi^\dagger \psi\right]}{\int [d\xi][d\psi]\exp\left[\Tr \xi\xi^\dagger\psi\psi^\dagger +\sqrt{N} z_0\Tr \xi^\dagger\xi+\sqrt{N} z_0^*\Tr \psi^\dagger \psi\right]}\\
=&\frac{\int [d\xi] [d\psi]\int_{\mathbb{C}^{k\times k}}[dA]\exp\left[-N\Tr A^\dagger A+i\sqrt{N}\Tr A \psi^\dagger\xi+i\sqrt{N}\Tr A^\dagger \xi^\dagger\psi +\Tr Z \xi^\dagger\xi+\Tr W^*\psi^\dagger \psi\right]}{\int [d\xi] [d\psi]\int_{\mathbb{C}^{k\times k}}[dA]\exp\!\left[-N\Tr A^\dagger A+i\sqrt{N}\Tr A \psi^\dagger\xi+i\sqrt{N}\Tr A^\dagger \xi^\dagger\psi +\sqrt{N} z_0\Tr \xi^\dagger\xi+\sqrt{N} z_0^*\Tr \psi^\dagger \psi\right]},
\end{split}
\end{equation}
where we used the anti-commutativity of the Grassmann variables and applied the Hubbard-Stratonovich transformation~\cite{Hubb} in the second equality. The linearisation in the dyadic matrices $\psi^\dagger\xi$ and $\xi^\dagger\psi$ allows us to integrate over all Grassmann variables again as they are simple Gaussians, i.e.,
\begin{equation}\label{D.N.k.Gin.b.0}
\begin{split}
\int [d\xi] [d\psi]\exp\left[i\sqrt{N}\Tr A \psi^\dagger\xi+i\sqrt{N}\Tr A^\dagger \xi^\dagger\psi +\Tr Z \xi^\dagger\xi+\Tr W^*\psi^\dagger \psi\right]=&{\det}\left[\begin{array}{cc} Z & i\sqrt{N} A^\dagger \\ i\sqrt{N} A & W^* \end{array}\right]^N,
\end{split}
\end{equation}
which leads to
\begin{equation}\label{D.N.k.Gin.b}
\begin{split}
D_{N,k}^{\rm A}(Z,W^*)=\frac{\int_{\mathbb{C}^{k\times k}}[dA]\exp\left[-N\Tr A^\dagger A\right]\det\left[ZW^*/N+A^\dagger (W^*)^{-1}AW^*\right]^N}{\int_{\mathbb{C}^{k\times k}}[dA]\exp\left[-N\Tr A^\dagger A\right]\det\left[|z_0|^2\mathbf{1}_k+A^\dagger A\right]^N}.
\end{split}
\end{equation}
To simplify the determinant we have exploited~\eqref{det-Id}.

When taking the large-$N$ limit, we need to maximise the leading order Lagrangian
\begin{equation}\label{Gin.Lag}
\mathcal{L}=-\Tr A^\dagger A+\Tr \log\left(|z_0|^2\mathbf{1}_k+A^\dagger A\right)
\end{equation}
in $A$. This is evidently uniquely given when
\begin{equation}
A^\dagger A=(1-|z_0|^2)\mathbf{1}_k\ ,
\end{equation}
which follows from differentiating $\mathcal{L}$ in $A^\dagger A$.

We expand about this solution by first going over to the polar decomposition $A=\widetilde{U}H$ with $\widetilde{U}\in{\rm U}(k)$ a unitary matrix and $H=H^\dagger\geq0$ a positive semi-definite matrix. The measure transforms (up to a normalisation) as follows
\begin{equation}
[dA]\propto\det(H\otimes\mathbf{1}_k+\mathbf{1}_k\otimes H)[dH]d\widetilde\mu(\widetilde{U}),
\end{equation}
where $d\widetilde\mu(\widetilde{U})$ is the normalised Haar measure on ${\rm U}(k)$. The Jacobian can be computed by performing first a singular value decomposition of $A$ and then reverse the eigenvalue decomposition of $H$. In this way, the term $\det(H\otimes\mathbf{1}_k+\mathbf{1}_k\otimes H)$ is reminiscent of the fact that the singular value decomposition yields a square of a Vandermonde determinant in the squared singular values of $A$ as well as a factor $\det H$. The singular values are, however, the eigenvalues of $H$ whose eigenvalue decomposition only needs a square of a Vandermonde determinant in the  eigenvalues,  and not their squares which is corrected by the non-trivial term $\det(H\otimes\mathbf{1}_k+\mathbf{1}_k\otimes H)$.

After this transformation we 
set $H=\sqrt{1-|z_0|^2}\mathbf{1}_k+\delta H/\sqrt{N}$ and expand in $1/\sqrt{N}$, showing that the term $\det(H\otimes\mathbf{1}_k+\mathbf{1}_k\otimes H)$ only yields a constant in leading order. For the determinant, we use
\begin{equation}
\begin{split}
\det\left[\frac{ZW^*}{N}+A^\dagger(W^*)^{-1}AW^*\right]=&{\det}\left[\begin{array}{cc} \displaystyle z_0\mathbf{1}_k+\frac{\chi}{\sqrt{N}} &\displaystyle  \left(\sqrt{1-|z_0|^2}\mathbf{1}_k+\frac{\delta H}{\sqrt{N}} \right)\widetilde{U}^\dagger \\ \displaystyle -\widetilde{U}\left(\sqrt{1-|z_0|^2}\mathbf{1}_k+\frac{\delta H}{\sqrt{N}} \right) & \displaystyle z_0^*\mathbf{1}_k+\frac{\eta^*}{\sqrt{N}} \end{array}\right]\\
=&\det\left[\mathbf{1}_{2k}+\frac{1}{\sqrt{N}}\!\left[\!\begin{array}{cc} z_0^*\mathbf{1}_k & -\sqrt{1-|z_0|^2}\widetilde{U}^\dagger \\ \sqrt{1-|z_0|^2}\widetilde{U} & z_0\mathbf{1}_k \end{array}\!\right]\!\!\left[\begin{array}{cc} \chi & \delta H\widetilde{U}^\dagger \\ -\widetilde{U}\delta H & \eta^* \end{array}\right]\!\right]
\end{split}
\end{equation}
which can be found by going back to the structure to the determinant in~\eqref{D.N.k.Gin.b.0} with the matrix in its block form. This  leads to the expansion
\begin{equation}\label{expansion.comp.Gin}
\begin{split}
&\exp[-N\Tr A^\dagger A]\det\left[\frac{ZW^*}{N}+A^\dagger(W^*)^{-1}AW^*\right]^N\\
=&\exp\biggl[-Nk(1-|z_0|^2)+\sqrt{N}\Tr \left(z_0^*\chi+z_0\eta^*\right)-\frac{1}{2}\Tr\left[(z_0^*\chi)^2+(z_0\eta^*)^2-2(1-|z_0|^2) \chi \widetilde{U}^\dagger \eta^* \widetilde{U}\right]\\
&\quad\quad-2(1-|z_0|^2)\Tr \delta H^2-2\sqrt{1-|z_0|^2}\Tr \delta H(z_0^*\chi+z_0\widetilde{U}^\dagger\eta^*\widetilde{U})+\mathcal{O}\left(\frac{||\delta H||^3}{\sqrt{N}}\right)\biggl],
\end{split}
\end{equation}
where $||X||=\sqrt{\Tr XX^\dagger}$ is the Hilbert-Schmidt norm of a matrix. This highlights that we obtain a Gaussian approximation in $\delta H$ uniformly when $||\delta H||\ll N^{1/6}$. One can, however, show that the integral restricted to $||\delta H||\gg N^\epsilon$ for any fixed $\epsilon>0$ is exponentially small compared to the main contribution at $\delta H=0$ so that a restriction to say $||\delta H||\leq N^{1/12}$ is sufficient. We will not work out these mathematical details as they 
do not illuminate the problem further.

The Gaussian integral in $\delta H$ we obtain after these expansions 
are integrated out and we arrive at
\begin{equation}\label{D.N.k.Gin.c}
\begin{split}
D_{N,k}^{\rm A}(Z,W^*)=\exp[\sqrt{N}\Tr \left(z_0^*\chi+z_0\eta^*\right)]\int_{{\rm U}(k)}d\widetilde\mu(\widetilde{U})\exp\left[\Tr(\chi \widetilde{U}^\dagger\eta^* \widetilde{U})\right]\left[1+\mathcal{O}\left(\frac{1}{\sqrt{N}}\right)\right].
\end{split}
\end{equation}  
We note that for $k=1$ it is trivially
\begin{equation}
\prod_{l=1}^kD_{N,1}^{\rm A}(z_l,w^*_l)\sim\exp\left[\sqrt{N}\sum_{j=1}^k\left(z_0^*\chi_j+z_0\eta_j^*\right)\right]\exp\left[\sum_{l=1}^k\chi_l\eta_l^*\right],
\label{DNkGin-bulk-prod}
\end{equation}
which shows how the exponential prefactor comes about for class $A$.
Finally, we bring it into the unifying form~\eqref{lim.bulk} by choosing $U={\rm diag}(\widetilde{U},\widetilde{U})$ and $d\mu(U)=d\widetilde\mu(\widetilde{U})$.

%%%%%%%%%%%%%%%%%%%%%%%%%%%%%%%%%%%%%%%%%%

{\bf Classes AI$^\dagger$ and AII$^\dagger$:}  We turn to the complex symmetric and complex self-dual random matrices  which can be dealt with in parallel. We recall that the scaling is for both ensembles $s=1/2$ while the dimension is $d_N=N$ for class AI$^\dag$ and $d_N=2N$ for class AII$^\dag$. Their corresponding symmetry can be written in a unifying form $J^T=TJT$, with
\begin{equation}
T=\left\{\begin{array}{clll} \mathbf{1}_N, & J\ {\rm is\ symmetric}, & d_N=N, & \mbox{class R=AI}^\dag
\\ \tau_2\otimes\mathbf{1}_N=\Sigma_y, & J\ \text{is\ self-dual}, &
 d_N=2N, & \mbox{class R=AII}^\dag
,\end{array}\right.
\end{equation}
where $\tau_2$ is the second Pauli matrix.
This matrix has a dual one of the form
\begin{equation}
\hat{T}=\left\{\begin{array}{cll} \tau_2\otimes\mathbf{1}_k, & J\ {\rm is\ symmetric}, &
\mbox{class R=AI}^\dag,
\\ \mathbf{1}_{2k}, & J\ \text{is\ self-dual}, & \mbox{class R=AII}^\dag,\end{array}\right.
\end{equation}
which are both of dimension $2k$.
Using these two matrices, we may rewrite the determinants in  terms of Pfaffians
\begin{equation}
\det[J\otimes\mathbf{1}_k-\mathbf{1}_{d_N}\otimes Z]\det[J^\dagger\otimes\mathbf{1}_k-\mathbf{1}_{d_N}\otimes W^*]=\pm{\rm Pf}[TJ\otimes\hat{T}-T\otimes \tilde{z}\hat{T}]{\rm Pf}[TJ^\dagger\otimes\hat{T}-T\otimes \tilde{w}\hat{T}],
\end{equation}
with $\tilde{z}={\rm diag}(Z,Z)$ and $\tilde{w}={\rm diag}(W^*,W^*)$. The overall sign is, however, not a problem as it is cancelled in the ratio~\eqref{D.N.k}.
The two Pfaffians can be expressed as Gaussian integrals over two $d_N\times 2k$ real Grassmann matrices
\begin{equation}
\xi=\left[\begin{array}{ccc} \xi_{11} & \cdots & \xi_{1\,2k} \\ \vdots & & \vdots \\ \xi_{d_N1} & \cdots & \xi_{d_N\,2k} \end{array}\right]\qquad{\rm and}\qquad \psi=\left[\begin{array}{ccc} \psi_{11} & \cdots & \psi_{1\,2k} \\ \vdots & & \vdots \\ \psi_{d_N1} & \cdots & \psi_{d_N\,2k} \end{array}\right],
\end{equation}
similar to~\eqref{GrassPf}, namely
\begin{equation}
\begin{split}
&{\rm Pf}[TJ\otimes\hat{T}-T\otimes \tilde{z}\hat{T}]{\rm Pf}[TJ^\dagger\otimes\hat{T}-T\otimes \tilde{w}\hat{T}]\\
=&\int [d\xi][d\psi]\exp[\Tr TJ\xi\hat{T}\xi^T+\Tr TJ^\dagger\psi\hat{T}\psi^T+\Tr\xi^TT\xi\tilde{z}\hat{T}+\Tr\psi^TT\psi\tilde{w}\hat{T}].
\end{split}
\end{equation}

Let the superscript R be either AI$^\dag$ or AII$^\dag$ to point out the ensemble in the ensuing computation.  Once we carry out the Gaussian average over $J$ we arrive at
\begin{equation}\label{D.N.k.A.Gin.a}
\begin{split}
D_{N,k}^{\rm R}(Z,W^*)=&\frac{\int [d\xi][d\psi]\exp\left[\Tr \xi\hat T\xi^TT\psi\hat T\psi^TT+\Tr\xi^TT\xi\tilde{z}\hat{T}+\Tr\psi^TT\psi\tilde{w}\hat{T}\right]}{\int [d\xi][d\psi]\exp\left[\Tr \xi\hat T\xi^TT\psi\hat T\psi^TT+\sqrt{\frac{d_N}{2}}z_0\Tr\xi^TT\xi\hat{T}+\sqrt{\frac{d_N}{2}}z_0^*\Tr\psi^TT\psi\hat{T}\right]}\\
=&
\frac{\int [d\xi][d\psi] \int_{\mathcal{M}}[dA]\exp\left[\widehat{\mathcal{L}}(A,\xi,\psi,\tilde{z},\tilde{w})\right]}{\int [d\xi][d\psi] \int_{\mathcal{M}}[dA]\exp\left[\widehat{\mathcal{L}}(A,\xi,\psi,z_0\mathbf{1}_{2k},z_0^*\mathbf{1}_{2k})\right]},
\end{split}
\end{equation}
with
\begin{equation}
\widehat{\mathcal{L}}(A,\xi,\psi,\tilde{z},\tilde{w})=-\frac{d_N}{2}\Tr AA^\dagger+i\sqrt{\frac{d_N}{2}}\left(\Tr A \hat{T} \psi^TT\xi+\Tr A^T\xi^TT\psi\hat{T} \right)+\Tr\xi^TT\xi\tilde{z}\hat{T}+\Tr\psi^TT\psi\tilde{w}\hat{T}.
\end{equation}
The Hubbard-Stratonovich transformation has been carried out with the help of the set $\mathcal{M}$ of $2k\times 2k$ quaternion (if $J$ has been symmetric) or real (if $J$ has been self-dual) matrices $A$. The different matrix spaces come into play as we need the property $A^*=\hat{T}A\hat{T}$ to guarantee an integrable Gaussian in $A$ where the completion of squares, especially the shift $A\to A+i\sqrt{\frac{2}{d_N}}\xi^TT\psi\hat{T}$, works out.

The integrals over $\xi$ and $\psi$ are again a Pfaffian determinant,
\begin{equation}
\begin{split}
&\int [d\xi][d\psi] \exp\left[i\sqrt{\frac{d_N}{2}}\left(\Tr A \hat{T} \psi^TT\xi+\Tr A^T\xi^TT\psi\hat{T} \right)+\Tr\xi^TT\xi\tilde{z}\hat{T}+\Tr\psi^TT\psi\tilde{w}\hat{T}\right]\\
=&{\rm Pf}\left[\begin{array}{cc} T\otimes\tilde{z}\hat{T} & \displaystyle i\sqrt{\frac{d_N}{2}}\,T\otimes\hat{T}A^T \\ \displaystyle i\sqrt{\frac{d_N}{2}}\,T\otimes A\hat{T} & T\otimes\tilde{w}\hat{T} \end{array}\right]=\det\left[\tilde{z}\tilde{w}+\frac{d_N}{2} A^\dagger \tilde{w}^{-1}A\tilde{w} \right]^{d_N/2}.
\end{split}
\end{equation}
The square root of the determinant is chosen such that the sign agrees with the Pfaffian.  The root is actually not a problem for the class AII$^\dag$ since $d_N=2N$ is even so that no root is taken. In contrast, for class AI$^\dag$ we understand the analytic structure of the right hand side in terms of the left hand side which is evidently a polynomial of the matrix entries.

Let us summarise so far. The ratio of averages is with the above input equals to
\begin{equation}\label{D.N.k.A.Gin.b}
\begin{split}
D_{N,k}^{\rm R}(Z,W^*)=&\frac{\int_{\mathcal{M}}[dA]\exp\left[-\frac{d_N}{2}\Tr AA^\dagger\right]\det\left[\frac{2}{d_N}\tilde{z}\tilde{w}+ A^\dagger \tilde{w}^{-1}A\tilde{w} \right]^{d_N/2}}{\int_{\mathcal{M}}[dA]\exp\left[-\frac{d_N}{2}\Tr AA^\dagger\right]\det\left[|z_0|^2\mathbf{1}_{2k}+ A^\dagger A \right]^{d_N/2}}.
\end{split}
\end{equation}
This duality formula for classes AI$^\dag$ and AII$^\dag$ for $k>1$ was derived independently in \cite{Liu}. 

To take the limit $N\to\infty$, we follow the ideas of class A and need to maximise  the leading order of the Lagrangian
\begin{equation}\label{AI.Lag}
\mathcal{L}=-\Tr A^\dagger A+\Tr \log\left(|z_0|^2\mathbf{1}_{2k}+A^\dagger A\right).
\end{equation}
The only difference to the complex Ginibre ensemble is that now $A$ is not complex but quaternion or real, respectively. 

Like in class A, the variation of $\mathcal{L}$ in $A^\dagger A$ gives the unique solution $A^\dagger A=(1-|z_0|^2)\mathbf{1}_{2k}$ so that a reasonable expansion could be done with the help of the substitution
 \begin{equation}
  A=U\left(\sqrt{1-|z_0|^2}\mathbf{1}_{2k}+\sqrt{\frac{2}{d_N}}\delta H\right).
 \end{equation}
Here, we made use of the polar decomposition for quaternion and real matrices, i.e., $U\in{\rm USp}(2k)$ is unitary symplectic and $\delta H\in{\rm Self}(2k)$ is quaternion self-dual when $J$ has been complex symmetric (class AI$^\dag$), and $U\in{\rm O}(2k)$ is orthogonal and $\delta H\in{\rm Sym}(2k)$ is real symmetric when $J$ has been complex self-dual (class AII$^\dag$). Apart from the different matrix spaces, the analysis  is completely analogous to class A, and we eventually arrive at
\begin{equation}\label{D.N.k.Gin.A}
\begin{split}
D_{N,k}^{\rm R}(Z,W^*)=e^{\sqrt{d_N}\Tr \left(z_0^*\chi +z_0\eta^*\right)}\int_{\mathcal{G}}d\mu(U)\exp\left[\frac{1}{2}\Tr(\mbox{diag}(\chi,\chi) U^\dagger\mbox{diag}(\eta^*,\eta^*) U)\right]\left[1+\mathcal{O}\left(\frac{1}{\sqrt{N}}\right)\right],
\end{split}
\end{equation}
with the group $\mathcal{G}={\rm USp}(2k)$ (class R=AI$^\dag$) or $\mathcal{G}={\rm O}(2k)$ (class R=AII$^\dag$). This also concludes the proof for these two classes.

%%%%%%%%%%%%%%%%%%%%%%%%%%%%%%%%%%%%%%%%%%%%%%%%%%%%%%%%
\subsection{Edge limit for $k>1$ pairs of characteristic polynomials}\label{edgek}

\begin{theorem}[Edge limit for $k>1$]\label{thm:edge}\

In the edge limit \eqref{zoomk} with $|z_0|=1$ 
we obtain the following 
expression
\begin{equation}\label{lim.edge}
\begin{split}
D_{{\rm edge}, k}^{\rm R}(\chi,\eta^*):=&\lim_{N\to\infty}\widehat{D}_{N,k}^{\rm R}(Z,W^*)\\
=&\frac{\int_{\mathcal{M}}[dA]\exp\left[-\frac{z_0^*}{\sqrt{2}}\Tr A^\dagger A{\rm diag}(\chi,\chi)-\frac{z_0}{\sqrt{2}}\Tr AA^\dagger {\rm diag}(\eta^*,\eta^*)-\frac{1}{2}\Tr( A^\dagger A)^2 \right]}{
\widetilde{Z}_{k}^{\rm R}\prod_{j=1}^k\exp\left[\frac{(z_0^*\chi_j +z_0\eta^*_j)^2}{2}\right]D_{{\rm edge}}^{\rm R}(\chi_j,\eta_j^*)
},
\end{split}
\end{equation}
where we recall the definitions~\eqref{D.N.k} and~\eqref{hatDNk}, the double scaling~\eqref{zoomk}, and the results for the average of a pair of characteristic polynomials at the edge in Theorem~\ref{Thm_lim_char_exp}.  The saddle manifold $\mathcal{M}$ is the set of complex $k\times k$ matrices $\mathbb{C}^{k\times k}$ (class R=A),  quaternion $2k\times 2k$ matrices $\mathbb{H}^{k\times k}$ (class R=AI$^\dag$), and real $2k\times 2k$ matrices $\mathbb{R}^{2k\times 2k}$ (class R=AII$^\dag$). All three sets are equipped with their corresponding Lebesgue measure $[dA]$. Similar to the bulk statistics, we have embedded the complex $k\times k$ matrices in the $2k\times 2k$ matrices, $A={\rm diag}(\widetilde{A},\widetilde{A})$ with $\widetilde{A}\in\mathbb{C}^{k\times k}$, to have a unifying result~\eqref{lim.edge} for all 
three ensembles.
The normalisation constants are for the
\begin{enumerate}
\item	complex Ginibre ensemble (class A):
\begin{equation}
\begin{split}
\widetilde{Z}_{k}^{\rm A}:=&\frac{\int_{\mathbb{C}^{k\times k}}[d\widetilde{A}]\exp\left[-\Tr( \widetilde{A}^\dagger \widetilde{A})^2 \right]}{
[D_{{\rm edge}}^{\rm A}(0,0)]^k
}=\frac{\pi^{k^2}}{\prod_{j=0}^{k-1}(j!)^2}\det\left[\Gamma\left(\frac{a+b-1}{2}\right)\right]_{a,b=1,\ldots,k},
\end{split}
\label{lim.edge.3Gin}
\end{equation}

\item complex symmetric random matrices (class AI$^\dag$):
\begin{equation}
\begin{split}
\widetilde{Z}_{k}^{\rm AI^\dag}:=&\frac{\int_{\mathbb{H}^{k\times k}}[dA]\exp\left[-\frac{1}{2}\Tr( A^\dagger A)^2 \right]}{
[D_{{\rm edge}}^{\rm AI^\dag}(0,0)]^k
}=\frac{\pi^{2k^2}(\pi/2)^{k/2}}{\prod_{j=0}^{k-1}[(2j+1)!]^2}\Pf\left[(b-a)\Gamma\left(\frac{a+b-1}{2}\right)\right]_{a,b=1,\ldots,2k},
\end{split}
\label{lim.edge.AI}
\end{equation}

\item complex self-dual random matrices (class AII$^\dag$):
\begin{equation}
\begin{split}
\widetilde{Z}_{k}^{\rm AII^\dag}:=&\frac{\int_{\mathbb{R}^{2k\times 2k}}[dA]\exp\left[-\frac{1}{2}\Tr( A^\dagger A)^2 \right]}{
[D_{{\rm edge}}^{\rm AII^\dag}(0,0)]^k}\\
=&\left(\frac{\sqrt{\pi}}{2^{5/4}\,{\rm arccoth}(\sqrt{2})}\right)^k\frac{(2\pi)^{2k^2}}{\prod_{j=0}^{k-1}[(2j)!]^2}\\
&\hspace*{-1cm}\times{\rm Pf}\left[\left(\tfrac{2^{b/2}}{2a-1}\,{ _2F_1}\left[\tfrac{5-2b}{4},\tfrac{2a-1}{4};\tfrac{2a+3}{4};\tfrac{1}{2}\right]-\tfrac{2^{a/2}}{2b-1}\,{ _2F_1}\left[\tfrac{5-2a}{4},\tfrac{2b-1}{4};\tfrac{2b+3}{4};\tfrac{1}{2}\right]\right)\Gamma\left(\tfrac{a+b-1}{2}\right)\right]_{a,b=1,\ldots,2k},
\end{split}
\label{lim.edge.AII}
\end{equation}
where ${ _2F_1}$ is Gauss hypergeometric function.

\end{enumerate} 
 
\end{theorem}

Anew, there is a translation invariance though only parallel to the boundary, meaning
\begin{equation}
D_{{\rm edge}, k}^{\rm R}(\chi,\eta^*)=D_{{\rm edge}, k}^{\rm R}(\chi+\lambda z_0,\eta^*-\lambda z_0^*)
\end{equation}
for an arbitrary $\lambda\in\mathbb{R}$. It only says that we can shift the point we are zooming into parallel to the boundary without changing the statistics. This symmetry is like in the bulk enhanced to  $(\chi,\eta^*)\to(\chi+\lambda z_0,\eta^*-\lambda z_0^*)$ with a complex $\lambda\in\mathbb{C}$ since there is an independent analyticity in $\chi$ and $\eta^*$.

The result for the complex Ginibre ensemble (class A) is again known, see~\cite[Theorem~1.14]{Simm1}. Like for the bulk, the corresponding integral
\begin{equation}\label{edge.Gin.int}
D_{{\rm edge}, k}^{\rm A}(\chi,\eta^*)=\frac{\int_{\mathbb{C}^{k\times k}}[d\widetilde{A}]\exp\left[-\sqrt{2}z_0^*\Tr \widetilde{A}^\dagger \widetilde{A}\chi-\sqrt{2}z_0\Tr \widetilde{A}\widetilde{A}^\dagger\eta^*-\Tr( \widetilde{A}^\dagger \widetilde{A})^2 \right]}{
\widetilde{Z}_{k}^{\rm A}\prod_{j=1}^k\exp\left[\frac{(z_0^*\chi_j +z_0\eta^*_j)^2}{2}\right]D_{{\rm edge}}(\chi_j,\eta_j^*)
},
\end{equation}
 can be carried out as it is related to a deformation of a Wishart ensemble. The result is
 \begin{equation}\label{edge-Gin-simp}
 D_{{\rm edge}, k}^{\rm A}(\chi,\eta^*)=\frac{\left(\prod_{j=0}^{k-1}2^{-j}\sqrt{\pi}\,(j!)^2\right)}{\det\left[\Gamma\left([a+b-1]/2\right)\right]_{a,b=1,\ldots,k}}\frac{\det\left[{\rm erfc}\left[(z_0^*\chi_a+z_0\eta^*_b)/\sqrt{2}\right]e^{\chi_a\eta^*_b}\right]_{a,b=1,\ldots,k}}{
\Delta_k(\chi)\Delta_k(\eta^*)\prod_{j=1}^k{\rm erfc}\left[(z_0^*\chi_j+z_0\eta_j^*)/\sqrt{2}\right]e^{\chi_j\eta^*_j}}.
 \end{equation}
It is compared to the derivation using orthogonal polynomials in Appendix \ref{App:Ak>1}. 
We present the derivation of the results for class A to be self-consistent and to prepare for the other two symmetry classes.

{\bf Class A:} We anew begin with the derivation for the complex Ginibre ensemble and can even start with~\eqref{D.N.k.Gin.b}  where $z_0$ is now a complex phase, i.e., $|z_0|=1$. The expansion is much simpler than in the bulk as the Lagrangian~\eqref{Gin.Lag}, which is still the same, takes its maximum at the origin in the current situation. Therefore, we only rescale as follows $A= (2/N)^{1/4}\widetilde{A}$ and, then, expand the determinant 
\begin{equation}\label{exp.edge}
\begin{split}
\det\left[\frac{ZW^*}{N}+\sqrt{\frac{2}{N}}\widetilde{A}^\dagger(w^*)^{-1}\widetilde{A}w^*\right]^N=&\det\left(\mathbf{1}_{2k}+\left[\begin{array}{cc} N^{-1/2}z_0^*\chi & (2/N)^{1/4}z_0^*\widetilde{A}^\dagger \\ -(2/N)^{1/4}z_0\widetilde{A} & N^{-1/2}z_0\eta^* \end{array}\right]\right)^N\\
&\hspace*{-1cm}=\exp\biggl[\sqrt{N}\Tr(z_0^*\chi+z_0\eta^*)-\frac{1}{2}\Tr[(z_0^*\chi)^2+(z_0\eta^*)^2]+\sqrt{2N}\Tr \widetilde{A}^\dagger \widetilde{A}\\
&\hspace*{-0.5cm}-\sqrt{2}z_0^*\Tr \widetilde{A}^\dagger \widetilde{A}\chi-\sqrt{2}z_0\Tr \widetilde{A}\widetilde{A}^\dagger \eta^*-\Tr( \widetilde{A}^\dagger \widetilde{A})^2+\mathcal{O}\left(\frac{||\widetilde{A}^\dagger \widetilde{A}||^3}{\sqrt{N}}\right)\biggl].
\end{split}
\end{equation}
The error shows that it becomes uniform when assuming a cut-off for the integral so that $||\widetilde{A}^\dagger \widetilde{A}||\ll N^{1/6}$. Like before this can be done as the integrand is exponentially suppressed compared to its maximal contribution at the origin whenever $||\widetilde{A}^\dagger \widetilde{A}||\geq N^{\epsilon}$ for any fixed $\epsilon>0$ so that Lebesgue's dominated convergence theorem can be applied. We omit the mathematical details here but only want to say that the maximum of the integrand in the region can be estimated at $\widetilde{A}^\dagger \widetilde{A}= N^{\epsilon}\mathbf{1}_k$.

Plugging the expansion~\eqref{exp.edge} into~\eqref{D.N.k.Gin.b} leads to
\begin{equation}\label{D.N.k.Gin.edge}
\begin{split}
D_{N,k}^{\rm A}(Z,W^*)=&\exp\biggl[\sqrt{N}\Tr(z_0^*\chi+z_0\eta^*)-\frac{1}{2}\Tr[(z_0^*\chi)^2+(z_0\eta^*)^2]\biggl]\\
&\times\frac{\int_{\mathbb{C}^{k\times k}}[d\widetilde{A}]\exp\left[-\sqrt{2}z_0^*\Tr \widetilde{A}^\dagger\widetilde{A}\chi-\sqrt{2}z_0\Tr \widetilde{A}\widetilde{A}^\dagger \eta^*-\Tr( \widetilde{A}^\dagger \widetilde{A})^2\right]}{\int_{\mathbb{C}^{k\times k}}[d\widetilde{A}]\exp\left[-\Tr( \widetilde{A}^\dagger \widetilde{A})^2\right]}\left[1+\mathcal{O}\left(\frac{1}{\sqrt{N}}\right)\right].
\end{split}
\end{equation}
Exploiting the result~\eqref{Dedge-A}  and dividing~\eqref{D.N.k.Gin.edge} by the product $\prod_{j=1}^k D_{\rm edge}^{\rm A}(z_j,w^*_j)$ we find the claim up to the normalisation constant. We underline that we have doubled the matrix dimension by the embedding $A={\rm diag}(\widetilde{A},\widetilde{A})$ and set $[dA]=[d\widetilde{A}]$, we abuse the notation and call the new matrix again $A$ to write the formula in a unifying way with the other two ensembles.

{\bf Normalisation of Class A:} We can compute the normalisation~\eqref{lim.edge.3Gin} by using~\eqref{edge.origin}  and computing the integral
\begin{equation}
\int_{\mathbb{C}^{k\times k}}[d\widetilde{A}]\exp\left[-\Tr( \widetilde{A}^\dagger \widetilde{A})^2 \right]=\int_{\mathbb{C}^{k\times k}}[d\widetilde{A}]\exp\left[-\Tr( \widetilde{A}^\dagger \widetilde{A}) \right]\frac{\int_{\mathbb{C}^{k\times k}}[d\widetilde{A}]\exp\left[-\Tr( \widetilde{A}^\dagger \widetilde{A})^2 \right]}{\int_{\mathbb{C}^{k\times k}}[d\widetilde{A}]\exp\left[-\Tr( \widetilde{A}^\dagger \widetilde{A}) \right]}.
\end{equation}
The extension of the fraction by the Gaussian integral allows us to perform the singular value decomposition $\widetilde{A}=U\sqrt{S}V^\dagger$ with $U,V\in{\rm U}(k)$ and $S={\rm diag}(s_1,\ldots,s_k)\geq1$, without worrying about the normalisation while the Gaussian prefactor can be readily computed,
\begin{equation}
\int_{\mathbb{C}^{k\times k}}[d\widetilde{A}]\exp\left[-\Tr( \widetilde{A}^\dagger \widetilde{A})^2 \right]=\pi^{k^2}\frac{\int_{\mathbb{R}_+^{k}}[dS]|\Delta_k(S)|^2\exp\left[-\Tr S^2 \right]}{\int_{\mathbb{R}_+^{k}}[dS]|\Delta_k(S)|^2\exp\left[-\Tr S \right]}.
\end{equation}
The integral in the denominator is a Selberg integral given in~\cite[Eq.~(17.6.5)]{Mehta}, and the numerator can be simplified with the help of Andr\'eief's integral indentity~\cite{Andreief} (sometimes also called Cauchy-Binet integral formula) so that we arrive at
\begin{equation}
\begin{split}
\int_{\mathbb{C}^{k\times k}}[d\widetilde{A}]\exp\left[-\Tr( \widetilde{A}^\dagger \widetilde{A})^2 \right]=&\frac{\pi^{k^2}}{\prod_{j=0}^{k-1}(j!)^2}\det\left[\int_0^\infty ds\,s^{a+b-2}e^{-s^2}\right]_{a,b=1,\ldots,k}\\
=&\frac{\pi^{k^2}}{2^k\prod_{j=0}^{k-1}(j!)^2}\det\left[\Gamma\left(\frac{a+b-1}{2}\right)\right]_{a,b=1,\ldots,k}.
\end{split}
\end{equation}
When combining this with the factor $2^k$ from~\eqref{edge.origin} we find~\eqref{lim.edge.3Gin}. 
 
Yet, the integral in the numerator of~\eqref{edge.Gin.int} can be carried out where we employ the same trick as for the normalisation constant
\begin{equation}
\begin{split}
&\int_{\mathbb{C}^{k\times k}}[d\widetilde{A}]\exp\left[-\sqrt{2}z_0^*\Tr \widetilde{A}^\dagger\widetilde{A}\chi-\sqrt{2}z_0\Tr \widetilde{A}\widetilde{A}^\dagger \eta^*-\Tr( \widetilde{A}^\dagger \widetilde{A})^2\right]\\
=&\int_{\mathbb{C}^{k\times k}}[d\widetilde{A}]\exp\left[-\sqrt{2}z_0^*\Tr \widetilde{A}^\dagger\widetilde{A}\chi-\sqrt{2}z_0\Tr \widetilde{A}\widetilde{A}^\dagger \eta^*\right]\\
&\times\frac{\int_{\mathbb{C}^{k\times k}}[d\widetilde{A}]\exp\left[-\sqrt{2}z_0^*\Tr \widetilde{A}^\dagger\widetilde{A}\chi-\sqrt{2}z_0\Tr \widetilde{A}\widetilde{A}^\dagger \eta^*-\Tr( \widetilde{A}^\dagger \widetilde{A})^2\right]}{\int_{\mathbb{C}^{k\times k}}[d\widetilde{A}]\exp\left[-\sqrt{2}z_0^*\Tr \widetilde{A}^\dagger\widetilde{A}\chi-\sqrt{2}z_0\Tr \widetilde{A}\widetilde{A}^\dagger \eta^*\right]}\\
=&\frac{\pi^{k^2}}{\prod_{a,b=1}^k(\sqrt{2}z_0^*\chi_a+\sqrt{2}z_0\eta^*_b)}\\
&\times\frac{\int_{\mathbb{R}_+^{k}}[dS]\int_{{\rm U}(k)^2}d\mu(U)d\mu(V)|\Delta_k(S)|^2\exp\left[-\sqrt{2}z_0^*\Tr(V^\dag SV\chi)-\sqrt{2}z_0\Tr(USU^\dag\eta^*)-\Tr S^2\right]}{\int_{\mathbb{R}_+^{k}}[dS]\int_{{\rm U}(k)^2}d\mu(U)d\mu(V)|\Delta_k(S)|^2\exp\left[-\sqrt{2}z_0^*\Tr(V^\dag SV\chi)-\sqrt{2}z_0\Tr(USU^\dag\eta^*)\right]}.
\end{split}
\end{equation}
The group integrals against the normalised Haar measures $d\mu(U)d\mu(V)$ are Harish-Chandra--Itzykson--Zuber integrals~\cite{HC,IZ} which can be carried out, and after we apply the Andr\'eief integration formula~\cite{Andreief} the expression simplifies to
\begin{equation}
\begin{split}
&\int_{\mathbb{C}^{k\times k}}[d\widetilde{A}]\exp\left[-\sqrt{2}z_0^*\Tr \widetilde{A}^\dagger\widetilde{A}\chi-\sqrt{2}z_0\Tr \widetilde{A}\widetilde{A}^\dagger \eta^*-\Tr( \widetilde{A}^\dagger \widetilde{A})^2\right]\\
=&\frac{\pi^{k^2}}{\prod_{a,b=1}^k(\sqrt{2}z_0^*\chi_a+\sqrt{2}z_0\eta^*_b)}\frac{\det\left[\int_0^\infty ds \exp\left(-\sqrt{2}(z_0^*\chi_a+z_0\eta^*_b)s-s^2\right)\right]_{a,b=1,\ldots,k}}{\det\left[\int_0^\infty ds \exp\left(-\sqrt{2}(z_0^*\chi_a+z_0\eta^*_b)s\right)\right]_{a,b=1,\ldots,k}}\\
=&\frac{\pi^{k^2}}{\prod_{a,b=1}^k(\sqrt{2}z_0^*\chi_a+\sqrt{2}z_0\eta^*_b)}\frac{\det\left[\frac{\sqrt{\pi}}{2}{\rm erfc}\left[\frac{z_0^*\chi_a+z_0\eta^*_b}{\sqrt{2}}\right]\exp\left(\frac{(z_0^*\chi_a+z_0\eta^*_b)^2}{2}\right)\right]_{a,b=1,\ldots,k}}{\det\left[\frac{1}{\sqrt{2}(z_0^*\chi_a+z_0\eta^*_b)}\right]_{a,b=1,\ldots,k}},
\end{split}
\end{equation}
where we have used the complementary error function~\eqref{erfc-def} once we have completed the squares. The determinant in the denominator is a Cauchy determinant~\cite{determinant} which can be expressed as follows
\begin{equation}
\det\left[\frac{1}{\sqrt{2}(z_0^*\chi_a+z_0\eta^*_b)}\right]_{a,b=1,\ldots,k}=\frac{\Delta_k(\sqrt{2}z_0^*\chi)\Delta_k(\sqrt{2}z_0\eta^*)}{\prod_{a,b=1}^k(\sqrt{2}z_0^*\chi_a+\sqrt{2}z_0\eta^*_b)}=2^{k(k-1)/2}\frac{\Delta_k(\chi)\Delta_k(\eta^*)}{\prod_{a,b=1}^k(\sqrt{2}z_0^*\chi_a+\sqrt{2}z_0\eta^*_b)},
\end{equation}
for which we exploited $|z_0|^2=1$ in the second equality. Finally, we put these results together with the normalisation~\eqref{lim.edge.3Gin}
and the normalisation from \eqref{hatDNk} given in \eqref{Dedge-A}, to obtain the simplification~\eqref{edge-Gin-simp}.

{\bf Classes AI$^\dagger$ and AII$^\dagger$:} The very same strategy as for class A to derive~\eqref{lim.edge} works out for both, complex symmetric and complex self-dual matrices.
The only difference is the rescaling, which is given by $A\to (\frac{2}{d_N})^{-1/4}A$, and the change of parameters, in particular we need to replace $N\to d_N/2$, $\chi\to {\rm diag}(\chi/\sqrt{2},\chi/\sqrt{2})$ and  $\eta^*\to{\rm diag}(\eta^*/\sqrt{2},\eta^*/\sqrt{2})$. When expanding along the lines of~\eqref{exp.edge}, we find
\begin{equation}\label{D.N.k.A.Gin.edge}
\begin{split}
D_{N,k}(z,w^*)=&\exp\biggl[\sqrt{d_N}\Tr(z_0^*\chi+z_0\eta^*)-\frac{1}{2}\Tr[(z_0^*\chi)^2+(z_0\eta^*)^2]\biggl]\\
&\hspace*{-1.5cm}\times\frac{\int_{\mathcal{M}}[dA]\exp\left[-\frac{z_0^*}{\sqrt{2}}\Tr A^\dagger A{\rm diag}(\chi,\chi)-\frac{z_0}{\sqrt{2}}\Tr AA^\dagger {\rm diag}(\eta^*,\eta^*)-\frac{1}{2}\Tr( A^\dagger A)^2\right]}{\int_{\mathcal{M}}[dA]\exp\left[-\frac{1}{2}\Tr( A^\dagger A)^2\right]}\left[1+\mathcal{O}\left(\frac{1}{\sqrt{N}}\right)\right],
\end{split}
\end{equation}
where $\mathcal{M}$ is the set of either the $2k\times 2k$ quaternion (for class AI$^\dag$) or real (for class AII$^\dag$) matrices. To render this limit mathematically rigorous one needs to cut-off from the original integration the region $||A^\dag A||\geq N^\epsilon$ for some $0<\epsilon<1/6$. Then, the expansion is uniform on the complement while for $||A^\dag A||\geq N^\epsilon$ one can find an exponentially suppressed bound compared to the main contribution about the origin.

Once we normalise along the definitions~\eqref{D.N.k} and~\eqref{hatDNk}, we arrive at~\eqref{lim.edge}. What is left to do is to compute the normalisation constants~\eqref{lim.edge.AI} and~\eqref{lim.edge.AII}.

{\bf Normalisation of Class AI$^\dagger$:} We compute along the same ideas as before to find the normalisation~\eqref{lim.edge.AI}. We use the result~\eqref{edge.origin} and evaluate the integral
\begin{equation}
\int_{\mathbb{H}^{k\times k}}[dA]\exp\left[-\frac{1}{2}\Tr( A^\dagger A)^2 \right]=\int_{\mathbb{H}^{k\times k}}[dA]\exp\left[-\frac{1}{2}\Tr( A^\dagger A) \right]\frac{\int_{\mathbb{H}^{k\times k}}[dA]\exp\left[-\Tr( A^\dagger A)^2/2 \right]}{\int_{\mathbb{H}^{k\times k}}[dA]\exp\left[-\Tr( A^\dagger A)/2 \right]}.
\end{equation}
We introduced the factor of $1/2$ in the Gaussian due to
\begin{equation}
\frac{1}{2}\Tr( A^\dagger A)=\sum_{l=1}^4\sum_{a,b=1}^k (A_{ab}^{l})^2,
\end{equation}
where $A_{ab}^{l}$ with $l=1,2,3,4$ are the four real components of a quaternion. Like for the complex Ginibre case, we integrate out the Gaussian in the prefactor and go over to a singular value decomposition $A=U{\rm diag}(\sqrt{S},\sqrt{S})V^\dagger$ with $U,V\in{\rm USp}(2k)$ and $S={\rm diag}(s_1,\ldots,s_k)\geq0$ without worrying about the normalisation constants resulting from such a change of variables as those in the numerator and denominator cancel each other,
\begin{equation}
\int_{\mathbb{H}^{k\times k}}[dA]\exp\left[-\frac{1}{2}\Tr( A^\dagger A)^2 \right]=\pi^{2k^2}\frac{\int_{\mathbb{R}_+^{k}}[dS]|\Delta_k(S)|^4\det S\exp\left[-\Tr S^2 \right]}{\int_{\mathbb{R}_+^{k}}[dS]|\Delta_k(S)|^4\det S\exp\left[-\Tr S \right]}.
\end{equation}
Let us recall that the singular values of a quaternion matrix are Kramers degenerate.
Anew, we notice that the integral in the denominator is a Selberg integral and employ the result of~\cite[Eq.~(17.6.5)]{Mehta}.  For the numerator we use the trick suggested in~\cite[Sec.~4]{Pfaffian} and double the number of integration variables by introducing first derivatives of Dirac delta functions,
\begin{equation}
\begin{split}
&\int_{\mathbb{R}_+^{k}}[dS]|\Delta_k(S)|^4\det S\exp\left[-\Tr S^2 \right]\\
=&\frac{2^k k!}{(2k)!}\int_{\mathbb{R}_+^{2k}} ds_1\cdots ds_{2k}\Delta_{2k}(s_1,\ldots,s_{2k}){\rm Pf}\left[\frac{1}{2}(\partial_a-\partial_b)\delta(s_a-s_b)\right]_{a,b=1,\ldots,2k} \prod_{j=1}^{2k}\sqrt{s_j}e^{-s_j^2/2}.
\end{split}
\end{equation}
The combinatorial prefactor  properly normalises the rest which can be checked when expanding the Pfaffian and  applying the derivatives. They effectively only act on terms of pairwise differences in the Vandermonde determinant as they would otherwise vanish when evaluating the Dirac delta function, i.e.,
\begin{equation}
\prod_{j=1}^k \frac{1}{2}(\partial_{2j}-\partial_{2j-1})\,\Delta_{2k}(s_1,\ldots,s_{2k})\bigl|_{s_{2j}=s_{2j-1}}=|\Delta_k(s_1,s_3,\ldots,s_{2k-1})|^4.
\end{equation}
The benefit of this way of writing the integral is that we can expand the Vandermonde determinant $\Delta_{2k}(s_1,\ldots,s_{2k})$ and apply de Bruijn's integral identity~\cite{deBruijn} which yields
\begin{equation}
\begin{split}
&\int_{\mathbb{H}^{k\times k}}[dA]\exp\left[-\frac{1}{2}\Tr( A^\dagger A)^2 \right]\\
=&\frac{2^k\pi^{2k^2}}{\prod_{j=0}^{k-1}(2j+2)!(2j+1)!}\frac{2^k k!}{(2k)!}(2k)!{\rm Pf}\left[\int_{\mathbb{R}_+^{2}}ds_1ds_2 s_1^{a-1/2}s_2^{b-1/2}e^{-(s_1^2+s_2^2)/2}\frac{1}{2}(\partial_1-\partial_2)\delta(s_1-s_2)\right]_{a,b=1}
^{2k}\\
=&\frac{\pi^{2k^2}}{\prod_{j=0}^{k-1}[(2j+1)!]^2}{\rm Pf}\left[(b-a)\int_{0}^\infty ds_1 s_1^{a+b-2}e^{-s_1^2}\right]_{a,b=1}
^{2k}.
\end{split}
\end{equation}
The remaining integral in the Pfaffian determinant gives $\Gamma[(a+b-1)/2]/2$. After multiplying this result with $(2\pi)^{k/2}$ originating from~\eqref{edge.origin} we arrive at~\eqref{lim.edge.AI}.

{\bf Normalisation of Class AII$^\dagger$:} Like before we start from
\begin{equation}
\begin{split}
\int_{\mathbb{R}^{2k\times 2k}}[dA]\exp\left[-\frac{1}{2}\Tr( A^\dagger A)^2 \right]=&\int_{\mathbb{R}^{2k\times 2k}}[dA]\exp\left[-\Tr( A^\dagger A) \right]\frac{\int_{\mathbb{R}^{2k\times 2k}}[dA]\exp\left[-\Tr( A^\dagger A)^2/2 \right]}{\int_{\mathbb{R}^{2k\times 2k}}[dA]\exp\left[-\Tr( A^\dagger A) \right]}\\
=&\pi^{2k^2}\frac{\int_{\mathbb{R}_+^{2k}}[dS]|\Delta_{2k}(S)|\det S^{-1/2}\exp\left[-\Tr S^2/2 \right]}{\int_{\mathbb{R}_+^{2k}}[dS]|\Delta_{2k}(S)|\det S^{-1/2}\exp\left[-\Tr S \right]},
\end{split}
\end{equation}
where the singular value decomposition is now $A=U\sqrt{S}V^\dagger$ with $U,V\in{\rm O}(2k)$ and $S={\rm diag}(s_1,\ldots,s_{2k})$. The integral in the denominator is another special case of the Selberg integral in~\cite[Eq.~(17.6.5)]{Mehta} and the numerator can be dealt with the identity~\cite{deBruijn},
\begin{equation}
|\Delta_{2k}(S)|=\Delta_{2k}(S){\rm Pf}[{\rm sign}(s_b-s_a)]_{a,b=1,\ldots,2k}
\end{equation}
with ${\rm sign}(s_b-s_a)=(s_b-s_a)/|s_b-s_a|$. Applying 
de Bruijn's integral identity~\cite{deBruijn} we arrive at
\begin{equation}
\begin{split}
&\int_{\mathbb{R}^{2k\times 2k}}[dA]\exp\left[-\frac{1}{2}\Tr( A^\dagger A)^2 \right]\\
=&\frac{\pi^{2k^2}(2k)!}{\prod_{j=0}^{2k-1}2\Gamma[(j+3)/2]\Gamma[(j+1)/2]/\sqrt{\pi}}{\rm Pf}\left[\int_{\mathbb{R}_+^2} ds_1ds_2\ {\rm sign}(s_2-s_1)s_1^{a-3/2}s_2^{b-3/2}e^{-(s_1^2+s_2^2)/2}\right]_{a,b=1}
^{2k}.
\end{split}
\end{equation}
The product of Gamma-functions can be simplified by Legendre's duplication formula by multiplying factors with odd and even $j$ together,
\begin{equation}
\prod_{j=0}^{2k-1}2\Gamma[(j+3)/2]\Gamma[(j+1)/2]/\sqrt{\pi}=\prod_{l=0}^{k-1}2^{-4l}(2l+2)!(2l)!
\end{equation}
 The integral in the Pfaffian is feasible when, first, ordering $s_1\geq s_2\geq0$ and, then, going over to polar coordinates $s_1+is_2=\sqrt{2r}e^{i\varphi}$ with $\varphi\in[0,\pi/4]$ to account for $s_1\geq s_2\geq0$. This gives
\begin{equation}
\begin{split}
&\int_{\mathbb{R}_+^2} ds_1ds_2\ {\rm sign}(s_2-s_1)s_1^{a-3/2}s_2^{b-3/2}e^{-(s_1^2+s_2^2)/2}\\
=&2^{(a+b-3)/2}\int_0^\infty \frac{dr}{r}\, r^{(a+b-1)/2}e^{-r}\ \int_0^{\pi/4}d\varphi \left[\cos^{b-3/2}(\varphi)\sin^{a-3/2}(\varphi)-\cos^{a-3/2}(\varphi)\sin^{b-3/2}(\varphi)\right].
\end{split}
\end{equation}
The integral over $r$ is anew the Gamma-function $\Gamma[(a+b-1)/2]$ while the integral over $\varphi$ yields essentially a hypergeometric function
\begin{equation}
\begin{split}
\int\limits_0^{\pi/4}d\varphi \cos^{b-3/2}(\varphi)\sin^{a-3/2}(\varphi)=&\frac{1}{2}\int_0^{1/2}dx(1-x)^{(2b-5)/2}x^{(2a-5)/4}\\
=&\frac{1}{2}\sum_{j=0}^\infty(-1)^j\binom{(2b-5)/4}{j}\frac{2^{-(2a-1)/4-j}}{(2a-1)/4+j}\\
=&\frac{2^{(5-2a)/4}}{2a-1}\,{ _2F_1}\left(\frac{2a-1}{4},\frac{5-2b}{4};\frac{2a+3}{4};\frac{1}{2}\right).
\end{split}
\end{equation}
We first substituted $x=\sin^2(\varphi)$ and then expanded $(1-x)^{(2b-5)/4}$ in $x$ leading to the binomial series which can be easily integrated.
The difference of this expression with $a$ and $b$ yields
\begin{equation}
\begin{split}
&\int_{\mathbb{R}^{2k\times 2k}}[dA]\exp\left[-\frac{1}{2}\Tr( A^\dagger A)^2 \right]=\frac{\pi^{2k^2}}{\prod_{l=0}^{k-1}2^{-4l}[(2l)!]^2}\\
&\hspace*{1cm}\times {\rm Pf}\left[2^{-1/4}\Gamma\left[\frac{a+b-1}{2}\right]\left(\frac{2^{b/2}}{2a-1}\,{ _2F_1}\left(\frac{2a-1}{4},\frac{5-2b}{4};\frac{2a+3}{4};\frac{1}{2}\right)-\{a\leftrightarrow b\}\right)\right]_{a,b=1}
^{2k}.
\end{split}
\end{equation}
When pulling the factor $2^{-1/4}$ out of the Pfaffian and multiplying it with $[\sqrt{4\pi}/\,{\rm arctanh}(1/\sqrt{2})]^k$ coming from~\eqref{edge.origin} we arrive at~\eqref{lim.edge.AII}.

%%%%%%%%%%%%%%%%%%%%%%%%%%%%%%%%%%%%%%%%%%%%
\section{Conclusions and open problems}\label{conclusio}

We provided analytical results for three ensembles of non-Hermitian random matrices, the complex Ginibre ensemble (Cartan class A), complex symmetric matrices (Cartan class AI$^\dag$) and complex self-dual random matrices (Cartan class AII$^\dag$).
Choosing
expectation values of $k$ pairs of conjugate characteristic polynomials as correlations functions, we proved that indeed all three ensembles lead to different local bulk and edge statistics for the complex eigenvalues. The explicit expressions for $k=1$ pair in class AI$^\dag$ and AII$^\dag$, see Proposition~\ref{Prop_char_exp},  show a remarkably similar structure to the well known respective answer for the complex Ginibre ensemble 
for finite-$N$, being given in terms of sums of truncated exponentials. This allowed us to derive asymptotic results in the local bulk and edge scaling limit, as well as for the global large-$N$ limit, with different results in all three limits for each ensemble, cf., Propositions~\ref{Prop_global} and~\ref{Prop_lim_char_exp} and Theorem~\ref{Thm_lim_char_exp}. 
In particular, the resulting properties at the edge are compatible with the expected behaviour of the spectral density based on an approximate Coulomb gas description with three different values of inverse temperature $\beta$.

For general $k$ we have identified the effective Lagrangians in the local bulk (Theorem~\ref{thm:bulk}) and in the edge scaling limit (Theorem~\ref{thm:edge}). In both limits the respective Lagrangians agree for all three classes. However, the integration domain differs for each ensemble. In the bulk it is given by the compact unitary, unitary symplectic and orthogonal group for class A, AI$^\dag$ and AII$^\dag$, respectively. At the edge, integrals over complex, quaternion and real valued matrices result. Only in class A these group integrals can be performed for $k>1$ and mapped to known results. Also here, from the known properties of these integrals it is clear that all three ensembles differ. We expect that these effective Lagrangians as well as Goldstone manifolds will be universal and should be shared with physical systems with the corresponding symmetries. 
The effective Lagrangians~\eqref{Lag.bulk} and~\eqref{Lag.edge} would then appear as the potential part of the corresponding non-linear $\sigma$-models.

Several open questions prevail for these three representatives of local bulk and edge statistics. Is it possible to access more detailed spectral information in the two classes AI$^\dag$ and AII$^\dag$, despite the lack of a joint eigenvalue density? 
Can we analytically prove universality in these two cases by deforming the Gaussian ensembles considered so far? 
More generally, is it possible to prove that all other known non-Hermitian symmetry classes fall into these three classes, away form their symmetry axes and symmetry points?
We 
plan 
to come back to some of these questions in future work.

\section*{Acknowledgments}
This work was partly funded by a Leverhulme Trust Visiting Professorship VP1-2023-007 (GA), by the 
German Research Foundation DFG with  grant SFB 1283/2 2021 -- 317210226 (GA, PP) 
and grant IRTG 2235 (NA), by the Australian Research Council via the
Discovery Project grant DP210102887 (MK), and by the Alexander-von-Humboldt Foundation (MK).
GA also acknowledges partial support through the Swedish Research Council grant 2016-06596 while visiting the Mittag-Leffler Institute during the programme Random Matrices and Scaling Limits 2024. 
He thanks the School of Mathematics at the University of Bristol for its kind hospitality during his sabbatical stay, as well as the Mittag-Leffler Institute, where part of these results were written up. We also thank Markus Ebke, Yan Fyodorov, Torben Kr\"uger and Nick Simm for fruitful discussions. MK is grateful for the hospitality of the Faculty of Physics at Bielefeld University and of the Centre for interdisciplinary Research  ZiF.

\begin{appendix}
%%%%%%%%%%%%%%%%%%%%%%%%%%%%%%%%%%%%%%%%%%%%%%%%%%%%%%%%%%%%%%%%%%%%%
\section{Differential equation for class AII$^\dag$ and the edge scaling limit}\label{App:fNdiff}

In this appendix, we derive a differential equation for the integral representation for the expected characteristic polynomials in class AII$^\dag$ at finite $N$, see Subsection~\ref{results-k1N},
\begin{equation}
D_N^{\rm AII^\dag}(z,w^*)\Big |_{x=zw^*}=\frac{1}{4^N}f_N(x),\qquad 
f_N(x)=\frac{1}{2}\int_0^\infty dv\ e^{-v}\int_{-v}^vdu
\left(4x^2+4xv+u^2\right)^{N},
\end{equation}
in the form \eqref{fNx-int1} which we need here.
We will also show how this could be employed to derive the edge scaling limit in this class.  

From the above integral we obtain for $x\neq0$ 
\begin{eqnarray}
\frac{\partial}{\partial x}f_N(x)&=& \frac{1}{2}\int_0^\infty dv\ e^{-v}\int_{-v}^vdu
N(8x +4v)\left(4x^2+4xv+u^2\right)^{N-1}
\nonumber\\
&=& \frac{1}{2}\int_0^\infty dv\ e^{-v}\int_{-v}^vdu
N\left(4x +\frac{4x^2+4vx+u^2}{x} - \frac{u^2}{x}\right)\left(4x^2+4xv+u^2\right)^{N-1}
\nonumber\\
&=& \frac{1}{2}\int_0^\infty dv\ e^{-v}\int_{-v}^vdu\left( \frac{\partial}{\partial v}+\frac{N}{x}-\frac{u}{2x}\frac{\partial}{\partial u}\right) \left(4x^2+4xv+u^2\right)^{N}.
\label{partialfN}
\end{eqnarray}
While the middle term immediately gives back $\frac{N}{x}f_N(x)$, the two terms with derivatives with respect to $v$ and $u$ can be simplified using integration by parts. 

For the third term differentiated with respect to $u$ we obtain
\begin{eqnarray}
&&\frac{1}{2}\int_0^\infty dv\ e^{-v}\left\{\left[-\frac{u}{2x}\left(4x^2+4xv+u^2\right)^{N}\right]_{-v}^v +\int_{-v}^vdu\frac{1}{2x}\left(4x^2+4xv+u^2\right)^{N}\right\}
\nonumber\\
&&=-\frac{1}{2x}\int_0^\infty dv\ e^{-v}v(2x+v)^{2N}+ \frac{1}{2x}f_N(x)
\nonumber\\
&&=-\frac{1}{2x}e^{2x}\int_0^\infty dv\ e^{-v-2x}\left((2x+v)^{2N+1}-2x(2x+v)^N\right)+ \frac{1}{2x}f_N(x)\nonumber\\
&&=-\frac{1}{2x}e^{2x}\left(\Gamma(2N+2,2x)-2x\Gamma(2N+1,2x)\right)+ \frac{1}{2x}f_N(x).
\label{partf1}
\end{eqnarray}
For the first term in \eqref{partialfN} we consider the following total derivative:
\begin{eqnarray}
&&\frac{d}{dv}\int_{-v}^vdu\ e^{-v}(4x^2+4xv+u^2)^N
\\
&&= 2e^{-v}(4x^2+4xv+v^2)^N-\int_{-v}^vdu\ e^{-v}(4x^2+4xv+u^2)^N+\int_{-v}^vdu\ e^{-v}\frac{\partial}{\partial v}(4x^2+4xv+u^2)^N.\nonumber
\end{eqnarray}
Integrating this expression over $v$ from zero to infinity, the boundary terms on the left hand side obviously vanish, and we arrive at
\begin{eqnarray}
0&=&\int_0^\infty dv\ e^{-v}(2x+v)^{2N} - f_N(x)+ \frac{1}{2}\int_0^\infty dv\int_{-v}^vdu\ e^{-v}\frac{\partial}{\partial v}(4x^2+4xv+u^2)^N\nonumber\\
&=& e^{2x}\Gamma(2N+1,2x)- f_N(x)+ \frac{1}{2}\int_0^\infty dv\int_{-v}^vdu\ e^{-v}\frac{\partial}{\partial v}(4x^2+4xv+u^2)^N.
\label{partf2}
\end{eqnarray}
Inserting this expression together with \eqref{partf1} into \eqref{partialfN} we obtain as a final result
\begin{equation}
\frac{\partial}{\partial x}f_N(x)= \left(1+\frac{N}{x}+\frac{1}{2x}\right) f_N(x)-\frac{1}{2x}e^{2x}\Gamma(2N+2,2x) .
\end{equation}
We note that two terms with the incomplete Gamma-function $\Gamma(2N+1,2x)$ cancel.

This can be translated into an equation for the  rescaled quantity
\begin{equation}
F_N(x)= \frac{1}{(2N)!}\ e^{-2x} f_N(x) = e^{-2zw^*}\frac{D_N^{\rm AII^\dag}(z,w^*)}{D_N^{\rm AII^\dag}(0,0)}\biggl|_{x=zw^*},
\label{FNdef}
\end{equation}
 see~\eqref{F-def}, in order to take the edge (and bulk) large-$N$ limit. It 
gives the following differential equation 
\begin{equation}
\frac{\partial}{\partial x}F_N(x)= \left(\frac{2N+1}{2x}-1\right) F_N(x)-\frac{1}{2x}\frac{\Gamma(2N+2,2x)}{\Gamma(2N+1)}.
\label{FdiffN}
\end{equation}

Let us briefly indicate how~\eqref{FdiffN} can be used to alternatively derive the edge scaling limit \eqref{zoom1}. Employing the scaling
\begin{equation}
x=N+\sqrt{N}y+\chi\eta^*/2,
\end{equation}
while keeping $y=(z_0^*\xi+z_0\eta^*)/\sqrt{2}$ fixed, the partial derivatives are related like
\begin{equation}
\frac{\partial}{\partial x}=\frac{1}{\sqrt{N}}\frac{\partial}{\partial y}.
\end{equation}
This implies for~\eqref{FdiffN} that
\begin{equation}
\frac{1}{\sqrt{N}}\frac{\partial}{\partial y}F_N(x)= \left[-\frac{y}{\sqrt{N}}+O(N^{-1})\right] F_N(x)
-\left[1+O(N^{-1/2})\right]\frac{\Gamma(2N+2,2x)}{\Gamma(2N+2)}.
\label{diffFN}
\end{equation}
Due to~\eqref{def-Q} and its asymptotic~\eqref{Q-lim} we have 
\begin{equation}
\frac{\Gamma(2N+2,2x)}{\Gamma(2N+2)}=Q(2N+2,2x)=\frac12\erfc(y)+O(N^{-1/2}). 
\end{equation}
From this we deduce that  the limit $F(y)=\lim_{N\to\infty}F_N(x)/\sqrt{2N}$ satisfies the differential equation
\begin{equation}
\frac{\partial}{\partial y}F(y)= -yF(y)-\frac{1}{\sqrt{8}}\erfc(y).
\label{diffF}
\end{equation}
The solution of the homogeneous solution $F_h(y)$ is  given by 
\begin{equation}
F_h(y)=Ce^{-y^2/2}.
\label{Fhom}
\end{equation}
In order to find a special solution $F_s(y)$ to the inhomogeneous equation \eqref{diffF}, we vary the integration constant, setting  $F_s(y)=g(y/\sqrt{2})e^{-y^2/2}$ which reduces~\eqref{diffF} to
\begin{equation}
g'(v)=-\frac12\erfc(\sqrt{2}\,v)\, e^{+v^2}\qquad{\rm with}\quad v=y/\sqrt{2}.
\end{equation}
It is identical to \eqref{gdiff} and further motivates why we have introduced this function. From here onwards we can proceed as in Subsection~\ref{edge-lim} to arrive at~\eqref{Dedge-QS}. The bulk scaling limit can be derived from~\eqref{diffFN} in a similar fashion.

%%%%%%%%%%%%%%%%%%%%%%%%%%%%%%%%%%%%%%%%%%%%%%%%%%%%%%%%%
\section{Relation to the orthogonal polynomial approach in class A}\label{App:Ak>1}

In this appendix, we make contact with an earlier result for the expectation value of products of characteristic polynomials and their complex conjugates derived in~\cite{AV03,determinant}. It is valid for any determinantal point process of orthogonal polynomial type, having a joint density of complex (or real) eigenvalues that includes the modulus square of the Vandermonde determinant, see~\eqref{Vand}. A special case of the theorem proven in \cite{AV03}, when the number $k$ of factors and conjugate factors agrees, is given by
\begin{equation}
\widetilde D_{N,k}^{\rm A}(Z,W^*):=\left\langle \prod_{j=1}^k\det[z_j\mathbf{1}_N-J]\det[w_j^*\mathbf{1}_N-J^*]\right\rangle
= \prod_{i=N}^{N+k-1}h_i\frac{\det_{j,l=1}^k[{K}_{N+k}(z_j,w_l^*)]}{\Delta_k(Z)\Delta_k(W^*)},
\label{AVthm}
\end{equation}
with $W=\mbox{diag}(w_1,\ldots,w_k)$. Here, the expectation value defined in~\eqref{vevdef} is taken with respect to class A, cf. eq.~\eqref{ZA}. 
The $h_j$ are the squared norms of the orthogonal polynomials $P_j(z)=z^j+\mathcal{O}(z^{j-1})$ in monic normalisation, and 
\begin{equation}
{K}_{N}(z,w^*)=\sum_{j=0}^{N-1}\frac{1}{h_j}P_j(z)P_j(w^*)
\label{polyKdef}
\end{equation}
 is the polynomial part of the kernel (also known as pre-kernel). In the particular case of the complex Ginibre ensemble we have
 \begin{equation}
 {K}_{N}(z,w^*)=E_{N-1}(zw^*), \quad P_j(z)=z^j, \quad h_j=j!
 \label{polyKGin}
 \end{equation}
In particular, setting $k=1$ in~\eqref{AVthm} for this case we obtain~\eqref{K-DN-A}. 

In Section~\ref{k>1}, explicit expressions in terms of $k$-dimensional group integrals were given for such expectation values for $k>1$ for all three ensembles, in the bulk and at the edge. For class A these group integrals can be solved, with the resulting expressions~\eqref{lim-bulk-Gin-det} and~\eqref{lim.edge.3Gin} to be rederived here. 

The asymptotic of the kernel in~\eqref{AVthm} in the bulk and at the edge were already discussed in Subsection~\ref{Ginrev} so that we can be very brief. In line with Section~\ref{k>1}, we recall the scaling of the $k$ arguments, 
\begin{eqnarray}
Z&=&\sqrt{N}z_0\mathbf{1}_k+\chi,\quad \chi=\mbox{diag}(\chi_1,\ldots, \chi_k),
\nonumber\\
W&=&\sqrt{N}z_0\mathbf{1}_k+\eta,\quad \ \eta=\mbox{diag}(\eta_1,\ldots, \eta_k).
\label{ZWscale}
\end{eqnarray}
Due to the translational invariance of the Vandermonde determinant~\eqref{Vand}, it immediately follows that $\Delta_k(Z)=\Delta_k(\chi)$, and likewise for $\Delta_k(W^*)=\Delta_k(\eta^*)$.

In bulk scaling for $|z_0|<1$, we insert~\eqref{Gin-bulk-kernel} into~\eqref{AVthm} to arrive at 
\begin{equation}
\widetilde D_{N,k}^{\rm A}(Z,W^*)\Big|_{|z_0|<1}\sim \left(\prod_{j=N}^{N+k-1}j!\right)
\exp\left[
{kN|z_0|^2+ \sqrt{N}\sum_{j=1}^k(z_0^*\chi_j+z_0\eta_j^*)}
\right]
\frac{\det_{i,j=1}^k\left[\exp[\chi_i\eta_j^*]\right]}{\Delta_k(\chi)\Delta_k(\eta^*)}
\label{Dkbulk-det}
\end{equation}
This has to be normalised by the product of the individual expectation values, for which it immediately follows
\begin{equation}
\prod_{j=1}^k\widetilde D_{N,1}^{\rm A}(\sqrt{N}z_0+\chi_j,\sqrt{N}z_0^*+\eta_j^*)\Big|_{|z_0|<1}\sim (N!)^k
\exp\left[\sum_{j=1}^k\left(N|z_0|^2+ \sqrt{N}(z_0^*\chi_j+z_0\eta_j^*)+\chi_j\eta_j^*\right)\right].
\label{Dkbulk-prod}
\end{equation}
After taking the ratio~\eqref{hatDNk} with $D_{N,k}^{\rm A}(Z,W^*)=\widetilde D_{N,k}^{\rm A}(Z,W^*)/\widetilde D_{N,k}^{\rm A}(z_0\mathbf{1}_k,z_0^*\mathbf{1}_k)$ all factorials and
most of the exponential pre-factors  cancel, leading to the result~\eqref{lim-bulk-Gin-det} in Section~\ref{k>1}.
 
When turning to the edge scaling limit, with $|z_0|=1$, the only difference resulting from the limiting kernel \eqref{Gin-edge-kernel} is the extra factor of the complementary error function,
\begin{eqnarray}
\widetilde D_{N,k}^{\rm A}(Z,W^*)\Big|_{|z_0|=1}&\sim& \left(\prod_{j=N}^{N+k-1}j!\right)
\exp\left[
{kN|z_0|^2+ \sqrt{N}\sum_{j=1}^k(z_0^*\chi_j+z_0\eta_j^*)}
\right]
\nonumber\\
&&\times
\frac{\det_{i,j=1}^k
\left[\exp[\chi_i\eta_j^*]
\erfc\left(\frac{1}{\sqrt{2}}(z_0^*\chi_i+z_0\eta_j^*)\right)
\right]}{\Delta_k(\chi)\Delta_k(\eta^*)}.
\label{Dkegde-det}
\end{eqnarray}
In comparison, the normalising product of the expectation values for $k=1$ are
\begin{eqnarray}
\prod_{j=1}^k\widetilde{D}_{N,1}^{\rm A}(\sqrt{N}z_0+\chi_j,\sqrt{N}z_0^*+\eta_j^*)\Big|_{|z_0|=1}&\sim& (N!)^k
\exp\left[kN|z_0|^2+ \sqrt{N}\sum_{j=1}^k(z_0^*\chi_j+z_0\eta_j^*)\right]
\nonumber\\
&&\times e^{\sum_{j=1}^k\chi_j\eta_j^*}
\prod_{l=1}^k\erfc\left(\frac{1}{\sqrt{2}}(z_0^*\chi_l+z_0\eta_l^*)\right).
\nonumber\\
\label{Dkegde-prod}
\end{eqnarray}
 Once again the factors in the first line cancel those in~\eqref{Dkegde-det} when taking the ratio. This agrees with~\eqref{edge-Gin-simp} in Section~\ref{k>1}.

\end{appendix}

%%%%%%%%%%%%%%%%%%%%%%%%%%%%%%%%%%%%%%%%%%%%%%%%%%%%%%

\end{document}